\documentclass{article}
\usepackage{fullpage}
\usepackage[table,xcdraw]{xcolor}
\usepackage{amsmath,amssymb}
\usepackage{subcaption}
\usepackage{hyperref}
\usepackage{multirow}
\usepackage{graphicx}

\newcommand{\la}{\lambda}
\newcommand{\ub}{\bar{u}}

\title{Dynamic soliton-mean flow interaction with nonconvex flux}

\author{Kiera van der Sande$^1$, Gennady A.~El$^2$, and Mark
  A.~Hoefer$^1$\thanks{Email address for correspondence:  hoefer@colorado.edu}}

\date{\normalsize $^1$Department of Applied Mathematics, University of
  Colorado, Boulder, CO 80309, USA, \\ $^2$Department of Mathematics,
  Physics and Electrical Engineering, Northumbria University,
  Newcastle upon Tyne NE1 8ST, UK}

\begin{document}

\maketitle

\begin{abstract}
  The interaction of localised solitary waves with large-scale,
  time-varying dispersive mean flows subject to nonconvex flux is
  studied in the framework of the modified Korteweg-de Vries (mKdV)
  equation, a canonical model for nonlinear internal gravity wave
  propagation in stratified fluids.  The principal feature of the
  studied interaction is that both the solitary wave and the
  large-scale mean flow---a rarefaction wave or a dispersive shock
  wave (undular bore)---are described by the same dispersive
  hydrodynamic equation.  A recent theoretical and experimental study
  of this new type of dynamic soliton-mean flow interaction has
  revealed two main scenarios when the solitary wave either tunnels
  through the varying mean flow that connects two constant asymptotic
  states, or remains trapped inside it.  While the previous work
  considered convex systems, in this paper it is demonstrated that the
  presence of a nonconvex hydrodynamic flux introduces significant
  modifications to the scenarios for transmission and trapping. A
  reduced set of Whitham modulation equations, termed the solitonic
  modulation system, is used to formulate a general, approximate
  mathematical framework for solitary wave-mean flow interaction with
  nonconvex flux. Solitary wave trapping is conveniently stated in
  terms of crossing characteristics for the solitonic
  system. Numerical simulations of the mKdV equation agree with the
  predictions of modulation theory.  The developed theory draws upon
  general properties of dispersive hydrodynamic partial differential
  equations, not on the complete integrability of the mKdV equation.
  As such, the mathematical framework developed here enables
  application to other fluid dynamic contexts subject to nonconvex
  flux.
\end{abstract}

\section{Introduction}

\begin{figure}
\centering		
\includegraphics[width=.6 \textwidth]{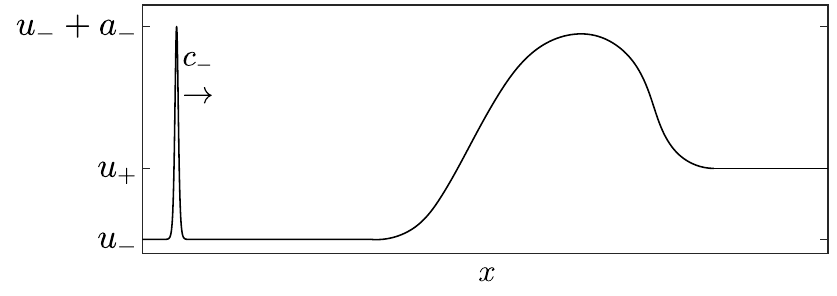}
\caption{Representative initial configuration for soliton-mean flow
  interaction.  The narrow soliton with amplitude $a_-$ on the uniform
  mean flow $\overline{u}_-$ transmits through the broad hydrodynamic
  flow if it reaches the uniform mean flow $\overline{u}_+$, otherwise
  it experiences trapping inside the mean flow. The mean flow
  generally exhibits expansion and compression waves.}
\label{fig:soli_hydro_config}
\end{figure}

The interaction of dispersive waves with slowly varying mean flows is
a fundamental and canonical problem of fluid mechanics with important
applications in geophysical fluid dynamics (see,
e.g.~\cite{mei_theory_2005,buhler_waves_2009} and references
therein). This multiscale problem is relevant for linear or weakly
nonlinear wavepackets and large amplitude solitons---in this work, we
do not distinguish between solitary waves and solitons. Traditionally,
the mean flow involved in the interaction is either prescribed
externally, e.g.~an external current, or is induced by amplitude
modulations of a nonlinear wave. A different class of wave-mean flow
interactions has recently been identified in
\cite{maiden_solitonic_2018}, where both the dynamic mean flow and the
propagating localised soliton are described by the same dispersive
hydrodynamic equation, a canonical example being the Korteweg-de Vries
(KdV) equation. However, the evolution of the field $u(x,t)$ occurs on two
well separated spatiotemporal scales, allowing for the distinct
identification of waves and mean flows. A prototypical configuration
of this (Fig.~\ref{fig:soli_hydro_config}) is the propagation of a
soliton through a dynamically evolving macroscopic flow, characterised
by different asymptotic states $u \to u_{\pm}$ as $x \to \pm \infty$.
We refer to such nonlinear wave interactions as \textit{soliton-mean
  flow interactions}.  The simplest mean flows are initiated by a
monotone transition or step between $u_-$ and $u_+$, which
asymptotically develops into either a rarefaction wave (RW) or a highly
oscillatory dispersive shock wave (DSW)
\cite{gurevich_nonstationary_1974, el_dispersive_2016}.  While the
former is slowly varying, the use of the term ``mean flow'' for the
latter implies some averaging over rapid oscillations.  We shall refer
to the step problem for dispersive hydrodynamics as a dispersive
Riemann problem.

Depending upon its initial position and amplitude, the soliton may
``tunnel'' or transmit through the large scale, expanding mean flow;
otherwise, it remains trapped within the mean flow.  Recent work has
investigated the interaction between solitons and mean flows
resulting from the evolution of an initial step.  Both fluid conduit
experiments and the theory for a rather general, single dispersive
hydrodynamic conservation law were described in
\cite{maiden_solitonic_2018}. A generalisation of soliton-mean flow
interaction to the bidirectional case for a pair of conservation laws
described by the defocusing NLS equation was explored in
\cite{sprenger_hydrodynamic_2018}. Soliton-mean flow interaction in
the focusing NLS equation was investigated in
\cite{biondini_nonlinear_2019}.  A similar problem involving the
interaction of linear wavepackets with mean flows arising from the
step problem in the KdV equation was studied using an analogous
modulation theory framework in \cite{congy_interaction_2019}.  Aside
from the focusing NLS case, for which mean flow evolution is described
by an elliptic system of equations, and the present work, the models
previously investigated in the context of soliton-mean flow
interaction were limited to dispersive conservation laws with
hyperbolic convex flux.

The focus of this work is the study of soliton-mean flow interaction
when the governing dispersive hydrodynamics exhibit a {\it nonconvex
  hydrodynamic flux}.  As we show, the presence of nonconvex flux,
e.g.~the cubic flux in the modified KdV (mKdV) equation or related
Gardner equation, introduces significant modifications to the
transmission and trapping scenarios realised in the KdV case.  First
of all, due to the nonconvex flux, the mKdV equation supports a much
broader family of solitons and large-scale mean flow solutions than
the KdV equation. In particular, it exhibits localised solutions in
the form of exponentially decaying solitons of both polarities and,
depending on the dispersion sign, kinks and algebraic solitons. The
mKdV nonconvex mean flow features include undercompressive DSWs (an
alternative interpretation of kinks), contact DSWs (CDSWs) and
compound two-wave structures \cite{kamchatnov_undular_2012,
  kamchatnov_transcritical_2013, el_dispersive_2017}. Here, we
investigate how the solution features that arise due to nonconvex flux
affect soliton-mean flow interactions. In particular, we show that
soliton transmission for the defocusing mKdV equation can be
accompanied by a soliton polarity change.  In the focusing case, there
is a soliton-mean flow interaction in which an exponential soliton is
asymptotically transformed into a trapped algebraic soliton. These are
just two examples of the rich catalogue of soliton-mean flow
interactions we describe in this paper.

Key to the study of soliton-mean flow interaction is scale separation,
whereby the characteristic length and time scales of the propagating
soliton are much shorter than those of the mean flow.  The rapidly
oscillating structure of dispersive hydrodynamic flows motivates the
use of multiscale asymptotic methods.  Here, we will make extensive
use of one such method known as Whitham modulation theory
\cite{whitham_linear_1974}, which is based on a projection of the
scalar dispersive hydrodynamics onto a three-parameter family of
slowly varying periodic travelling wave solutions to the governing
equation.  The projection is achieved, equivalently, by averaging
conservation laws, an averaged variational principle, or multiple
scales perturbation methods.  The dispersive hydrodynamics are then
approximately described by a system of three first order quasilinear
partial differential equations (PDEs)---the Whitham modulation
equations---for the periodic travelling wave's parameters such as the
wave amplitude, the wavenumber and the period-mean.  Within the
framework of Whitham modulation theory, the original dispersive
Riemann problem is posed as a special Riemann problem, sometimes
called the Gurevich-Pitaevskii (GP) problem
\cite{gurevich_nonstationary_1974}, for the modulation equations
subject to piecewise constant initial data with a single discontinuity
at the origin.  Continuous, self-similar solutions of the GP problem
describe RW and DSW mean flow modulations.

Classical DSW modulation theory has been developed for the KdV
equation \cite{gurevich_nonstationary_1974} and other ``KdV-like''
equations, both integrable and non-integrable
\cite{el_resolution_2005, el_dispersive_2016}.  It is useful to
identify this class of KdV-type equations, or \textit{classical,
  convex} dispersive hydrodynamic equations, as those equations whose
associated Whitham modulation equations are strictly hyperbolic and
genuinely nonlinear.  In this case, the generic solution of the GP
problem is either a DSW or a RW.  More broadly, even nonconvex
equations such as mKdV can exhibit convex dispersive hydrodynamics for
a restricted subset of modulation parameters in which the Whitham
modulation equations remain strictly hyperbolic and genuinely
nonlinear.  Therefore, we shall call the DSWs generated within the
framework of convex dispersive hydrodynamics {\it convex DSWs}.

It was shown in \cite{maiden_solitonic_2018} that the interaction of a
soliton with a RW is described by an exact, soliton limit reduction of
the Whitham modulation system, which we call the {\it solitonic
  modulation system}.  Two integrals or adiabatic invariants of the
solitonic modulation system were identified that determine the
amplitude and phase shift of the soliton when transmitted through the
variable mean flow.  The nonexistence of a transmitted soliton (zero
or negative transmitted amplitude) signifies soliton trapping within
the mean flow. The soliton-DSW transmission/trapping conditions were
shown to be equivalent to those for the soliton-RW interaction by the
fundamental property of hydrodynamic reciprocity of the modulation
solution, which is related to time reversibility of the original
dispersive hydrodynamics.

In this paper, we investigate the effects of a flux's nonconvexity on
the transmission and phase conditions.  One of the main, general
outcomes of our work is the identification of the condition for
soliton trapping with the coalescence of characteristics of the
solitonic modulation system, a signature of nonstrict hyperbolicity.
Analysis of the solitonic modulation system for the mKdV equation
shows that, in contrast to the convex case, the characteristic
coalescence and, consequently, soliton trapping can occur even for
nonzero soliton amplitude. This new type of soliton trapping is
accompanied by the asymptotic transformation of a conventional,
exponentially decaying soliton into an algebraic soliton of the mKdV
equation. Another notable effect is the dynamic reversal of soliton
polarity upon its transmission through a kink mean flow, which
resembles but is different from the well-known soliton polarity
reversal due to KdV soliton propagation in a variable medium caused
for example, by internal wave propagation through variable fluid
stratification and/or variable depth.  Upon passage through a critical
point where the coefficient for the quadratic flux vanishes, nonconvex
mKdV/Gardner dynamics emerge \cite{shroyer_observations_2008,
  li_observation_2015}.  Modulation theory predicts a zero---or more
accurately, vanishing in the zero dispersion limit---phase shift of
soliton transmission through a nonconvex mean flow such as a kink or
contact DSW.  Although all concrete calculations and numerical
simulations are performed for the mKdV equation, the developed
solitonic modulation system framework for the analysis of nonconvex
soliton-mean flow interactions is general and can be applied to other
dispersive hydrodynamic equations with nonconvex flux, both integrable
and non-integrable.
 
Perhaps the most prominent application of the present work is to
internal waves in the ocean and atmosphere where solitons are known to
arise and may interact with large scale mean flows modelled by the
unidirectional mKdV and Gardner equations
\cite{kakutani_solitary_1978, holloway_generalized_1999,
  grimshaw_internal_2002, helfrich_long_2006, apel_internal_2007}.  A
more sophisticated nonconvex model describing long-wave
potential-vorticity dynamics of coastal fronts was recently introduced
in \cite{jamshidi_long-wave_2020}.  Fully nonlinear, bidirectional
internal waves are described by nonconvex dispersive models such as
the Miyata-Camassa-Choi system
\cite{miyata_internal_1985,choi_fully_1999} for fully nonlinear
internal waves.  Nonconvex dispersive dynamics modelled by the mKdV
and Gardner equations also occur in the physics of multicomponent
superfluids \cite{ivanov_solution_2017} and collisionless plasma
\cite{chanteur_formation_1987, ruderman_dynamics_2008}.

The structure of this paper is as follows. In Section
\ref{sec:disp_hyd} we introduce the notion and nomenclature of
dispersive hydrodynamics, and provide a review of the effects of
nonconvex hydrodynamic flux on mKdV solutions.  In Section
\ref{sec:soliton-mean-flow}, we detail the general modulation theory
framework for soliton-mean flow interaction problems, originally
introduced in \cite{maiden_solitonic_2018}, and extend it to the case
of nonconvex flux. Then, in Section \ref{sec:mKdV_trav_mod}, we narrow
our focus to the modulation description of mKdV dispersive
hydrodynamics and in Section \ref{sec:class_mkdv} to the
classification of mean flows realised in the mKdV regularisation of
Riemann step data. In the next Section
\ref{sec:mKdVsolitonichydrodynamics}, we formulate the soliton-mean
flow problem for mKdV and determine admissibility conditions for
soliton transmission (tunnelling) through the mean flow. We then
partition our classification of mKdV soliton-mean flow interaction
into soliton-convex mean flow interactions (Section
\ref{sec:soliton-convex}), soliton-nonconvex mean flow interactions
(Section \ref{sec:soliton-nonconvex}), and the special case of
kink-mean flow interactions (Section \ref{sec:kink-mean}).  Finally,
we generalise our analysis from convex mean flows generated by the GP
problem to a much broader class of convex mean flows generated from
slowly varying initial conditions. Throughout this paper, we confirm
the predictions of our asymptotic analysis using numerical experiments
with a spectral integrating factor Fourier method (described in
Appendix B of \cite{el_dispersive_2017}) in space and fourth order
Runge-Kutta time integration. Discussion, conclusions and future
outlooks are given in Section \ref{sec:conclusion}.

\section{Nonconvex dispersive hydrodynamics}
\label{sec:disp_hyd}

Dispersive hydrodynamics are modeled by hyperbolic conservation laws
modified by dispersive terms \cite{el_dispersive_2016}.  We express a
single one-dimensional dispersive hydrodynamic conservation law in the
general form
\begin{equation}
  u_t + f(u)_x = D[u]_x, \label{eqn:scalarconservationlaw}
\end{equation}
where $f(u) \in C^2(\mathbb{R})$ is the hydrodynamic (or hyperbolic) flux function.
The term $D[u]$ is an integro-differential operator acting on $u(x,t)$
that gives rise to a real-valued linear dispersion relation
\begin{equation}
  \label{eq:1}
  \omega_0(k,\overline{u}) = f'(\overline{u})k + \Omega(k,\overline{u}) ,
  \quad k \in \mathbb{R} 
\end{equation}
for vanishingly small amplitude travelling wave solutions
$\propto e^{i(kx-\omega_0 t)}$ of the PDE
(\ref{eqn:scalarconservationlaw}) linearized about the constant
solution $u(x,t) = \overline{u} \in \mathbb{R}$.  We assume that
$\Omega(k,\overline{u}) = o(k)$ as $k \to 0$ and
$\Omega_{kk}(k,\overline{u})$ is not identically zero in order to
separate the long-wave hydrodynamic flux from short-wave dispersive
effects.  The field of dispersive hydrodynamics encompasses multiscale
nonlinear wave solutions of initial and boundary value problems for
eq.~\eqref{eqn:scalarconservationlaw} (possibly with perturbations) in
which at least two length and time scales play a prominent role: the
oscillatory scale (e.g., the width of a soliton or the
wavelength/period of a periodic travelling wave) and a longer,
hydrodynamic scale (e.g., the slowly varying oscillatory amplitude of
a wavepacket or DSW).  One canonical dispersive hydrodynamic problem
for eq.~\eqref{eqn:scalarconservationlaw} is the so-called
Gurevich-Pitaevskii (GP) problem \cite{gurevich_nonstationary_1974} in
which $u(x,0)$ for $x \in \mathbb{R}$ exhibits a sharp, monotone
transition between two distinct far-field boundary conditions.  The
solution of the GP problem then describes the long-time asymptotic
behaviour for more general initial data with distinct far-field
equilibrium states.

When $f''(u)$ in \eqref{eqn:scalarconservationlaw} is sign definite,
we say that the hydrodynamic flux---or just ``flux'' for short---is
convex, not distinguishing between convex and concave associated with
different signs.  Similarly, when $\Omega_{kk}(k,\overline{u})$ in
\eqref{eq:1} is sign definite for $k > 0$, we say that the dispersion
is convex.  A necessary condition for
\eqref{eqn:scalarconservationlaw} to be a classical, convex dispersive
hydrodynamic equation is the convexity of both the flux and the
dispersion \cite{el_dispersive_2016}.  Consequently, when $f''$ or
$\Omega_{kk}$ are sign indefinite, the dispersive hydrodynamics are
nonconvex.  In this paper, we focus on the nonconvex flux case and
assume convex dispersion throughout.

Nonconvexity is known to introduce new types of dispersive
hydrodynamic solutions. The simplest generic model of dispersive
hydrodynamics with nonconvex flux is the modified Korteweg-de-Vries
(mKdV) equation
\begin{equation}
  u_t + (u^3)_x = \mu u_{xxx}  \label{eqn:mKdV}.
\end{equation}
The mKdV equation with $\mu > 0$ is often referred to as defocusing
and with $\mu <0$ as focusing.  The review \cite{el_dispersive_2017}
presents a full classification of mKdV solutions to the GP problem
associated to the Riemann initial data
\begin{equation}
  \label{eq:2}
  u(x,0) =
  \begin{cases}
    u_- & x < 0 \\ u_+ & x > 0
  \end{cases}
\end{equation}
for both signs of $\mu$.  Due to its cubic flux, the mKdV Riemann
problem exhibits nonclassical solutions that were first studied in
\cite{chanteur_formation_1987, kamchatnov_dissipationless_2004-2,
  marchant_undular_2008, leach_initial-value_2012,
  leach_initial-value_2013}.  The full classification was carried out
in \cite{kamchatnov_undular_2012, kamchatnov_transcritical_2013}
within the framework of the Gardner equation that combines the
quadratic and cubic fluxes of the KdV and mKdV equations,
respectively. The classification is presented in Section
\eqref{sec:class_mkdv}; see Fig.~\ref{fig:mKdVRiemannsolns}.

The complete integrability of the mKdV equation was utilised in
\cite{kotlyarov_riemannhilbert_2010,kotlyarov_step-initial_2012,grava_long_2019}
to obtain detailed asymptotics of exact solutions to \eqref{eqn:mKdV}
with $\mu < 0$ and step-like initial data for far-field asymptotic
states satisfying $u_-> u_+ > 0$.  The method of analysis is the
formulation and asymptotic solution of a Riemann-Hilbert problem
within the inverse scattering transform formalism.

The properties of the mKdV equation for $\mu>0$ and $\mu<0$ are very
different with respect to the evolution of Riemann data \eqref{eq:2}.
In addition to convex DSWs and RWs exhibited by both mKdV
incarnations, there are new types of nonclassical, nonconvex solutions that do
not exist for convex dispersive hydrodynamic equations and depend on
the sign of $\mu$.  These features occur for initial steps satisfying
$u_- u_+<0$, i.e. when the initial data include the inflection point
$u = 0$ of the cubic flux $f(u) = u^3$. When $\mu > 0$, monotonic,
heteroclinic travelling wave solutions, commonly known as kinks, were
identified in \cite{kluwick_near-critical_2007} as undercompressive
DSWs analogous to discontinuous, undercompressive shock wave solutions
in conservation law theory that do not satisfy the Lax entropy
condition \cite{lefloch_hyperbolic_2002}.  The solutions of the mKdV
Riemann problem involving kinks were analyzed in
\cite{chanteur_formation_1987} using the inverse scattering transform
and in \cite{leach_initial-value_2012} using matched asymptotic
expansions.

When $\mu < 0$, a family of contact DSWs (CDSWs) exist whose
modulation solution coincides with a nonstrictly hyperbolic double
characteristic of the Whitham modulation system.  Contact DSWs are
analogous to contact discontinuities in conservation law theory that
propagate with characteristic velocity
\cite{dafermos_hyperbolic_2016}.  Contact DSWs were first described in
\cite{marchant_undular_2008} as ``sinusoidal undular bores'', then
later as trigonometric bores which were studied in
\cite{leach_initial-value_2013} using matched asymptotic expansions.

While convex DSWs are a continuous, two-parameter family of solutions
to the GP problem depending on both $(u_-,u_+)$, undercompressive and
contact DSWs are a continuous, one-parameter family of solutions.  For mKdV
(\ref{eqn:mKdV}), the undercompressive and CDSWs exhibit the
additional restriction $u_+ = -u_-$.  As a result, undercompressive
and contact DSWs resulting from the GP problem are typically
accompanied by a convex RW or DSW in the form of a double wave
structure.  Representative numerical simulations for each type of
solution to the mKdV GP problem are shown in
Fig.~\ref{fig:mKdVRiemannsolns}.  In the context of soliton
interaction with dispersive hydrodynamic structures, we shall refer to
solutions of the GP problem generally as \textit{mean flows}.  The DSW
modulations in this context are further specified as \textit{DSW mean
  flows}.

In addition to the already mentioned Gadner equation, which can be
reduced to mKdV by a simple transformation, there are a number of
dispersive PDEs modelling bidirectional flows in the nonconvex regime.
These manifest as a coupled pair of dispersive hydrodynamic
conservation laws whose dispersionless limit is not genuinely
nonlinear for a subset of field values.  Prominent examples of
nonconvex flux for bidirectional flows include the Landau-Lifshitz
equation modelling two-component Bose-Einstein condensates and thin,
easy-plane ferromagnetic materials \cite{ivanov_solution_2017} and the
Miyata-Camassa-Choi equations modelling two-layer, stratified fluids
\cite{esler_dispersive_2011}.  The GP problem for these equations
includes double waves involving contact DSWs (Landau-Lifshitz
equation) and undercompressive DSWs (Miyata-Camassa-Choi equation).
Nonconvex mean flows have also been studied for variants of the
nonlinear Schr\"odinger (NLS) equation including the defocusing
complex mKdV equation \cite{kodama_whitham_2009}, the derivative NLS
equation \cite{kamchatnov_evolution_2018}, the discrete NLS equation
from which mKdV can be derived
\cite{kamchatnov_dissipationless_2004-2}, and NLS with self-steepening
terms \cite{ivanov2017_riemann,ivanov2020_chenleeliu}, all of which
include contact DSW solutions.

\section{Modulation Theory for Soliton-Mean Interaction}
\label{sec:soliton-mean-flow}

We now review the general approach to the mathematical study of
soliton-mean interaction via Whitham modulation theory
\cite{whitham_linear_1974}. This approach, termed solitonic dispersive
hydrodynamics, was introduced in \cite{maiden_solitonic_2018}. The key
feature of solitonic dispersive hydrodynamics is that both the soliton
and the mean flow are described by the same equation, albeit the
variations of the wave field occur on disparate scales.  This is in
sharp contrast with the traditional wave-mean flow interaction setting
where the mean flow is prescribed externally; see,
e.g.~\cite{buhler_waves_2009}. We shall first follow the general
description introduced in \cite{maiden_solitonic_2018} for convex
systems and then consider the implications of a nonconvex flux, not
explored previously.

\subsection{Solitonic modulation system}
\label{sec:sol_mod}

The analytical description of solitonic dispersive hydrodynamics is
based on considering the soliton reduction of the Whitham modulation
equations.  Having the mKdV equation in mind, we present the general
theory for the unidirectional, scalar case.

For a periodic travelling wave solution parametrised by three
independent constants (as in the case of KdV or mKdV equations, third
order PDEs), the modulation equations can be written in terms of the
physical wave parameters: the mean flow $\bar{u}$, the amplitude $a$,
and the wavenumber $k$. Allowing $\bar{u}$, $a$, $k$ to be slow
functions of $x,t$, the requirement for the modulated periodic wave to
be an asymptotic solution to the dispersive hydrodynamic equation
\eqref{eqn:scalarconservationlaw} results in the quasilinear
modulation system,
\begin{equation}\label{general_whitham}
  {\bf u}_t + \mathrm{A} ({\bf u})  {\bf u}_x =0,\end{equation}
where ${\bf u}= ( \bar{u}, a, k)^T$  and $\mathrm{A}( {\bf u}) $ is a
$3 \times 3$ modulation matrix.   We call the dispersive hydrodynamics
convex if the associated Whitham modulation system
\eqref{general_whitham} is strictly hyperbolic and genuinely
nonlinear. If at least one of these conditions is violated, the system
is nonconvex. Strict hyperbolicity requires that the eigenvalues 
$v_i ({\bf u}), \ i=1,2,3$ of matrix $\mathrm{A}( {\bf u})$ are real
and distinct, $v_1 <v_2 <v_3$ for all admissible $\bf u$ in the
admissible set
\begin{equation}
  \label{eq:13}
  {\bf u} \in
  \mathcal{A} = \left \{ (\bar{u},a,k)^T ~ | ~ \bar{u} \in \mathbb{R}, ~
    a > 0, ~ k \in \mathbb{R} \setminus \{ 0 \} \right \} .
\end{equation}
The genuine nonlinearity condition then reads
$\nabla_{\bf u} v_i \cdot {\bf r}_i \ne 0, \ i=1,2,3$ for all
${\bf u} \in \mathcal{A}$, where ${\bf r}_i({\bf u})$ is the right
eigenvector corresponding to the eigenvalue $v_i$
\cite{lax_hyperbolic_1973}. If the system is nonstrictly hyperbolic
(the eigenvectors ${\bf r}_i$ span $\mathbb{R}^3$ but multiple
eigenvalues are admissible), then it is not genuinely nonlinear either
\cite{dafermos_hyperbolic_2016}. The converse is generally not
true. Nevertheless, a nonconvex system can exhibit convex properties
in a restricted domain of ${\bf u} \in \mathcal{D} \subset \mathcal{A}$.

The KdV-Whitham modulation system \eqref{general_whitham} is strictly
hyperbolic and genuinely nonlinear for all admissible
${\bf u} \in \mathcal{A}$ \cite{levermore_hyperbolic_1988}, while for
the mKdV equation, the properties of strict hyperbolicity and genuine
nonlinearity depend on the sign of $\mu$ and on the range of initial
Riemann data \eqref{eq:2} \cite{el_dispersive_2017}.

An important ingredient for modulation theory is the equation for $k$
in \eqref{general_whitham}. It has the universal form 
\begin{equation}\label{wave_cons1}
k_t+ [\omega(\bar{u},  k, a)]_x=0,
\end{equation}
known as the conservation of waves, where $\omega(\bar{u}, k, a)$ is
the travelling wave frequency.

Being a $3 \times 3$ system, the quasi-linear equation
\eqref{general_whitham} is generally not diagonalisable, although
Riemann invariants are available for modulation systems associated
with integrable dispersive hydrodynamics such as KdV or mKdV
equations. It is important to stress that the general soliton-mean
interaction theory developed in \cite{maiden_solitonic_2018} is not
reliant on the availability of Riemann invariants for or the
integrability of the modulation system.

Soliton-mean interaction theory is based on the fundamental property
of Whitham modulation systems that we postulate here in a general form
and later explicitly justify for mKdV: {\it in the $k \to 0$ soliton
  limit, the modulation system \eqref{general_whitham} admits the
  following exact reduction} \cite{gurevich_nonlinear_1990}:
\begin{equation}
  \begin{bmatrix} \bar{u}\\a \end{bmatrix}_t
  + \begin{bmatrix}f'(\bar{u}) & 0 \\  g(a,\bar{u}) & c(a,\bar{u})    \end{bmatrix} \begin{bmatrix}
    \bar{u}\\a  \end{bmatrix}_x =   \begin{bmatrix} 0 \\ 0 \end{bmatrix},
  \label{eqn:2by2modulationsystem}
\end{equation}	
where $c(a,\bar{u})= \lim_{k \to 0} (\omega/k)$ is the soliton
amplitude-speed relation for propagation on the background $\bar{u}$
and $g(a,\bar{u})$ is a coupling function that is system dependent.
Equation \eqref{eqn:2by2modulationsystem} is called the
\textit{solitonic modulation system}.

The third modulation equation \eqref{wave_cons1} is identically
satisfied for $k=0$ while for $0 < k \ll 1$, it assumes at leading
order the form
\begin{equation}
  k_t + [c(a,\bar{u})k]_x = 0. \label{eqn:conservationofwaves}
\end{equation} 
Equation \eqref{eqn:conservationofwaves} can be added to the solitonic
modulation system \eqref{eqn:2by2modulationsystem} to give an
approximate modulation system for a train of noninteracting solitons
propagating on a variable mean flow. Equation
\eqref{eqn:conservationofwaves} then signifies the conservation of the
number of solitons in the train.  We shall refer to the combined
system \eqref{eqn:2by2modulationsystem} and
\eqref{eqn:conservationofwaves} as the \textit{augmented solitonic
  modulation system}.  Note that a particular case of this system was
derived in \cite{grimshaw_slowly_1979} for slowly varying soliton
solutions of the variable coefficent KdV equation.
 
The soliton train interpretation of the modulation system
\eqref{eqn:2by2modulationsystem} is instrumental for solitonic
dispersive hydrodynamics as it enables the description of a single
modulated soliton by treating the soliton amplitude $a(x,t)$ as a
spatiotemporal field, in contrast to standard soliton perturbation
theory where the soliton's parameters evolve temporally along its
trajectory in the $x,t$-plane; see,
e.g.~\cite{kivshar_dynamics_1989}. Additionally, as we will show, the
introduction of the fictitious wavenumber field $k(x,t)$ for a single
soliton enables the determination of the soliton phase shift due to
interaction with the mean flow.

The characteristic velocities of the system
\eqref{eqn:2by2modulationsystem} are $f'(\bar{u})$ and $c(a,\bar{u})$.
The right eigenvectors ${\bf r}_{1,2}$ of the Jacobian matrix in
\eqref{eqn:2by2modulationsystem} for each characteristic velocity are
\begin{align}
  \label{eq:11}
  v_1 = f'(\bar{u}), \qquad &{\bf r}_1 = \begin{bmatrix} f'- c \\
    g
  \end{bmatrix}, \\
  \label{eq:12}
  v_2 = c(a,\bar{u}), \qquad
  &{\bf r}_2 = \begin{bmatrix}
	0 \\ 1
  \end{bmatrix}.
\end{align}
Thus, the system \eqref{eqn:2by2modulationsystem} is strictly
hyperbolic if $f' \ne c$ for all
$(\bar{u},a)\in \mathcal{A}_0$, where
\begin{equation}
  \label{eq:14}
  \mathcal{A}_0 = \mathbb{R} \times (0,\infty)
\end{equation}
is the set of admissible states.  The system
\eqref{eqn:2by2modulationsystem} is genuinely nonlinear in the $j$th
characteristic field if $\nabla v_j \cdot {\bf r}_j \ne 0$ for all
$(\bar{u},a)\in \mathcal{A}_0$. For the first characteristic field,
\begin{equation}\label{gennon1}
  \nabla f'(\bar{u}) \cdot {\bf r}_1  \ne 0 
  \implies f''(\bar{u})(f'(\bar{u})-c(a,\bar{u})) \ne 0,
\end{equation}
which holds, provided the characteristic velocities are distinct
(strict hyperbolicity) and the flux $f$ of the original scalar
evolution equation \eqref{eqn:scalarconservationlaw} is convex. Thus,
when two characteristic velocities merge (nonstrict hyperbolicity),
the corresponding characteristic field is not genuinely nonlinear.

The genuine nonlinearity of the second characteristic field requires
\begin{equation}\label{gennon2}
  \nabla c(a,\bar{u}) \cdot {\bf r}_2  \ne 0 \quad
  \implies c_{a}(a,\bar{u}) \ne  0.
\end{equation}
To summarise, the quasi-linear system \eqref{eqn:2by2modulationsystem}
is strictly hyperbolic when $f'(\bar{u}) \neq c(a,\bar{u})$ and is
genuinely nonlinear when additionally $f''(\bar{u})\neq 0$ and
$c_a(a,\bar{u})\neq 0$ for all
$(\bar{u},a) \in \mathbb{R} \times (0,\infty)$.  Negation of any of
these three conditions gives rise to nonconvex solitonic dispersive
hydrodynamics.

Since the exact soliton reduction \eqref{eqn:2by2modulationsystem} is
a $2\times2$ quasi-linear hyperbolic system, it can be reduced to
Riemann invariant form.  We refer to the mean flow $\bar{u}$ as the
``hydrodynamic'' Riemann invariant and the other is found by
integrating the differential form
$g\, \mathrm{d}\bar{u} + (c - f')\,\mathrm{d}a$ provided $c \ne f'$
\cite{whitham_linear_1974}.  Denoting the second, ``solitonic''
Riemann invariant as $q = q(a,\bar{u})$, the diagonalised system can
be written as
\begin{equation}
  \begin{bmatrix} \bar{u}\\	q \end{bmatrix}_t
  + \begin{bmatrix}f'(\bar{u}) & 0 \\ 0 &
    C(q,\bar{u})\end{bmatrix} \begin{bmatrix}
    \bar{u}\\q \end{bmatrix}_x = \begin{bmatrix} 0 \\ 0 \end{bmatrix}, 
  \label{eqn:2by2modulationsystemdiagonal}
\end{equation}	
where $C\big (q(a,\bar{u}),\bar{u} \big) \equiv c(a,\bar{u})$.  In
terms of the diagonal system \eqref{eqn:2by2modulationsystemdiagonal},
the condition of strict hyperbolicity reads $f'(\ub) \ne C(q,\ub)$ and
the conditions of genuine nonlinearity of the first and second
characteristic field are written respectively as
\begin{equation}
  f''(\ub) \ne 0, \quad \hbox{and} \quad C_q \ne 0.
\end{equation}

It is important to stress that the existence of the solitonic Riemann
invariant $q$ is not reliant on the diagonalisability of the full
quasi-linear system \eqref{general_whitham} in Riemann invariants.  In
fact, as was shown in \cite{el_resolution_2005}, this Riemann
invariant can be obtained directly, as the integral
$q=Q(\tilde k, \ub) = const$ on $\mathrm{d}x/\mathrm{d}t=C$, of the
following characteristic ODE
\begin{equation}\label{char_ODE}
  \frac{\mathrm{d} \tilde k}{d \ub} = \frac{\partial _{\ub} \tilde
    \omega_0}{f'(\ub) - \partial_{\tilde k} \tilde \omega_0},
\end{equation}
where $\tilde k$ and $\tilde \omega_0 $ are called the conjugate
wavenumber and conjugate frequency, respectively.  They are defined in
terms of the soliton amplitude-speed relation $c(a, \ub)$ and the
linear dispersion relation \eqref{eq:1} $\omega_0(k, \ub)$ by
\begin{equation}\label{conjug}
  \tilde \omega_0 (\tilde k, \ub) = -i \omega_0 (i \tilde k, \ub );
  \quad c(a, \ub) = \frac{\tilde \omega_0}{\tilde k}.
\end{equation}
We note that the Riemann invariant $q=Q(\tilde k, \ub)$ is not defined
uniquely, as any smooth function of a Riemann invariant is also a
Riemann invariant.  In the case of convex systems, a convenient
normalisation is suggested by the requirement to maintain strict
hyperbolicity
of the solitonic system in the limit of vanishing amplitude where the
long-wave speed $f'(\ub)$ and soliton speed $c(a, \ub)$ must
coincide. The variable $\tilde k$ can be identified as an
amplitude-type variable \cite{el_resolution_2005}, so that
$\tilde k =0 \Longleftrightarrow a =0$, and require that the
hydrodynamic and solitonic Riemann invariants coincide when
$\tilde k \to 0$, i.e. $Q(0, \ub)=\ub$. As a result, the system
\eqref{eqn:2by2modulationsystemdiagonal} reduces to a single
hyperbolic equation $\ub_t + f'(\ub)\ub_x=0$.  The situation is
different for nonconvex systems, where two or more distinct Riemann
invariants associated with the same characteristic speed may exist.
For example, for cubic flux $f(\ub)=\ub^3$, the mean flow equation
$\ub_t +3 \ub^2 \ub_x=0$ is invariant with respect to the
transformation $\ub \to - \ub$ so another possible normalization is
$Q(0, \ub)=-\ub$.  To avoid ambiguity, we will be using the
normalization
\begin{equation}\label{norm_q}
 Q(0, \ub)=\ub
\end{equation} 
for the initial configuration.  For the case of a general nonconvex
flux, we assumed, without loss of generality, that it satisfies
$f''(0)=0$.  Then, if the solution curve crosses $\ub=0$, the
normalisation of the Riemann invariant should be changed to
$Q(0, \ub)=-\ub$ across this point to maintain smoothness of $Q$.

The two Riemann invariants $\bar{u}$ and $q$ for the $2\times 2$
system \eqref{eqn:2by2modulationsystemdiagonal} are also Riemann
invariants for the $3 \times 3$ augmented solitonic modulation system
\eqref{eqn:2by2modulationsystemdiagonal} and
\eqref{eqn:conservationofwaves}.  But the latter quasi-linear system
is not hyperbolic because its corresponding Jacobian matrix is
deficient, with just two eigenvalues and two linearly independent
eigenvectors.  Nevertheless, it has another hyperbolic subsystem, in
addition to \eqref{eqn:2by2modulationsystemdiagonal}, which is
obtained by setting $q \equiv q_0$ constant, as will be the case for
the soliton-mean flow interaction problems we consider.  Then the
remaining simple wave equation $\bar u_t + f(\bar u)_x=0$, together
with the approximate equation \eqref{eqn:conservationofwaves}, where
we replace $c(a, \bar u)$ with $C(q_0, \bar u)$, form a hyperbolic
subsystem.  Equation (\ref{eqn:conservationofwaves}) is diagonalised
by the quantity $kp(q_0,\bar{u})$, where
\begin{equation}
  p(q_0,\bar{u}) = \exp\left(-\int_{\bar{u}_0}^{\bar{u}}
    \frac{C_u(q_0,u)}{f'(u)-C(q_0,u)}\mathrm{d}u\right), \quad \bar{u}_0
  \in \mathbb{R} .
  \label{eqn:pintegral} 
\end{equation}
In other words, if $q=q_0$ is constant, we can use
\begin{equation}
  \label{kp}
  (kp)_t + C(q_0,\bar{u})(kp)_x = 0
\end{equation}
instead of \eqref{eqn:conservationofwaves}.  The quantities $q$ and
$kp$ have been identified in \cite{maiden_solitonic_2018} as adiabatic
invariants of soliton-mean flow interaction.
 
\subsection{Soliton-mean interaction}

Solutions to the solitonic modulation system can now be sought subject
to an initial mean flow $\bar{u}(x,0)=\bar{u}_0(x)$ and an initial
soliton with amplitude $a_0$ located at $x=x_0$. However, we need an
initial amplitude and wavenumber field $a(x,0)$, $k(x,0)$ defined for
all $x$. This is obtained by invoking the soliton train description
and asserting that the required solution of the augmented solitonic
system \eqref{eqn:2by2modulationsystem},
\eqref{eqn:conservationofwaves} is a simple wave (to be justified),
meaning all but one Riemann invariant are constant. The nonconstant
Riemann invariant is $\bar{u}$, in order to satisfy the initial
condition. Then $a(x,0)$ is selected to maintain constant $q$:
\begin{equation}
  \label{eq:6}
  q \big (a(x,0),\bar{u}_0(x) \big ) = q_0 \equiv q \big (a_0,\bar{u}_0(x_0) \big
  ) .
\end{equation}
Since $q$ constant is a solution, this reduces the augmented solitonic
modulation system to the hyperbolic subsystem consisting of two
diagonalised equations for $\bar u$ and $kp(q_0, \bar u)$.  In order
to define the initial wavenumber field $k(x,0)$, we set the latter
Riemann invariant to also be constant
\begin{equation}
  \label{eq:7}
  k(x,0) = k_0 \frac{p_0}{p \big (q_0,\bar{u}_0(x) \big)},
\end{equation}
where $p_0 \equiv p\big ( q_0 , \bar{u}_0(x_0) \big )$ and $k_0 \equiv
k(x_0,0) \ll 1$ is a small, positive quantity whose particular value is not important for our consideration  since we  assume
the limit $k_0 \to 0$ in the  soliton number conservation equation \eqref{eqn:conservationofwaves}, and, therefore in \eqref{kp}.

The soliton-mean interaction problem can now be formulated and solved.
Given the initial mean flow profile $\ub(x,0) =
\bar{u}_0(x)$, the soliton amplitude $a_0$ and location $x_0$,
$\bar{u}(x,t)$ is the simple wave solution
\begin{equation}
  \label{eq:8}
  x - f'(\bar{u}) t = H(\bar{u}), \quad H = u_0^{-1}
\end{equation}
and the soliton amplitude and  wavenumber fields satisfy 
\begin{equation}
  \label{eq:9}
  q(a,\bar{u}) = q_0, \quad k = k_0 \frac{p_0}{p(q_0,\bar{u})} .
\end{equation}

\begin{figure}
  \centering		
  \includegraphics{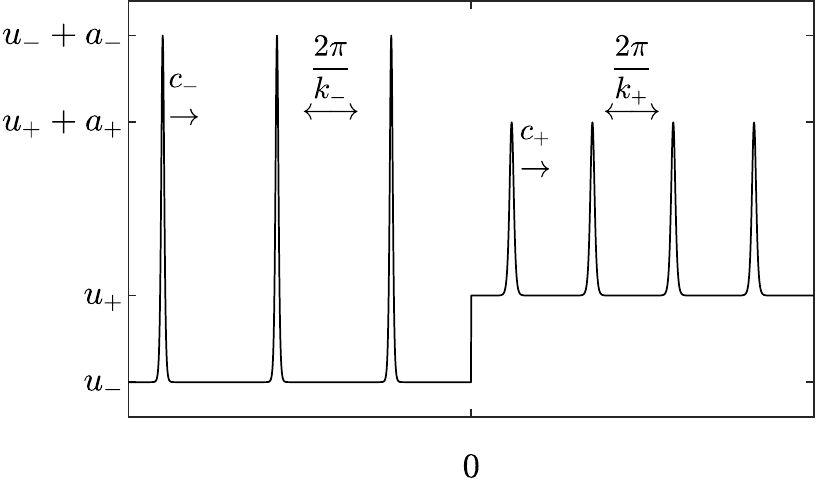}
  \caption{Sketch of the generalised GP problem.}
  \label{fig:gen_GP}
\end{figure}

We will focus our analysis on a generalised GP problem, in which
initial conditions for the mean flow are given as in the original
Riemann problem \eqref{eq:2}
\begin{equation}
  \bar{u}(x,0) = \left\{ \begin{array}{ll}
      u_-, & x<0, \\ u_+, & x> 0,
	\end{array}\right .\label{eqn:riemannproblem} 
\end{equation}
and the amplitude and wavenumber fields exhibit step variations
\begin{equation}
  \label{eq:4}
  a(x,0) =
  \begin{cases}
    a_- & x < 0 \\ a_+ & x > 0
  \end{cases}, \qquad
  k(x,0) = 
  \begin{cases}
    k_- & x < 0 \\ k_+ & x > 0.
  \end{cases} 
\end{equation}
A sketch illustrating the generalised GP problem is shown in Fig.~\ref{fig:gen_GP}

Depending on the initial location $x_0$ of the soliton relative to the
mean flow discontinuity at $x=0$, either the left $(a_-, k_-)$ or
right $(a_+, k_+)$ part of the initial wave field $a(x,0), k(x,0)$ is
prescribed with the other part to be determined as described below.

Due to the scaling invariance of both the quasilinear solitonic
modulation system (\ref{eqn:2by2modulationsystem}),
(\ref{eqn:conservationofwaves}) and the step initial data
(\ref{eqn:riemannproblem}), (\ref{eq:4}), the soliton-mean interaction
problem is solved by a simple wave solution of the Riemann problem,
thus justifying the constant Riemann invariant assumption for $q$ and
$kp$ expressed by equations \eqref{eq:6}, \eqref{eq:7}. Therefore, the
amplitude and wavenumber fields in the soliton-mean flow interaction
must satisfy (\ref{eq:9}), yielding the relations between admissible
values of $a_{\pm}$ and $k_{\pm}$ in \eqref{eq:4}. These are
formulated as {\it transmission and phase conditions}:
\begin{align}
  \label{eq:3}
  q(a_-,u_-) &= q(a_+,u_+), \\
  \label{eqn:transmissionphasecondgeneral}
  k_-  p(q_0,u_-)&= k_+ p(q_0,u_+),
\end{align}
where $p(q, \ub)$ is defined by \eqref{eqn:pintegral}. It is important
to stress that the existence of the simple wave solution leading to
the conditions in eqs.~\eqref{eq:3} and
\eqref{eqn:transmissionphasecondgeneral} requires convexity (genuine
nonlinearity) of the characteristic field \eqref{eq:11} along the
integral curve so that the conditions \eqref{gennon1} are not
violated.

In the context of a single soliton interacting with a varying mean
flow connecting two equilibrium states $\ub=u_-$ and $\ub=u_+$, the
conditions \eqref{eq:3} and \eqref{eqn:transmissionphasecondgeneral}
should be interpreted as follows.  The initial discontinuity
\eqref{eqn:riemannproblem} initiates the varying mean flow that is
generally confined to the bounded, expanding region
$s_-t < x < s_+ t$.  There is an exception to this for the
undercompressive DSW mean flow, which is a nonexpanding travelling
wave and requires a separate treatment. Then two basic scenarios of
soliton-mean interaction can be realised that we describe by assuming
positive polarity of the propagating soliton.  The generalisation to
negative polarity (dark) solitons is straightforward.

\bigskip
(i) {\it Forward (left to right) transmission/trapping}. 

\medskip Assuming that the soliton with amplitude $a_->0$ is initially
placed at $x_0=x_-<0$ on the left, background mean flow state
$\ub=u_-$, then if the soliton velocity satisfies $c(a_-, u_-)> s_-$,
soliton-mean flow interaction occurs for times
$t>t_1=|x_-/(c(a_-, u_-)-s_-)|$. As a result, the soliton either (a)
gets transmitted (tunnels) through the variable mean flow and emerges
on the right state $\ub = u_+$ with the new amplitude $a_+>0$
determined by the condition \eqref{eq:3} or (b) gets trapped within
the variable mean flow. The trapping occurs if the transmitted soliton
amplitude defined by \eqref{eq:3} is negative or zero, $a_+ \le 0$.

For this case of forward transmission, the trajectory of the soliton
post interaction is given by $x = c(a_+, u_+) t + x_+$, where
generally $x_+ \ne x_-$.  This implies that soliton-mean flow
transmission is accompanied by both an amplitude change and a soliton
phase shift $\Delta=x_+-x_-$, which can be determined from the
condition \eqref{eqn:transmissionphasecondgeneral}.  To relate the
$x$-intercepts $x_\pm$ of the soliton characteristic pre and post mean
flow interaction we note that the conservation of the number of
solitons in the fictitious modulated train of non-interacting solitons
implies
\begin{equation}
  \label{eq:5}
  k_- x_- = k_+ x_+ .
\end{equation}
Given $x_-$, only the ratio of $k_+/k_-$ is needed to determine $x_+$,
so, by virtue of the linear relationship between $k_+$ and $k_-$, the
particular value of $k_-$ in \eqref{eq:9} is irrelevant. The soliton
phase shift $\Delta = x_+ - x_-$ due to interaction with the mean flow
is then given by
\begin{equation}
  \label{phase_shift1}
  \frac{\Delta}{x_-} = \left (\frac{k_-}{k_+} -1 \right ) =
  \left ( \frac{p_+}{p_-} -1 \right ),
\end{equation}
where we have used the shorthand notation
$p_{\pm} \equiv p(q_0,u_\pm)$.

\bigskip {(ii)} \ {\it Backward (right to left)
  transmission/trapping}.

\medskip If the soliton with amplitude $a_+$ is initially placed at
$x_0=x_+>0$ on the right background $\ub=u_+$ and $c(a_+, u_+) < s_+$,
then soliton-mean flow interaction occurs for times
$t > t_{2}=x_+/(s_+-c(a_+, u_+))$. If the soliton eventually emerges
from mean flow interaction onto the opposite constant background
$\bar{u}=u_-$, its amplitude $a_->0$ and phase shift
$\Delta = x_- - x_+ = x_+(p_-/p_+ -1)$ are determined by the same
transmission and phase conditions \eqref{eq:3},
\eqref{eqn:transmissionphasecondgeneral}. Otherwise, if the
transmitted amplitude $a_- \le 0$, the soliton remains trapped within
the mean flow.

The generalisation to negative (dark) soliton interaction with mean
flow is straightforward. For this, it is convenient to introduce a
signed amplitude $a$, which enables the representation of both bright
$a > 0$ and dark $a < 0$ solitons.  Assuming negative initial
amplitude $a_\pm <0$, forward/backward transmission requires that the
transmitted amplitude $a_\mp$ maintains the same, negative, sign.
Generally, the condition $a_+a_->0$ is the sufficient condition for
transmission in both bright and dark soliton cases. Its negation
implies trapping.

In all cases of forward/backward transmission/trapping, the soliton
trajectory for $t>0$ is given by the characteristic,
\begin{equation}
  \frac{\mathrm{d}x}{\mathrm{d}t} = c(a(x,t),\bar{u}(x,t)), \qquad x(0) = x_0, \label{eqn:solitontrajectory}
\end{equation}	
where $|x_0| \gg 1$ so that the soliton is initially well-separated
from the initial step in the mean flow at $x=0$.

In the present work, we consider the implications of a nonconvex
solitonic modulation system \eqref{eqn:2by2modulationsystem} on the
above soliton transmission and trapping scenarios. As described in
Section~\ref{sec:sol_mod}, nonconvexity enters when strict
hyperbolicity and/or genuine nonlinearity is lost via one of the three
conditions: $f''(\ub)=0$, $f'(\ub)=c(a, \ub)$, or $c_a(a, \ub)=0$ for
any $(\ub,a) \in \mathcal{A}_0$.

In \cite{maiden_solitonic_2018}, positivity of the transmitted
amplitude (one of $a_\pm$) was proposed as a necessary and sufficient
condition for bright soliton tunnelling to occur through a mean flow
for convex dispersive hydrodynamics.  In fact, this condition
coincides with a less restrictive definition of strict hyperbolicity
for (\ref{eqn:2by2modulationsystem}) where $a = 0$ is included in the
set of admissible states
$\mathcal{A}_0' = \mathbb{R}\times [0,\infty)$.  Generally, the
soliton speed coincides with the long wave speed when its amplitude
vanishes, $c(0,\bar{u}) = f'(\bar{u})$, which signifies the onset of
soliton trapping.  Within the context of Whitham modulation theory,
states in which $a = 0$ or $k = 0$ are not considered admissible when
assessing strict hyperbolicity and genuine nonlinearity of the
modulation equations because they coincide with a degeneracy in which
the number of modulation equations is reduced; see, e.g.,
\cite{levermore_hyperbolic_1988,bikbaev_korteweg-vries_1989}.  We will
utilise the traditional definition in which $a = 0$ is not included in
the set of admissible states \eqref{eq:14}.

In the more general nonconvex case, we find that in order for the
soliton to tunnel through the mean flow, we must require the
additional condition that the modulation system
(\ref{eqn:2by2modulationsystem}) remain strictly hyperbolic along the
entire soliton trajectory for all admissible states
$(\ub,a) \in \mathcal{A}_0$. If the characteristic speeds
$f'(\bar{u})$ and $c(a,\bar{u})$ coincide for nonzero $a$, then strict
hyperbolicity is lost and the soliton is trapped inside the mean
flow. If the speeds remain separated, the soliton amplitude on the
transmitted side is non-zero and the phase shift is well-defined
according to \eqref{eqn:transmissionphasecondgeneral}.  In summary,
the necessary and sufficient conditions for tunnelling in a nonconvex
solitonic modulation system \eqref{eqn:2by2modulationsystemdiagonal}
with initial data \eqref{eqn:riemannproblem}, \eqref{eq:4} is
\begin{equation}
  \label{eq:16}
  q(a_-,u_-) = q(a_+,u_+), \quad a_+a_- > 0, \quad
  f'(\bar{u}(x,t)) \ne c(a(x,t),\bar{u}(x,t)),
\end{equation}
where $x = x(t)$ is the characteristic \eqref{eqn:solitontrajectory}
and $t \ge 0$.

\subsection{Hydrodynamic reciprocity}
\label{sec:hydro_recipr}
	
So far, we have assumed that the mean flow satisfies the simple wave
equation $\bar{u}_t + f'(\bar{u})\ub_x = 0$.  For step initial data
(\ref{eqn:riemannproblem}), the only candidate continuous solution is
a RW
\begin{equation}
  \label{eq:10}
  \bar{u}(x,t) =
  \begin{cases}
    u_- & x < f'(u_-)t, \\
   (f')^{-1}(x/t) & f'(u_-) t < x < f'(u_+) t, \\
    u_+ & f'(u_+) t < x.
  \end{cases}
\end{equation}
so long as the admissibility criterion $f'(u_-) < f'(u_+)$ holds,
corresponding to expansive initial data.  As will be shown in the next
section, there is a much richer variety of dispersive mean flows
generated by the mKdV GP problem when the initial data is compressive.
Thus, we need soliton-mean flow modulation theory to be flexible
enough to accommodate a wide class of mean flows.

The solitonic modulation equations (\ref{eqn:2by2modulationsystem}),
(\ref{eqn:conservationofwaves}) directly apply for expansive mean flow
initial data, yielding a description of soliton-RW interaction. For
compressive initial data (\ref{eqn:riemannproblem}), rather than form
a discontinuous shock solution, a DSW is formed that occupies the
space-time region $A \subset \mathbb{R} \times (0,\infty)$ where the
solution is described by the full system of Whitham modulation
equations for a slowly varying nonlinear periodic wave. As a result,
the Riemann invariant $q$ and secondary invariant $kp$ of the
augmented solitonic system (\ref{eqn:2by2modulationsystem}),
(\ref{eqn:conservationofwaves}) are not conserved in $A$, and our
arguments leading to the transmission and phase conditions
\eqref{eq:3}, \eqref{eqn:transmissionphasecondgeneral} do not apply to
the soliton interaction with the DSW mean flow.

To address this, we invoke an important property of the dispersive
conservation law (\ref{eqn:scalarconservationlaw}): time
reversibility. A consequence of time reversibility is the continuity
of the modulation solution for all $(x,t) \in \mathbb{R}^2$.  For
compressive data, we consider the solution for $t < 0$ that consists
of a simple wave described by \eqref{eq:10}, i.e., the expansive mean
flow case.  Then, since $q$ and $kp$ are constant for all
$x \in \mathbb{R}$ and $t < 0$, they remain constant by continuity for
$(x,t)$ in the complement of $A$, outside of the oscillatory region,
where the augmented solitonic system (\ref{eqn:2by2modulationsystem}),
(\ref{eqn:conservationofwaves}) remains valid.  Note that for the
Riemann data \eqref{eqn:riemannproblem}, \eqref{eq:4}, the solution
remains continuous outside $\mathbb{R}^2 \setminus \{(0,0)\}$, which
is justified by taking the limit of smooth solutions. This property
was called {\it hydrodynamic reciprocity} in
\cite{maiden_solitonic_2018} and has been used previously in the
characterization of DSWs for a single or pair of dispersive
hydrodynamic conservation laws \cite{el_resolution_2005}.  Since the
transmission and phase conditions (\ref{eq:3}),
(\ref{eqn:transmissionphasecondgeneral}) hold outside the oscillatory
region, hydrodynamic reciprocity allows us to predict the transmitted
amplitude and phase shift $\Delta$ of a soliton interacting with DSW
mean flows entirely within the framework of the augmented solitonic
modulation system (\ref{eqn:2by2modulationsystem}),
(\ref{eqn:conservationofwaves}).

The details of the modulation dynamics for the soliton within the
interior of the oscillatory region $A$ can, in principle, be described
by a degenerate two-phase solution (see
\cite{flaschka_multiphase_1980} for multiphase modulation theory of
the KdV equation). However, as we will show, this rather technical
approach can be partially, approximately circumvented by replacing
$f(\ub)$ in the characteristic equation \eqref{char_ODE} by an
appropriate choice of the mean flow variation and effectively defining
a new adiabatic invariant $q$ holding within $A$.
			
\section{Modified Korteweg--de Vries equation: travelling wave
  solutions and modulation equations}
\label{sec:mKdV_trav_mod}
	
As the simplest example of dispersive hydrodynamics with nonconvex
flux, we study the mKdV equation (\ref{eqn:mKdV}). The mean flow
behaviours that arise when solving \eqref{eqn:mKdV} subject to
\eqref{eqn:riemannproblem} depend on the sign of the dispersive term
$\text{sgn}(\mu)$.  The mKdV hyperbolic flux $f(u)=u^3$ exhibits the
inflection point $f''(0) = 0$ so that nonconvexity affects the
solutions whenever the initial data contain an open interval including
the point $u=0$.  For either sign of $\mu$, the mKdV equation allows
for solitons of both polarities by the symmetry $u \to -u$.  The
linear dispersion relation is
\begin{equation}
  \omega_0 = 3\bar{u}^2k + \mu k^3 . \label{eqn:mKdVdispersionrelation}
\end{equation}

In what follows, we present a compendium of the results from
\cite{el_dispersive_2017} necessary for the development in this paper.
 
\subsection{Travelling wave solutions}
\begin{figure}
  \centering
  \begin{subfigure}{0.47\textwidth}
    \centering
    \includegraphics[width=0.8\textwidth]{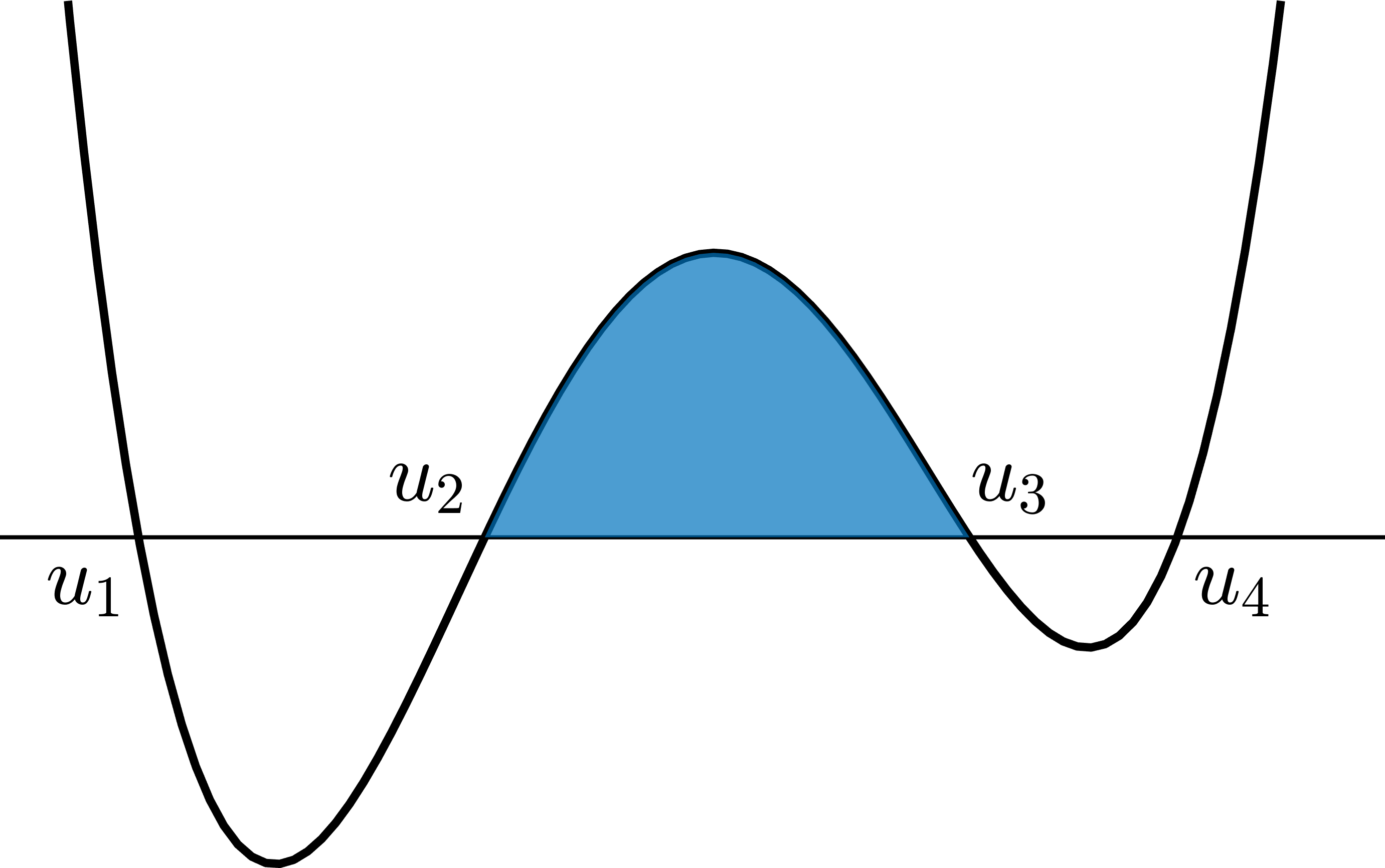}
    \caption{$\mu > 0$}
  \end{subfigure}
  \begin{subfigure}{0.47\textwidth}
    \centering
    \includegraphics[width=0.8\textwidth]{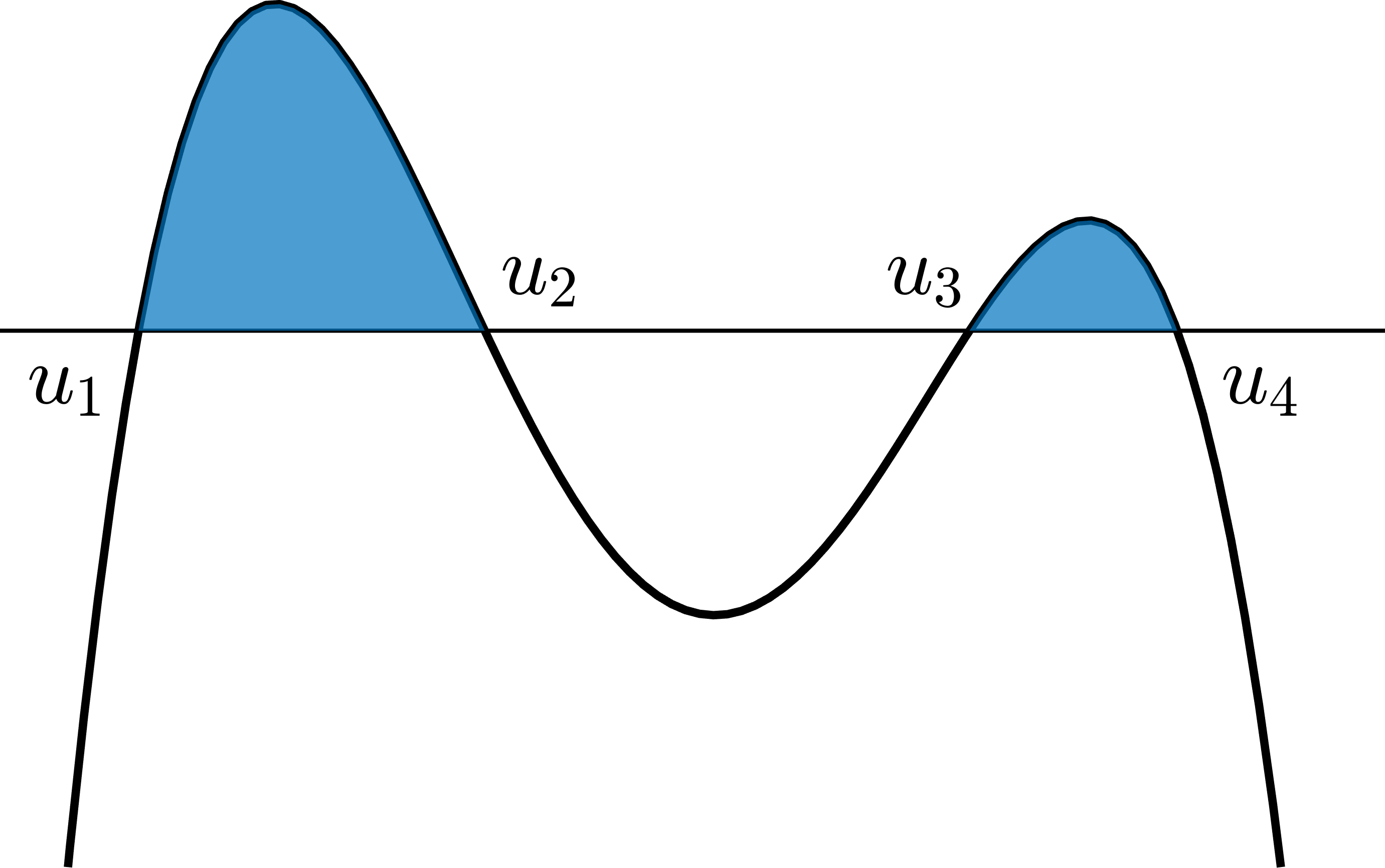}
    \caption{$\mu < 0$}
  \end{subfigure}
  \caption{Potential curve $Q(u)$ of the nonlinear oscillator equation
    \eqref{eqn:mKdVtravelingwaveODE}. Travelling wave solutions exist
    in the shaded regions.}
  \label{fig:Potential}
\end{figure}

The mKdV travelling wave solutions $u = u(\eta)$,
$\eta = \frac{x-Ut}{\sqrt{2|\mu|}}$ are described by the ODE
\begin{equation}
  (u_\eta)^2 = \text{sgn}(\mu) (u-u_1)(u-u_2)(u-u_3)(u-u_4) \equiv Q(u) \label{eqn:mKdVtravelingwaveODE},
\end{equation}
subject to the constraint $\sum_{i=1}^4 u_i = 0$ and ordering of the
roots $u_1 \leq u_2 \leq u_3 \leq u_4$.  We only consider the
modulationally stable case in which all roots are real. The phase
velocity $U$ is given by
\begin{equation}\label{U}
  U = -\frac{1}{2}(u_1 u_2 + u_1 u_3 + u_1 u_4 + u_2 u_3 + u_2 u_4 + u_3 u_4) .
\end{equation}
Equation \eqref{eqn:mKdVtravelingwaveODE} is a nonlinear oscillator
equation in the potential $-Q(u)$. Figure \ref{fig:Potential} shows
representative potential curves $Q(u)$ for both signs of the
dispersion coefficient $\mu$. Travelling wave solutions exist in the
regions where $Q(u)>0$ (shaded regions) and can be obtained by
integrating \eqref{eqn:mKdVtravelingwaveODE} in terms of Jacobi
elliptic functions.  The cases $\mu<0$ and $\mu>0$ are treated
separately.
	
\medskip (i) \ For $\mu > 0$, the travelling wave solution is
expressed in terms of Jacobi elliptic functions as
\begin{equation}
  u = u_2 +
  \frac{(u_3-u_2)\text{cn}^2(\theta,m)}
  {1-\frac{u_3-u_2}{u_4-u_2}\text{sn}^2(\theta,m)}. \label{eqn:mupos_periodicsoln}
\end{equation}
with $\theta = \sqrt{(u_3-u_1)(u_4-u_2)}\eta$ and modulus
$m=m_+ = \frac{(u_3-u_2)(u_4-u_1)}{(u_4-u_2)(u_3-u_1)}$.  The
wavenumber of the travelling wave is given by
\begin{equation}
  k = \frac{\pi \sqrt{(u_3-u_1)(u_4-u_2)}}{2K(m)\sqrt{2|\mu|}}. \label{eqn:kexact}
\end{equation}

When $u_2 \to u_1$ ($m_+ \to 1$), the solution becomes a bright
(positive polarity) soliton with amplitude $a = u_3-u_1$, mean
background $\bar{u} = u_1<0$,
\begin{equation}
  u = u_1 + \frac{u_3-u_1}{\cosh^2 \theta -
    \frac{u_3-u_1}{u_4-u_1}\sinh^2\theta}, \label{eqn:mupos_brightsoliton} 
\end{equation}
which travels with the velocity $U = c_+$
\begin{equation}\label{c+}
  c_+(a,\bar{u}) = \frac{1}{2}a^2 + 2a\bar{u}+ 3\bar{u}^2.
\end{equation}
Due to the root ordering, these bright solitons exist only for a
certain range of positive amplitudes and a negative background, given
by the constraint
\begin{equation} 
  0 < a < -2\bar{u}. \label{eqn:mupos_solitonexistence}
\end{equation}
	
Dark (negative polarity) soliton solutions occur when $u_3 \to u_4$
instead. In this case, $\ub =u_4>0$, $a=u_4-u_2>0$ and the soliton
velocity $U = c_-$
\begin{equation}\label{c-}
  c_-(a,\bar{u}) = \frac{1}{2}a^2 - 2a\bar{u}+ 3\bar{u}^2.
\end{equation}	
with negative amplitudes $a$ satisfying
\begin{equation}\label{amp_neg_sol}
  -2 \ub < a <0 .
\end{equation}

When $u_2 \to u_1$ and additionally, $u_3 \to u_4$, the travelling
wave becomes a kink,
\begin{equation}
  u = \pm \bar{u} \tanh(\bar{u}\eta), \label{eqn:mKdVkink}
\end{equation}
a heteroclinic smooth transition connecting two equilibria
$ \bar{u} = u_1<0 $ and $ -\bar{u}=u_4>0 $ (note that the constraint
$\sum u_j=0$ becomes $u_4+u_1=0$ in this limit) and travelling with
speed $U = \bar{u}^2$, which matches the classical shock speed
determined by the Rankine-Hugoniot condition.

\medskip (ii) \ For $\mu < 0$, travelling wave solutions can occur
between $u_1$ and $u_2$ or between $u_3$ and $u_4$. Between $u_3$ and
$u_4$, the travelling wave solution is
\begin{equation}
  u = u_3 + \frac{(u_4-u_3)\text{cn}^2(\theta,m)}
  {1+\frac{u_4-u_3}{u_3-u_1}\text{sn}^2(\theta,m)}, 
  \label{eqn:muneg_periodicsoln}
\end{equation}
with $m=m_- = \frac{(u_4-u_3)(u_2 - u_1)}{(u_4-u_2)(u_3-u_1)}$. The
wavenumber is given by the same formula
\eqref{eqn:mupos_brightsoliton}.  When $u_3 \to u_2$ ($m_- \to 1$) the
solution becomes a bright exponential soliton with amplitude
$a = u_4-u_2$ and background $\bar{u} = u_2$
\begin{equation}
  u = u_2 + \frac{u_4-u_2}{\cosh^2 \theta + \frac{u_4-u_2}{u_2-u_1}\sinh^2\theta}. \label{eqn:muneg_brightsoliton}
\end{equation}
This soliton solution travels according to the same soliton
amplitude-speed relation \eqref{eqn:mKdVsolitonspeedamplitude} as in
the case $\mu > 0$. Due to the root ordering, valid bright soliton
amplitudes for the solution to exist are given by
\begin{equation}
  a > \max(0,-4\bar{u}), \label{eqn:muneg_solitonexistence}
\end{equation}
with no constraint on the background $\bar{u}$.
	
For $\mu<0$, there is a special type of travelling wave solution
expressible in terms of trigonometric functions. Again, these
solutions occur either between $u_1$ and $u_2$ or between $u_3$ and
$u_4$ but under the additional constraint that $u_3=u_4$ in the first
case and $u_1=u_2$ in the second case. For $u_3 \le u \le u_4$,
$u_1=u_2$ the solution is given by
\begin{equation}\label{trig_sol}
  u=u_3+\frac{u_4-u_3}{1+\frac{u_4-u_1}{u_3-u_1}\tan^2 \theta},
\end{equation}
The nonlinear trigonometric solution \eqref{trig_sol} has no analogue
in KdV theory.  When $u_3 \to u_2 = u_1 \equiv \ub$, the solution
\eqref{trig_sol} becomes an algebraic bright soliton described by
\begin{equation}
  u = u_1 + \frac{u_4 - u_1}{1 + (u_4-u_1)^2\eta^2/4}, \label{eqn:mKdValgebraicsoliton}
\end{equation}
with amplitude $a=u_4-u_1= - 4\ub$ and travelling at speed
$U = 3u_1^2=3 \ub^2$, which is the characteristic speed of the
dispersionless mKdV equation. Dark algebraic solitons can be obtained
by the transformation \eqref{eqn:mKdVinvarianttransform} below.

The solution oscillating between $u_1$ and $u_2$, can be obtained by
applying the invariant transformation
\begin{equation}
  u \to -u, \qquad u_i \to -u_{5-i}, \qquad i=1,2,3,4. \label{eqn:mKdVinvarianttransform}
\end{equation}	
In this region, both the exponential and algebraic soliton solutions
have negative polarity with amplitude $a=u_3-u_1$ and background
$\ub=u_3$ satisfying
\begin{equation}
  \label{eq:15}
  a < \min(0,- 4 \ub) .
\end{equation}
In both cases, the soliton amplitude-speed relation is given by
\eqref{c-}.  Heteroclinic kink solutions of mKdV do not exist if
$\mu<0$.
	
Summarising, the mKdV equation differs from the KdV equation in that
it supports solitons of both polarities for either sign of the
dispersion $\mu$. For $\mu >0$, bright soliton solutions occur when
$u_1 \to u_2$ and dark soliton solutions occur when $u_3 \to u_4$. For
$\mu < 0$, solitons arise when $u_2 \to u_3$ with bright solitons as
solutions between $u_3$ and $u_4$ while dark solitons occur between
$u_1$ and $u_2$. The amplitude-speed relations \eqref{c+} and
\eqref{c-} for bright and dark exponential solitons, respectively, can
be combined into a single relation by introducing the convention that
$a>0$ for bright solitons and $a<0$ for dark solitons. Then, the
general formula
\begin{equation}
  c(a,\bar{u}) = \frac{1}{2}a^2 + 2a\bar{u}+ 3\bar{u}^2, \quad a \in
  \mathbb{R} \label{eqn:mKdVsolitonspeedamplitude}
\end{equation}	
holds, covering all cases: $\mu \lessgtr 0$, dark and bright
exponential solitons. Note that this formula also includes kinks
($a= - 2\ub, c= \ub^2$) and algebraic solitons
($a=-4\ub, c= 3 \ub^2$). From now on, we will be assuming the
generalised amplitude $a \in \mathbb{R}$.
	
\subsection{mKdV modulation equations and solitonic reductions}
\label{sec:mKdV_mod_sol_red}
	
The purpose of this section is twofold: (i) to obtain the augmented
solitonic modulation system \eqref{eqn:2by2modulationsystem},
\eqref{eqn:conservationofwaves} by direct computation for the
mKdV-Whitham system and (ii) to explore the implications of mKdV's
nonconvex flux on the structure of the augmented solitonic modulation
system.
	
The system of modulation equations for the mKdV equation
\eqref{eqn:mKdV} was first derived in
\cite{driscoll_modulational_1975} following Whitham's original
averaging procedure \cite{whitham_non-linear_1965}, and reduced to
diagonal form.  The focus of the research in
\cite{driscoll_modulational_1975} was on the modulational stability of
nonlinear periodic solutions.
	
A detailed derivation of the travelling wave solutions and the
respective modulation equations for the Gardner equation (an extended
version of mKdV), revealing the differences between various
modulationally stable DSW structures arising in the $\mu>0$ and
$\mu < 0$ cases was performed in \cite{kamchatnov_undular_2012} and
then utilised in \cite{el_dispersive_2017} for the analysis of
modulated mKdV solutions in the zero-viscosity limit of the
mKdV-Burgers equation.  Following \cite{el_dispersive_2017}, the full
diagonalised mKdV modulation system is
\begin{equation}
  \frac{\partial \lambda_i}{\partial t} + W_i(\boldsymbol{\lambda})
  \frac{\partial \lambda_i}{\partial x} = 0, \qquad i = 1,2,3,  \label{eqn:mKdVWhitham}
\end{equation}
where $\lambda_i$ are the Riemann invariants expressed in terms of the
roots of the potential function $Q(\bar{u})$,
\begin{equation}\label{lau}
  \lambda_1 = \frac{1}{2}(u_1+u_2), \qquad \lambda_2 =
  \frac{1}{2}(u_1+u_3), \qquad \lambda_3=\frac{1}{2}(u_2+u_3). 
\end{equation}
	
The characteristic velocities $W_i(\boldsymbol{\la})$ can be written
as
\begin{equation}\label{Wj-}
  \begin{aligned}
	\mu < 0: \quad & &W_1 &= U + \frac{2}{3}(r_3 - r_2)\frac{K(m)}{E(m)} \\
	& &W_2 &= U - \frac{2}{3}(r_2 - r_1)\frac{(1-m)K(m)}{E(m)-(1-m)K(m)} \\
	& &W_3 &= U + \frac{2}{3}(r_2 - r_1)\frac{K(m)}{E(m)-K(m)}
  \end{aligned}
\end{equation}	

\begin{equation}\label{Wj+}
  \begin{aligned}	
	\mu > 0: \quad & &W_1 &= -U - \frac{2}{3}(r_2 - r_1)\frac{K(m)}{E(m)-K(m)} \\
	& &W_2 &= -U + \frac{2}{3}(r_2 - r_1)\frac{(1-m)K(m)}{E(m)-(1-m)K(m)} \\
	& &W_3 &= -U - \frac{2}{3}(r_3 - r_2)\frac{K(m)}{E(m)}, 
  \end{aligned}
\end{equation}
where $U=\frac{1}{3}(r_1 + r_2 + r_3)$, $m = (r_2 - r_1)/(r_3 - r_1)$
is the modulus, $0 \le m \le 1$, and $K(m)$ and $E(m)$ are complete
elliptic integrals of the first and second kind, respectively.  The
parameters $r_1, r_2, r_3$ are related to the Riemann invariants
$\la_1, \la_2, \la_3$ by
\begin{equation}
  \begin{aligned}
	\mu < 0: & & && & & r_1 &= 3\lambda_3^2, & r_2 &= 3\lambda_2^2, &
    r_3 &=  3\lambda_1^2;\\
	\mu > 0: & & & & && r_1 &= -3\lambda_1^2, & r_2 &= -3\lambda_2^2,
    & r_3 &=  -3\lambda_3^2.
	\label{eq:rtolambda}
  \end{aligned}
\end{equation}
The mapping ${\bf r} \mapsto \boldsymbol{\lambda}$ specified by
\eqref{eq:rtolambda} is multivalued, which implies that the mKdV
modulation system \eqref{eqn:mKdVWhitham} with characteristic
velocities \eqref{Wj-}, \eqref{Wj+} is neither strictly hyperbolic nor
genuinely nonlinear in both cases $\mu<0$ and $\mu >0$. However,
within the restricted subset in which $\lambda_j \ne 0$, $j = 1,2,3$,
the mKdV modulation system is strictly hyperbolic and genuinely
nonlinear. The relevant modulation solutions are subject to the
ordering $\lambda_1 \leq \lambda_2 \leq \lambda_3$ and
$r_1 \leq r_2 \leq r_3$.
	
We note here that the expressions \eqref{Wj-}, \eqref{Wj+} for the
mKV-Whitham characteristic velocities $W_j(\boldsymbol{\lambda})$ are
related to the characteristic velocities $V_j({\bf r})$ of the
diagonal KdV-Whitham system \cite{whitham_non-linear_1965} as
\begin{equation}
  \begin{aligned}
	\mu < 0: & &W_i(\boldsymbol{\lambda}) &= V_{4-i}({\bf r}), \\
	\mu > 0: & &W_i(\boldsymbol{\lambda}) &= -V_{i}({\bf r}).
	\label{eqn:KdVtomKdV}
  \end{aligned}
\end{equation}
	
The quadratic transformations \eqref{eq:rtolambda} can be viewed as a
modulation theory counterpart of the celebrated Miura transform
\cite{miura_kortewegvries_1968}.
	
We now obtain the soliton reduction of the mKdV-Whitham system.
First, note that the soliton limit of the mKdV travelling wave
solutions described in Section~\ref{sec:mKdV_trav_mod} is achieved by
letting either $m_+ \to 1$ ($\mu >0$) or $m_- \to 1$ ($\mu <0$). Using
the relations \eqref{lau}, \eqref{eq:rtolambda}, we find that both
cases correspond to the limit $m \to 1$ in the respective modulation
systems specified by \eqref{Wj-} ($\mu <0$) and \eqref{Wj+}
($\mu >0$).
	
For $\mu >0$, bright soliton solutions occur when $u_1 \to u_2$, which
coincides with $\la_2 \to \la_3$ by \eqref{lau}.  Furthermore,
$r_2 \to r_3$ and $m \to 1$ in \eqref{Wj+}, yielding the limiting
characteristic velocities
\begin{align}
  W_1(\boldsymbol{\lambda}) & = 3\lambda_1^2 ,\\
  W_2(\boldsymbol{\lambda}) &=  \lambda_1^2 + 2\lambda_3^2 =
                              W_3(\boldsymbol{\lambda}).
\end{align}
Substituting into \eqref{eqn:mKdVWhitham} gives the reduced diagonal
system
\begin{equation}\label{eq:la13}
  \begin{aligned}
    &\frac{\partial \lambda_1}{\partial t} + 3\lambda_1^2
    \frac{\partial \lambda_1}{\partial x} = 0,   \\ 
    &\frac{\partial \lambda_3}{\partial t} +
    (\lambda_1^2+2\lambda_3^2) \frac{\partial \lambda_3}{\partial x} =
    0.  \end{aligned}
\end{equation}
Using $\bar{u} = u_1 = \la_1$ and $a = u_3 - u_1=2 (\la_3 - \la_1)$
(see \eqref{eqn:mupos_brightsoliton}), we can now write
\eqref{eq:la13} as
\begin{equation}
  \begin{aligned}
    &\bar{u}_t + 3\bar{u}^2 \bar{u}_x = 0, \\
    &a_t + \left(\frac{1}{2}a^2 + 2a\bar{u} + 3\bar{u}^2\right)a_x +
    \left(a^2 + 4a\bar{u}\right)\bar{u}_x =0.
  \end{aligned}
  \label{eqn:2by2mKdVWhitham}
\end{equation}
The system \eqref{eqn:2by2mKdVWhitham} represents the mKdV realisation
of the general solitonic modulation system
\eqref{eqn:2by2modulationsystem} with the hyperbolic flux
$f(\ub)=\ub^3$, the soliton amplitude-speed relation
\eqref{eqn:mKdVsolitonspeedamplitude} and the coupling function
$g(a, \ub) = a^2 + 4a\bar{u}$. Comparing the diagonal form
\eqref{eq:la13} of the mKdV solitonic modulation system with the
general representation \eqref{eqn:2by2modulationsystemdiagonal}, we
identify the Riemann invariant $\lambda_1$ in \eqref{eq:la13} with
$\ub$ and $\lambda_3$ with $q=\frac{1}{2}a+\bar{u}$, and the
characteristic velocity $W_3(\lambda_1, \lambda_3, \lambda_3)$ with
$C(q, \ub)=\ub^2 + 2q^2$.

The dark soliton limit is achieved when $u_3 \to u_4$, which
translates to $\la_2 \to -\la_3$, so, due to the quadratic dependence
\eqref{eq:rtolambda} of ${\bf r}$ on $\boldsymbol \lambda$, we arrive
at the same system \eqref{eq:la13} for $\la_1, \la_3$ and,
equivalently, the system \eqref{eqn:2by2mKdVWhitham} for $\ub=-\la_1$
and $q=\la_3 =\ub+\frac12 a$, $C(q, \ub)=\ub^2 + 2q^2$, where we used
the extended notion of the signed amplitude,
$a=u_2-u_3=2(\la_1 - \la_2)<0$.

The derivation for $\mu <0$ is analogous and also results in the
solitonic modulation system \eqref{eqn:2by2mKdVWhitham} for both
bright and dark soliton cases with the identification of the merged
Riemann invariant $\lambda_1=\lambda_2=\ub + \frac12 a=q$.

As described in Section~\ref{sec:soliton-mean-flow}, the Riemann
invariant $q(a,\ub)$ can be obtained directly, bypassing the
derivation of the full mKdV modulation system and the evaluation of
its soliton limit. This is achieved by integrating the characteristic
ODE \eqref{char_ODE} with the mKdV conjugate dispersion relation
\eqref{conjug} given by
$\tilde \omega_0 = 3\tilde k \ub^2 - \mu \tilde k^3$. The ODE then
assumes the form
$\mathrm{d}\tilde k/ \mathrm{d}\ub = 2 \ub/ (\mu \tilde k)$. Its
integral $Q(\tilde k, \ub)=const$ is found as
$Q= \pm \sqrt{\ub^2 - \mu \tilde k^2/2}$.  The conjugate wavenumber
$\tilde k$ is related to the soliton amplitude $a$ via equation
$c(a, \ub)=\tilde \omega_0/\tilde k$ \eqref{conjug}, where $c(a, \ub)$
is given by \eqref{eqn:mKdVsolitonspeedamplitude}. This yields
$\mu \tilde k^2 =-\frac12 a^2 - 2 a \ub$. Substituting in the
expression for $Q$ and applying the normalisation \eqref{norm_q}
yields the Riemann invariant
\begin{equation}
  \label{q}
  q=\ub + \frac12 a,
\end{equation}
in full agreement with the previous identification of the Riemann
invariant of the solitonic modulation system
\eqref{eqn:2by2mKdVWhitham}.

Thus, for both signs of $\mu$ and for both bright and dark solitons,
the diagonalised mKdV solitonic modulation system assumes the form
\begin{equation}\label{eqn:ubarWhitham}
  \begin{aligned}
    \bar{u}_t + 3\bar{u}^2\bar{u}_x &= &0, \\
    q_t + (\bar{u}^2 + 2q^2)q_x &= &0.
  \end{aligned}
\end{equation}
The system \eqref{eqn:ubarWhitham} is augmented by the approximate
equation \eqref{eqn:conservationofwaves} for conservation of waves
(solitons), which assumes the form
\begin{equation}
  \label{eqn:consofwaves1}
  k_t + \left((\bar{u}^2 + 2q^2)k\right)_x = 0. 
\end{equation}
As a matter of fact, equation \eqref{eqn:consofwaves1} can be derived
as a consequence of the full modulation system \eqref{eqn:mKdVWhitham}
by considering the pair $k$, $U$ given by eqs.~\eqref{eqn:kexact},
\eqref{U} expressed in terms of $\la_1, \la_2, \la_3$ for $\mu > 0$ as
\begin{equation}
  \label{kUm}
  k=\frac{\pi \sqrt{\la_1^2 -\la_3^2}}{4 K(m) \sqrt{2|\mu|}}, \quad
  m=\frac{\la_1^2 - \la_2 ^2}{\la_1^2 - \la_3^2},  
  \quad U=\la_1^2 + \la_2^2 + \la_3^2. \quad 
\end{equation}
Expanding \eqref{kUm} for $1-m \ll 1$ ($\la_2 \to \la_3$), evaluating
$k_t+(kU)_x=0$ at leading order, and using \eqref{eqn:mKdVWhitham} we
arrive at \eqref{eqn:consofwaves1} with $q^2=\la_1^2$ for $\mu>0$. A similar analysis for $\mu < 0$ arrives at the same result with $q^2=\la_3^2$.  The approximate conservation of waves
equation \eqref{eqn:consofwaves1} is subject to corrections of order $ke^{-\alpha k}$ where $\alpha = \pi\sqrt{2(\lambda_1^2-\lambda_3^2)} = \pi\sqrt{-a(a+4\bar{u})/2}$ as $k\to 0$.

The solitonic modulation system \eqref{eqn:ubarWhitham} loses strict
hyperbolicity when $3 \ub^2=\ub^2 + 2q^2$---corresponding to
$f'(\bar{u}) = c(a,\bar{u})$ in the general notation of
\eqref{eqn:2by2modulationsystem} and is consistent with the negation
of the genuine nonlinearity condition \eqref{gennon1}---which yields
$q^2 = \ub^2$ and implies via \eqref{q} that either
\begin{equation}
  a = 0 ,  \quad  \text{ or } \quad a = -4\bar{u} .
  \label{eqn:mKdVhyperbolicitycondition}
\end{equation}
As mentioned earlier, the $a = 0$ case corresponds to a reduction in
the order of the solitonic modulation system \eqref{eqn:ubarWhitham}
to the mean flow equation $\bar{u}_t + 3\ub^2 \ub_x = 0$.  So,
strictly speaking, it does not correspond to the loss of strict
hyperbolicity as traditionally defined for Whitham modulation systems,
but it is relevant for the general tunnelling conditions
\eqref{eq:16}.

Genuine nonlinearity is lost when
\eqref{eqn:mKdVhyperbolicitycondition} holds or, alternatively, if
$f''(\bar{u}) = 0$, or $c_a = 0$, cf \eqref{gennon1}, \eqref{gennon2},
i.e.
\begin{equation}
  \bar{u} = 0  \ \  \text{ or }  \ \ a = -2\bar{u} \ \ \Leftrightarrow
  \ \ q = 0. \label{eqn:mKdVnonlinearityconditions} 
\end{equation}
	
In all cases, the soliton speed in terms of the Riemann invariants is
given by
\begin{equation}
  C(q,\bar{u}) = \bar{u}^2 + 2q^2 > 0, \text{ for } a \neq 0. \label{eqn:C}
\end{equation}	
	
As shown in Section~\ref{sec:soliton-mean-flow}, for modulations with
constant $q$, the wave conservation equation \eqref{eqn:consofwaves1}
is diagonalised by the variable $kp$, where $p(q, \ub)$ is given by
\eqref{eqn:pintegral}. Using \eqref{eqn:C} and $f'(u) = 3u^2$ in
\eqref{eqn:pintegral}, we determine $p(q,\bar{u})$ for mKdV solitonic
modulations,
\begin{equation}\label{p}
  \begin{aligned}
	p(q,\bar{u}) &=
    \exp\left(-\int_{\bar{u}_0}^{\bar{u}}\frac{u}{u^2-q^2}du\right) \\ 
	&= |q^2 - \bar{u}^2|^{-1/2}, \quad q^2 \neq \bar{u}^2 ,
  \end{aligned}
\end{equation}
where we have chosen $\bar{u}_0^2 = q^2 + 1$ for convenience.
	
\section{Classification of mean flows in the mKdV GP problem}
\label{sec:class_mkdv}

\begin{figure}[h]
  \centering
  \begin{subfigure}{0.45\textwidth}
    \centering
    \includegraphics[width=\textwidth]{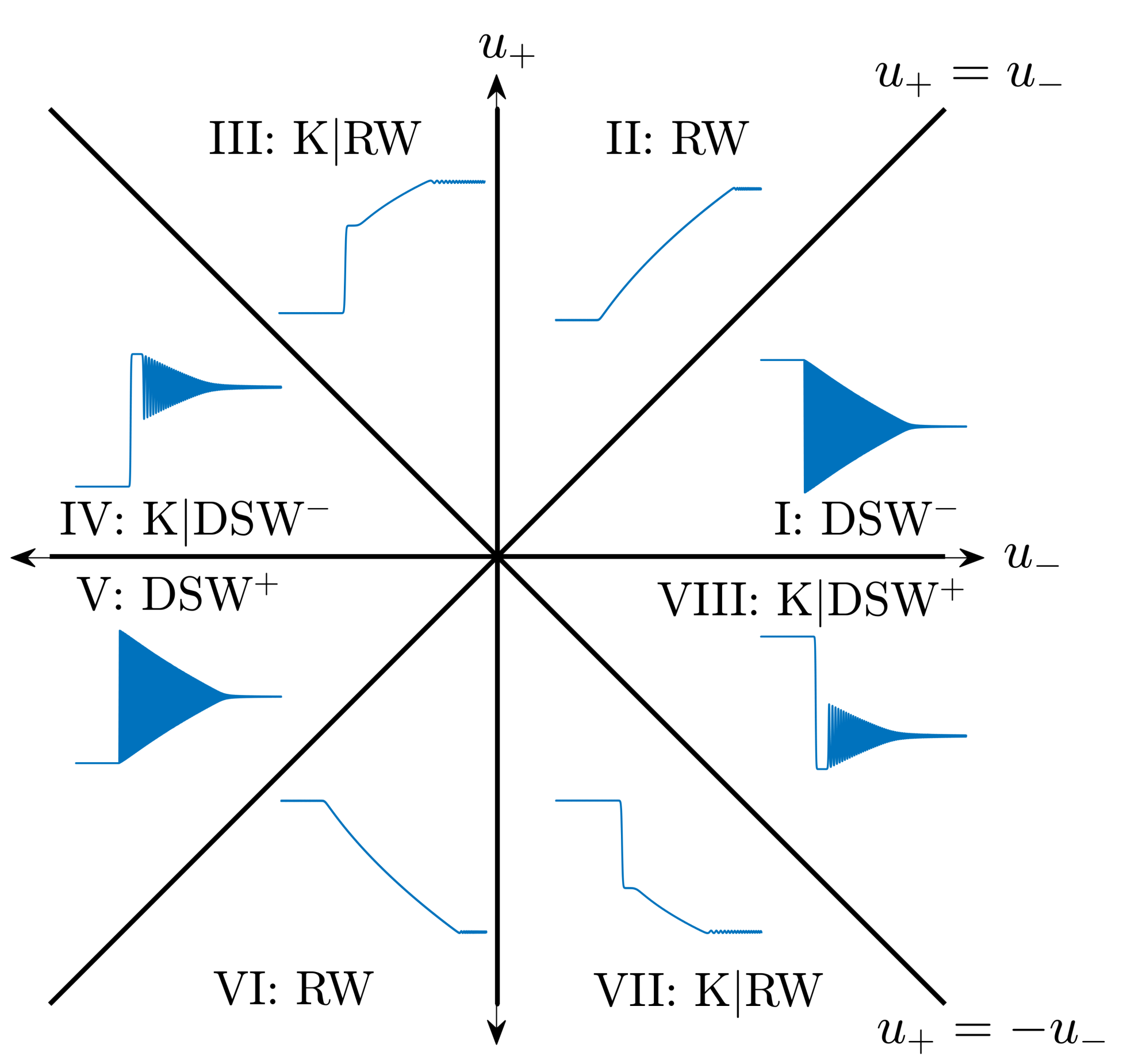}
    \caption{$\mu > 0$}
  \end{subfigure}
  \begin{subfigure}{0.45\textwidth}
    \centering
    \includegraphics[width=\textwidth]{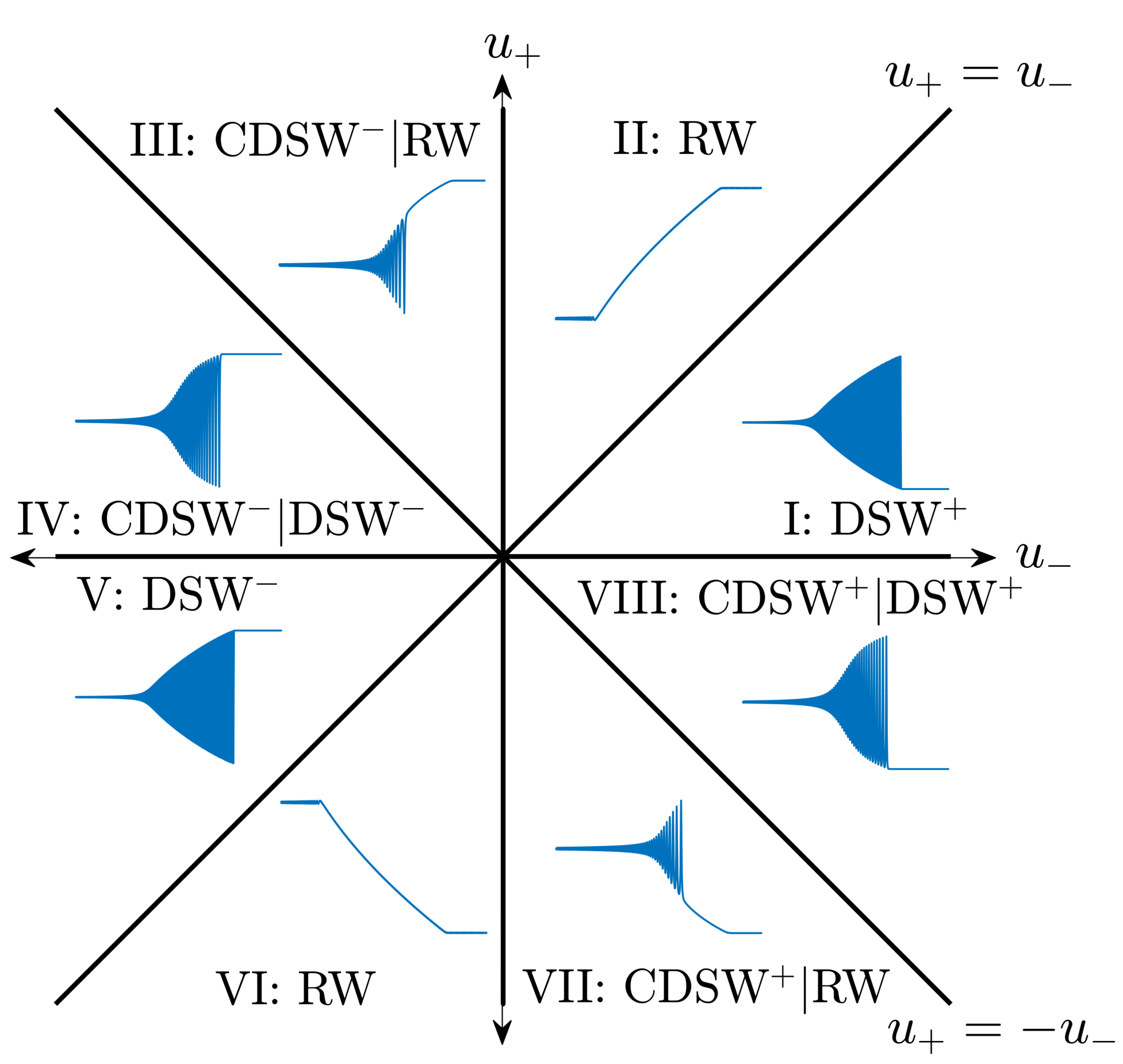}
    \caption{$\mu < 0$}
  \end{subfigure}
  \caption{Classification of the mKdV GP problem in terms of the
    initial values $u_\pm$ with representative numerical solutions,
    see \cite{kamchatnov_undular_2012},
    \cite{el_dispersive_2017}. Legend: (RW) rarefaction wave,
    (DSW$^{+/-}$) bright/dark dispersive shock wave, (K) kink,
    (CDSW$^{+/-}$) bright/dark contact DSW.}
  \label{fig:mKdVRiemannsolns}
\end{figure}

The solution to the GP problem for mKdV was classified in
\cite{el_dispersive_2017} by combining previous work on the Riemann
problem for either sign of dispersion \cite{chanteur_formation_1987,
  marchant_undular_2008} and elaborating on the GP problem
classification for the Gardner equation
$u_t + 6uu_x - 6\alpha u^2u_x + u_{xxx}$
\cite{kamchatnov_undular_2012}. The wave behaviour that emerges from
the GP problem depends on the sign of $\mu$ and relative sign and
magnitude of $u_-$ and $u_+$, as shown in the classification diagram
of Fig.~\ref{fig:mKdVRiemannsolns}. We refer to the octants in this
figure, as regions I to VIII, counted in a counterclockwise fashion.
Owing to its universality as a model of weakly nonlinear, long
dispersive waves \cite{el_dispersive_2017}, the mKdV equation provides
a fundamental description of the GP problem for other PDEs with
nonconvex flux.
	
Rarefaction waves and DSWs solve the GP problem in certain convex and
nonconvex cases.  Dispersive shock waves are classified as DSW$^+$ and
DSW$^-$ according to the polarity of the solitary wave generated at
one of the edges---leading or trailing, depending on the DSW
orientation.  In the nonconvex case, we see the emergence of
additional wave structures. These occur when the hydrodynamic flux
$f(u) = u^3$ exhibits an inflection point $u=0$ within the range of
step data \eqref{eq:2} so that $u_+ u_-<0$. Particularly, when
$\mu > 0$, and $u_- = -u_+$, the long-time asymptotic solution is a
kink, which is an undercompressive shock in the limit $\mu \to 0^+$.
When $\mu < 0$ and $u_- = -u_+$, the long-time asymptotic solution is
a contact DSW whose leading, algebraic soliton edge is a
dispersionless characteristic with velocity $3\bar{u}^2$ as
$\mu \to 0^-$. For other configurations with steps that pass through
0, the solution develops into a hybrid double wave structure as seen
in Fig.~\ref{fig:mKdVRiemannsolns}. We stress that in the DSW case,
the mean flow is interpreted as the local period average of the DSW's
oscillations.
	
We now present explicit expressions for the basic mean flows occurring
in the mKdV Riemann problem, distinguishing between convex and
nonconvex solutions. For brevity, we shall call them convex and
nonconvex mean flows, respectively.

\subsection{Convex mean flows}
	
{\it RW mean flows (Regions II and VI)}

\medskip RWs that emerge from the Riemann problems in Regions II and
VI of Fig.~\ref{fig:mKdVRiemannsolns} are described to leading order
by
\begin{equation}
  \bar{u}(x,t) = \left\{
    \begin{array}{ll} 
      u_-, & x<3u_-^2 t \\
      \text{sgn}(u_+-u_-) \sqrt{\frac{x}{3t}},
           &
             3u_-^2t < x < 3u_+^2t
      \\ 
      u_+, & x>3u_+^2 t.
    \end{array}\right. \label{eqn:RWsoln}
\end{equation}
This is the long-time approximation of the full mKdV solution that
includes dispersive regularisation of weak discontinuities at
$x = 3 u_\pm^2 t$.
	
\medskip
\noindent{\it DSW mean flows (Regions I and V)}

\medskip The GP modulation solution describing a DSW depends on the sign
of the dispersion coefficient.
	
According to \cite{el_dispersive_2017} for $\mu>0$ a DSW$^+$ is realized
as the solution to the Riemann problem with $u_- < u_+ < 0$, see
quadrant V in Fig.~\ref{fig:mKdVRiemannsolns}(a). The relevant GP
solution to the mKdV modulation equations \eqref{eqn:mKdVWhitham} is a
centred simple wave given by
\begin{equation}\label{GP1}
  \lambda_1 = u_-, \quad \lambda_3 = u_+, \quad
  W_2(u_-,\lambda_2,u_+) = \frac{x}{t}, 
\end{equation}
where the characteristic speed $W_2$ is given by \eqref{Wj+},
\eqref{eq:rtolambda} so
\begin{equation}
  W_2(u_-,\lambda_2,u_+) = u_-^2 + u_+^2 + \lambda_2^2 +
  2(u_-^2-\lambda_2^2)\frac{(1-m)K(m)}{E(m) - (1-m)K(m)}, \quad
  m=\frac{\la_2^2 - u_-^2  }{u_+^2 -u_-^2} .
  \label{eqn:mupos_W2}
\end{equation}	
	
To obtain the mean flow $\bar{u}$ through a DSW, we average the mKdV
periodic solution for $u$ over a period. Integrating
\eqref{eqn:mupos_periodicsoln} over the period $2K(m)$ and writing the
solution in terms of the Riemann invariants gives
\begin{equation}
  \bar{u} = -(\lambda_1+\lambda_2+\lambda_3)+
  2\frac{(\lambda_2+\lambda_3)}{K(m)}\Pi\left(\frac{\lambda_1 -
      \lambda_2}{\lambda_1+\lambda_3},m\right), \label{eqn:mupos_dswmeanflow} 
\end{equation}
where $\Pi$ is the complete elliptic integral of the third kind. The
dependence $\bar{u}(x,t)$ is obtained by inserting the modulation
solution \eqref{GP1}, \eqref{eqn:mupos_W2} in
\eqref{eqn:mupos_dswmeanflow}.

For $\mu<0$, a similar averaging over a period of
\eqref{eqn:muneg_periodicsoln} gives
\begin{equation}\label{mean2}
  \bar{u} = \lambda_1 + \lambda_2 - \lambda_3 -
  2\frac{(\lambda_1+\lambda_2)}{K(m)}\Pi \left ( \frac{\lambda_2+\lambda_3}
  {\lambda_3-\lambda_1},m \right ) .
\end{equation}
For a DSW$^+$ with $u_->u_+>0$ (quadrant I in
Fig.~\ref{fig:mKdVRiemannsolns}(b)), the GP solution to the modulation
equations is
\begin{equation}\label{GP2}
  \lambda_1 = -u_-, \quad \lambda_3 = u_+, \quad
  W_2(-u_-,\lambda_2,u_+) = \frac{x}{t},
\end{equation}
where the characteristic speed is given by \eqref{Wj-},
\eqref{eq:rtolambda} is
\begin{equation}
  W_2 = u_-^2 + u_+^2 + \lambda_2^2 + 2(u_+^2 -
  \lambda_2^2)\frac{(1-m)K(m)}{E(m) - (1-m)K(m)}. \label{eqn:muneg_W2} 
\end{equation}
Either \eqref{eqn:mupos_W2} or \eqref{eqn:muneg_W2} gives a
parameterisation of the DSW mean flow in terms of
$\lambda_2 \in (u_-,u_+)$, yielding $W_2(\bar{u})$.  This behaviour is
not affected by the sign of the dispersion coefficient $\mu$.
	
For solutions between the roots $u_1$ and $u_2$, the DSW$^-$ mean flow
can be found by applying the transformation
\eqref{eqn:mKdVinvarianttransform}.
	
For application to soliton-DSW mean flow interaction, it is
instructive to write down the evolution equation for the DSW mean flow
$\ub(x,t)$, the simple-wave equation
\begin{equation}
  \label{ubarPDE}
  \ub_t + W_2(\ub) \ub_x=0.
\end{equation}
As a matter of fact, $\ub(x,t)$ given by \eqref{eqn:mupos_dswmeanflow},
\eqref{GP1} (or \eqref{mean2}, \eqref{GP2}) satisfies equation
\eqref{ubarPDE}. The advantage of using the PDE \eqref{ubarPDE},
instead of the explicitly prescribed mean flow $\ub(x,t)$ will become
clear later, in Section~\ref{sec:soliton-convex}, where we shall use
it instead of the original mean flow equation $\ub_t + f'(\ub)\ub_x=0$
as part of the solitonic modulation system
\eqref{eqn:2by2modulationsystemdiagonal}.
	
\subsection{Nonconvex mean flows}
\label{sec:noncon_mean_flow}

As we have mentioned, nonconvex mean flows are generated if the
Riemann data \eqref{eq:2} satisfy $u_-u_+<0$. In contrast to
two-parameter convex mean flows, nonconvex mean flows are constrained,
one-parameter families of mKdV solutions and, because of this,
generally occur in combination with a convex mean flow---either a RW
or DSW, see regions III, IV, VII, VIII in
Fig.~\ref{fig:mKdVRiemannsolns}. The two classes of ``pure'' nonconvex
mKdV mean flows are kinks described by \eqref{eqn:mKdVkink} for
$\mu>0$ and contact DSWs (CDSWs) when $\mu<0$ described by modulated
trigonometric solutions \eqref{trig_sol} that exhibit an algebraic
soliton \eqref{eqn:mKdValgebraicsoliton} at one of its edges.

\medskip
\noindent {\it Kink mean flows ($\mu>0$, $u_+=-u_-$)}.

Unlike other mean flows that solve the mKdV GP problem, kinks are
localised, {\it steady} transitions between antisymmetric means
$\ub(-\infty)=u_-$ and $\ub(+\infty)=u_+=-u_-$ described by
\eqref{eqn:mKdVkink}. It has been shown in
\cite{leach_initial-value_2012} that kinks dominate the long-time
asymptotic solution of defocusing mKdV Riemann problems with
antisymmetric data.  Kinks are special in the sense that, in addition
to considering them as mean flows, we can also treat them as localised
soliton solutions that interact with convex mean flows such as RWs and
DSWs.
	
In the limit $\mu \to 0^+$, kinks are the weak discontinuous solutions
\begin{equation}
  \bar{u}(x,t) =
  \left\{ \begin{array}{ll}
            u_-, & x <u_-^2t, \\ -u_-, & x> u_-^2 t,
          \end{array}\right .\label{eq:kink_weak} 
\end{equation}	
of the hydrodynamic modulation equation $\ub_t + (\ub^3)_x=0$ for the
solitonic modulation system. The weak solution \eqref{eq:kink_weak}
represents an undercompressive shock
\cite{hayes_undercompressive_1999, lefloch_hyperbolic_2002} since the
hydrodynamic characteristic velocity $c=3u_-^2=3u_+^2$ is the same on
both sides of the shock.

\medskip
\noindent{\it  CDSW mean flows ($\mu <0$, $u_+ = - u_-$)}
	
A CDSW is a modulated trigonometric solution \eqref{trig_sol}
connecting antisymmetric states $u_-$ and $-u_+$, the negative
dispersion counterpart of the kink solution. The CDSW mean flow is
given by \eqref{mean2} with $\lambda_2=\lambda_3$ for CDSW$^+$ and
$\lambda_3=-\lambda_2$ for CDSW$^-$.
	
For CDSW$^+$, realised when $u_->0$, we have
\begin{equation}
  \label{eq:mean_CDSW}
  \ub= \lambda_1 + 2 \sqrt{\lambda_1^2 - \lambda_3^2},
\end{equation}
where the modulations of $\lambda_1$ and $\lambda_3$ are given by
\cite{el_dispersive_2017}
\begin{equation}\label{cdsw_mod}
  \lambda_1 = -u_-, \quad W_2=W_3= -3u_-^2 + 6 \lambda_3^2
  =\frac{x}{t}. 
\end{equation}
As earlier, the mean flow variations satisfy equation \eqref{ubarPDE}.

Although we call the CDSW mean flow nonconvex because its existence
necessitates nonstrict hyperbolicity of the mKdV-Whitham modulation
equations, the mean flow characteristic velocity $W_2(\ub)$ in a CDSW
is monotone along the simple wave solution curve \eqref{cdsw_mod}.
	
\section{mKdV soliton-mean flow interaction: transmission and phase
  conditions}  
\label{sec:mKdVsolitonichydrodynamics}

In order to obtain solutions to the soliton-mean interaction problem,
we seek simple wave solutions to the augmented solitonic modulation
system \eqref{eqn:ubarWhitham}, \eqref{eqn:consofwaves1} in which $q$
and $kp(q,\bar{u})$ are constant while the remaining Riemann
invariant, the mean flow $\bar{u}$, varies.
	

The ordering of the roots $u_i$ leads to constraints on the background
and amplitudes of the initial solitons. To simplify our analysis, we
will consider initial bright solitons in the tunnelling problem. The
solution for dark solitons can be obtained using the fact that the
mKdV equation is invariant under the transformation
\eqref{eqn:mKdVinvarianttransform}. For $\mu >0$, the amplitude of an
initial bright soliton must satisfy
\eqref{eqn:mupos_solitonexistence}: $\ub < 0$ and $0 < a < -2\ub$. For
$\mu < 0$, an initial bright soliton must satisfy
\eqref{eqn:muneg_solitonexistence}: $\ub \in \mathbb{R}$ and
$a > - 4\ub$.
	
For both signs of $\mu$, the transmission and phase conditions can be
determined from \eqref{eq:3},
\eqref{eqn:transmissionphasecondgeneral}, \eqref{q}, \eqref{p} as
\begin{equation}
  \frac{a_+}{2} +u_+ = \frac{a_-}{2} + u_- , \qquad \frac{k_-}{k_+} =
  \sqrt{\frac{q_-^2 - u_-^2}{q_+^2 - u_+^2}} =
  \sqrt{\frac{\frac{1}{4}a_-^2 + a_-u_-}{\frac{1}{4}a_+^2+a_+u_+}}.  
  \label{eqn:transmissionphasecond}
\end{equation}
Notably, these transmission and phase conditions are exactly the same
as those for the KdV equation $u_t + (u^2)_x = \mu u_{xxx}$ with
convex flux \cite{maiden_solitonic_2018}.  Although, for mKdV, the
conditions apply for both positive and negative soliton amplitudes.

A tunnelling condition \eqref{eq:16} fails when the characteristic
speeds $f'(\bar{u})$ and $C(q,\bar{u})$ cross, which occurs when (see
\eqref{eqn:mKdVhyperbolicitycondition})
\begin{equation}
  q^2 = \bar{u}^2 \implies a \in \{0, -4\bar{u}\}.
\end{equation}
Crossing through $a = 0$ gives the same condition as in the convex
case, where for bright solitons, $a > 0$ on the transmitted side
implies tunnelling, and $a \leq 0$ means the soliton is trapped. For
dark solitons, the inequalities must be reversed.
	
The additional tunnelling condition resulting from nonconvexity when
$\bar{u} < 0$ is the constraint
\eqref{eqn:muneg_solitonexistence} that the amplitude does not pass
through $-4\bar{u}$. When $a \to -4\bar{u}$, we again have trapping,
but with a non-zero amplitude and speed $3\bar{u}^2$. This limit
corresponds to an algebraic soliton. When $\mu>0$, the initial
amplitude will be less than $-4u_\pm$ since
\eqref{eqn:mupos_solitonexistence} holds for valid bright solitons,
and so the transmitted amplitude must also be smaller than
$-4u_\mp$. For $\mu<0$, initial amplitudes must satisfy
\eqref{eqn:muneg_solitonexistence}, so the transmitted amplitude must
also be greater than $-4u_\mp$.
	
Considering the intersection of the characteristic speeds is the most
direct way to verify the admissibility criteria \eqref{eq:16} for the
soliton to tunnel through the mean flow. However, we can also see how
the phase and transmission conditions
\eqref{eqn:transmissionphasecond} are affected. The phase shift
$\Delta$ can be obtained from the relation \eqref{phase_shift1},
yielding for mKdV
\begin{equation}\label{mkdv_shift}
  \frac{\Delta}{x_-}= \sqrt{\frac{\frac{1}{4}a_-^2 +
      a_-u_-}{\frac{1}{4}a_+^2+a_+u_+}} -1 
\end{equation}
for the forward (left to right) soliton transmission through a mean
flow.  If $a_+ = -2u_+$ as when strict hyperbolicity is lost, then
from \eqref{eqn:transmissionphasecond} we have
$\frac{k_-}{k_+} \to \infty$. For the backward (right to left)
transmission one simply interchanges $``+"$ and $``-"$ in
eq.~\eqref{mkdv_shift}.
		
\section{Soliton-convex mean flow interaction}
\label{sec:soliton-convex} 
	
First, we consider the tunnelling problem in the classical case of
convex mean flows: rarefaction waves (RWs) and dispersive shock waves
(DSWs). However, nonconvexity of the mKdV equation makes the problem
novel in the sense that both bright and dark solitons exist. We see
that the additional admissibility criterion of strict hyperbolicity
leads to more restrictive tunnelling conditions than in the convex
case. When trapping occurs in situations with soliton amplitudes
$a>0$, we find that an exponentially decaying soliton limits to an
algebraic soliton when $a \to -4\bar{u}>0$.
		
\subsection{Soliton tunnelling through RWs: Regions II and VI}
	
Rarefaction waves that emerge from the GP problems in Regions II and
VI of Fig.~\ref{fig:mKdVRiemannsolns} are described to leading order
by eq.~\eqref{eqn:RWsoln}. The RW is confined to the interval
$3u_-^2 t < x < 3u_+^2 t$.

To determine the admissible directionality for soliton-mean
interaction, an admissible soliton's velocity must be compared to the
edge velocity of the RW. For $\mu > 0$, it is only possible for
solitons to travel from right to left, implying backward soliton-mean
interaction while for $\mu < 0$, solitons can only go from left to
right. Soliton tunnelling occurs in either case if the system
maintains strict hyperbolicity and $a \ne 0$. The tunnelling
parameters are determined by the transmission conditions
\eqref{eqn:transmissionphasecond}.
	
\begin{figure}
  \centering
  \begin{subfigure}{0.27\textwidth}
    \centering
    \includegraphics[width=\textwidth]{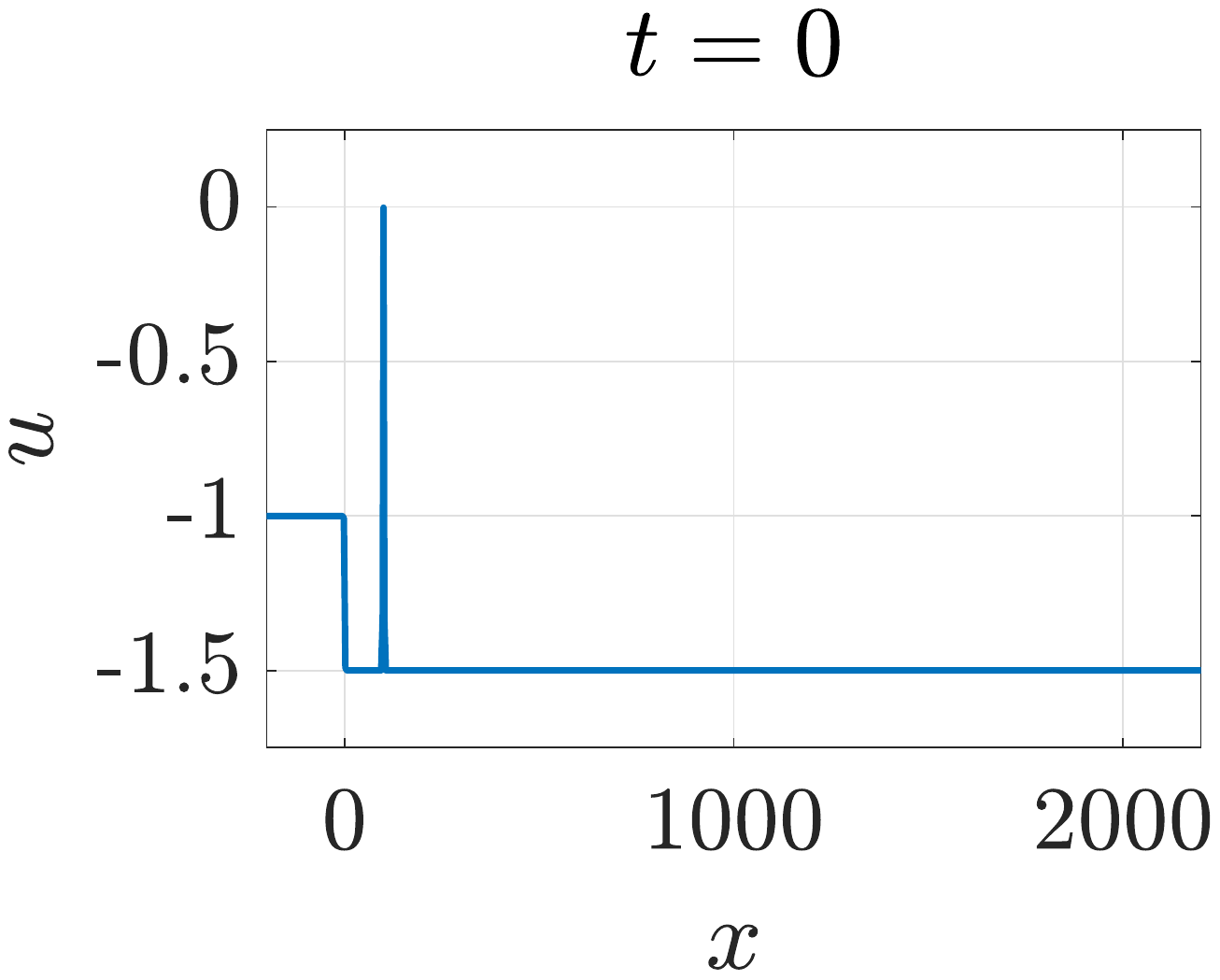}
  \end{subfigure}
  \begin{subfigure}{0.43\textwidth}
    \centering
    \includegraphics[width=0.92\textwidth]{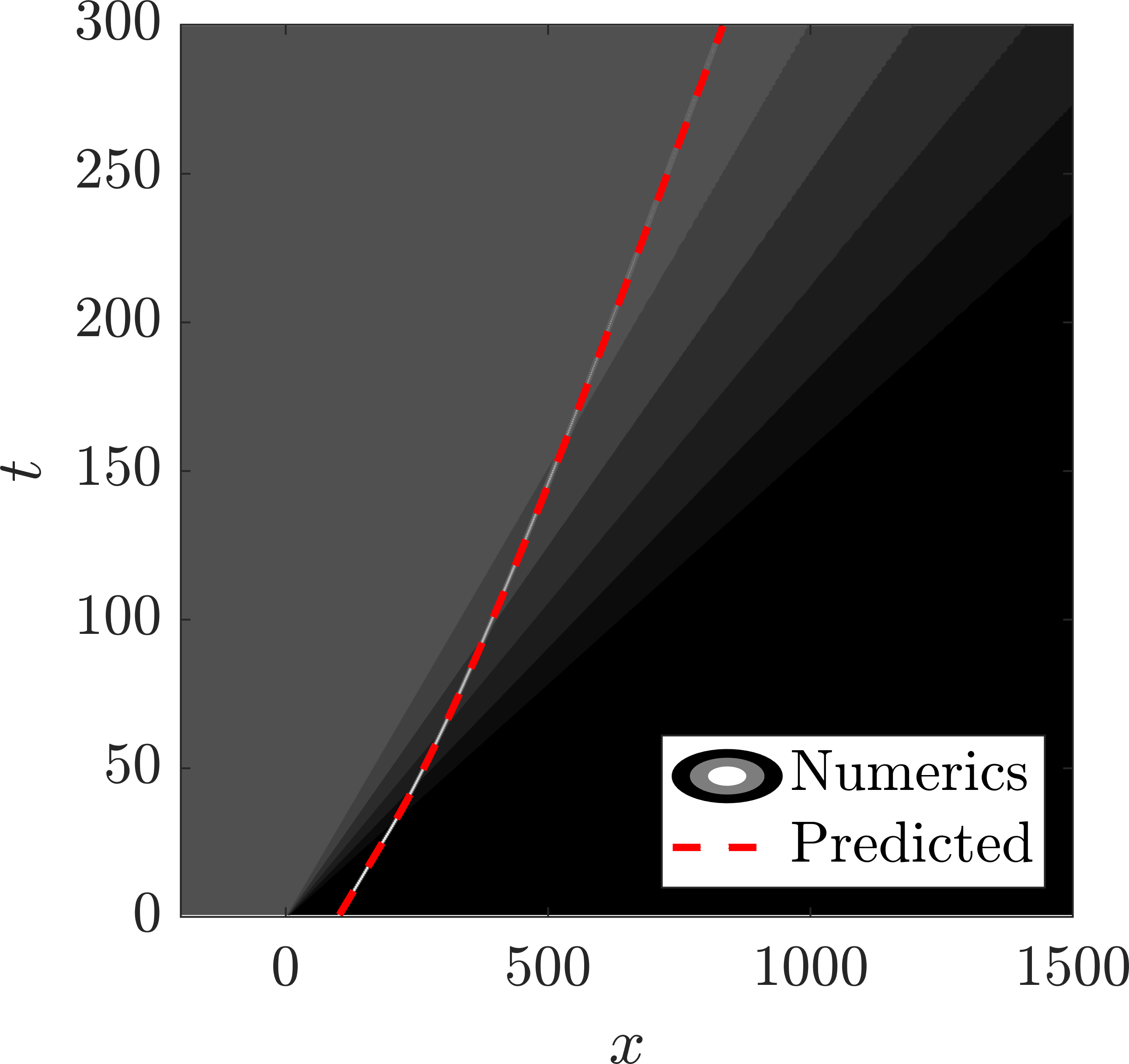}
  \end{subfigure}		
  \begin{subfigure}{0.27\textwidth}
    \centering
    \includegraphics[width=\textwidth]{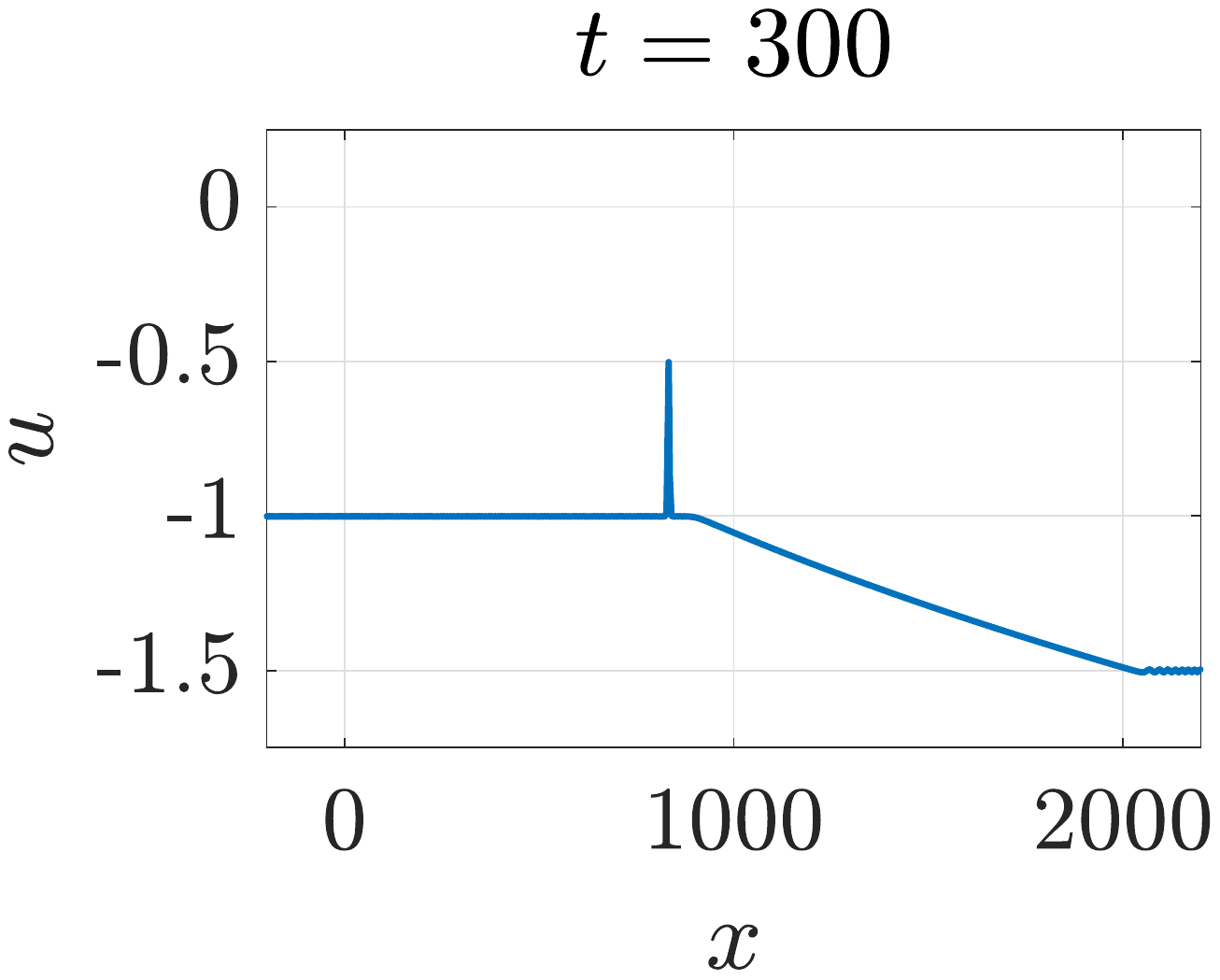}
  \end{subfigure}
  \caption{Soliton-RW interaction with $\mu = 1$, $a_+ = 1.5$,
    $x_+ = 100$, $u_- = -1$, $u_+ =-1.5$. The amplitude $a_-$
    predicted by the transmission condition
    \eqref{eqn:transmissionphasecond} is 0.5, and the predicted phase shift $\Delta$ is 96.40. The numerical
    solution gives $a_- = 0.4981$ and $\Delta = 95.02$ at $t = 300$.}
  \label{fig:case1_region6}
\end{figure}

For $\mu > 0$, we focus on RWs in region VI of
Fig.~\ref{fig:mKdVRiemannsolns} because bright solitons are
admissible.  The case of region II with dark soliton-RW interaction
can be obtained by the transformation
\eqref{eqn:mKdVinvarianttransform}.  Initialising $x_0 = x_+$,
$u_+ < 0$, $0 < a(x_+,0) = a_+ < -2u_+$, we only need to check that
$0 < a_- < -4u_-$ for $u_+ < u_- < 0$ to prove that the characteristic
speeds did not cross because the RW transitions continuously and
monotonically from $u_-$ to $u_+$. We use the transmission condition
\eqref{eqn:transmissionphasecond} to express this inequality in terms
of the initial soliton amplitude $a_+$ for a bright soliton starting
on the right,
\begin{equation}
  2(u_- -u_+) < a_+ < -2(u_+ + u_-).
\end{equation}
The second inequality is automatically satisfied by any valid initial
soliton. Hence, the first is a sufficient condition for tunnelling,
\begin{align}
  \label{eq:17}
  a_+ > a_{cr} = 2(u_- - u_+).
\end{align}
Since $u_+ < u_- < 0$ in region VI, \eqref{eq:17} gives a positive
critical amplitude for transmission to occur. A numerical example of a
soliton tunnelling through a RW in this case is shown in
Fig.~\ref{fig:case1_region6}.

The soliton trajectory is specified by
$\mathrm{d}x/\mathrm{d}t = C(q,\ub)$, where $C(q, \ub)$ is given by
\eqref{eqn:C}. Integrating this equation we obtain
\begin{equation}
  x(t) = \left\{
	\begin{array}{ll} 
      (u_-^2 + 2q^2)t + E, & x<3u_-^2 t \\
      3q^2t + Dt^{1/3}, & 3u_-^2 t < x < 3u_+^2 t \\
      (u_+^2 + 2q^2)t + x_+, & x>3u_+^2 t
	\end{array}
  \right. \label{eqn:RWsolitontraject}
\end{equation}
where $D$ and $E$ are obtained by continuity of $x(t)$
\begin{align}
  D &= \frac{3}{2}x_+^{2/3}(2u_+^2-2q^2)^{1/3} \\
  E &= x_+ \sqrt{\frac{u_+^2-q^2}{u_-^2-q^2}}.
\end{align}
The phase shift is $\Delta = E - x_+$, which matches the condition
given by \eqref{eqn:transmissionphasecond}.
	
A similar analysis can be carried out for each region in
Fig.~\ref{fig:mKdVRiemannsolns} to determine the tunnelling
criterion. We summarise the remaining results without detailing the
analysis for each case in Table \ref{table:summaryRW} for either sign
of $\mu$ in Regions II and VI. Note that for $\mu < 0$ in Region VI,
the tunnelling criterion is different than the condition that
$a_+ > 0$. This is because there are cases for valid initial soliton
amplitudes $a_-$ where the amplitude crosses $-4\bar{u}$ during
interaction with the RW, causing the soliton to become trapped and
travel too slowly to reach the other side. In the limit as
$t \to \infty$, the trapped soliton becomes an algebraic soliton
travelling at constant hydrodynamic characteristic velocity
$3\bar{u}^2$.
	
\begin{figure}
  \centering
  \begin{subfigure}{0.45\textwidth}
    \includegraphics[width=\textwidth]{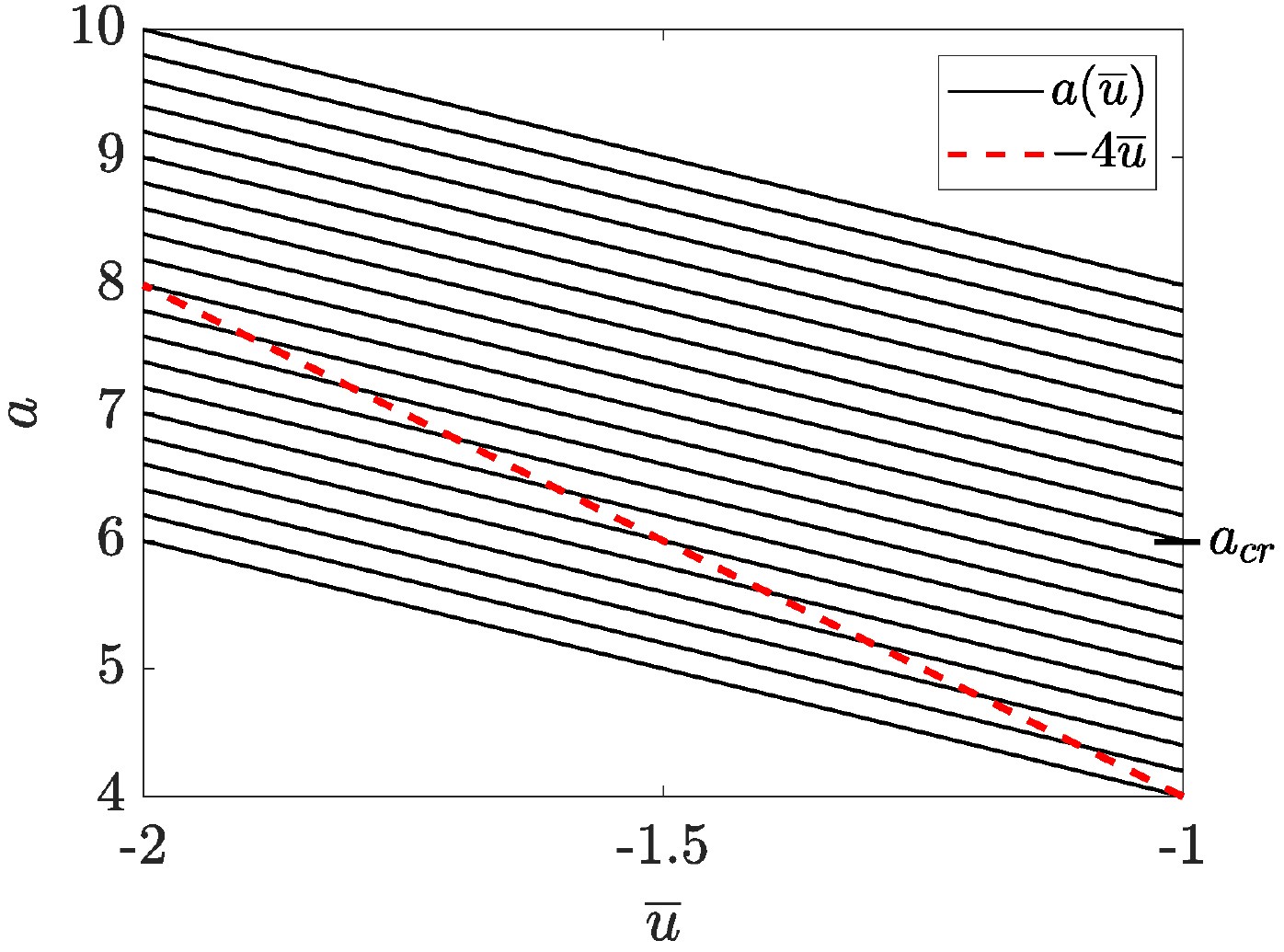}
    \caption{Soliton amplitudes.}
    \label{fig:case3region6_avsubar}
  \end{subfigure}
  \begin{subfigure}{0.45\textwidth}
    \centering
    \includegraphics[width=0.95\textwidth]{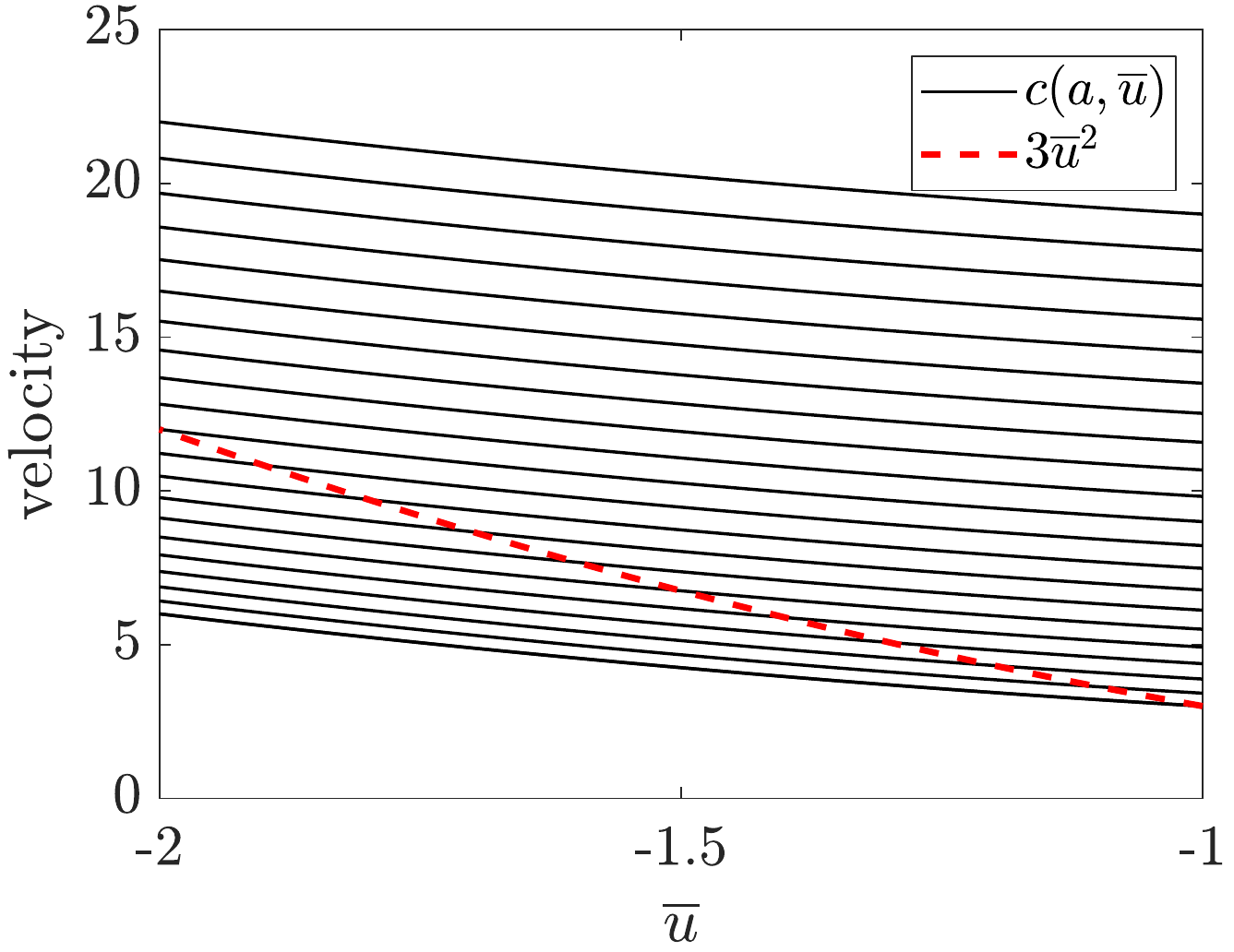}
    \caption{Soliton velocities.}
    \label{fig:case3region6_w2cs}
  \end{subfigure}
  \caption{Simple wave curves of constant $q(a,\bar{u})$ (solid) and
    curves of nonstrict hyperbolicity (dashed) illustrating
    interaction with a RW with $u_- = -1$, $u_+ = -2$ and $\mu = -1$
    for various initial amplitudes $a_-$. Left: soliton amplitudes as
    a function of $\bar{u}$; right: soliton speeds. When travelling
    from $u_-$ to $u_+$ with $a_-<a_{cr}$, the solid and dashed curves
    intersect, corresponding to loss of strict
    hyperbolicity.}
  \label{fig:case3region6}
\end{figure}

Figure \ref{fig:case3region6} illustrates the loss of strict
hyperbolicity when $\mu < 0$ for nonzero amplitudes by depicting the
wave curves $a(\ub)$ corresponding to constant $q(a,\ub)$ and the
corresponding soliton speed $c(a(\ub),\ub)$. For interaction to occur
in region VI, solitons are initialised on the left at $x_0 = x_-$ with
mean flow $u_- < 0$ (we take $u_- = -1$ for definiteness) and
amplitudes satisfying \eqref{eqn:mupos_solitonexistence}. 
For initial amplitudes $-4 u_- < a_- < a_{cr} = -2(u_+ + u_-)$,
solitons pass through the mean flow from $u_-$ to $u_+ < u_-$
($u_+ = -2$ for definiteness) maintaining positive amplitude.  But
these wave curves eventually intersects the critical line $-4\bar{u}$
(shown in red).  In Fig.~\ref{fig:case1region5_w2cs}, the
corresponding soliton speeds are plotted. Intersection of $a(\bar{u})$
with $-4\bar{u}$ corresponds to the intersection of the soliton
velocity $c(a,\bar{u})$ with the characteristic velocity $3\bar{u}^2$,
also shown in red.  As the two coincide, the soliton asymptotically
limits to a trapped algebraic soliton propagating with characteristic
velocity.
\begin{table}
  \centering
  \begin{tabular}{c|c|l|l}
    Dispersion & Direction & \begin{tabular}[c]{@{}c@{}}Region
                               II - RW\\ ($u_+ > u_- >
                               0$)\end{tabular}
               & \begin{tabular}[c]{@{}c@{}}Region VI - RW\\
                   ($u_+ < u_- < 0$)\end{tabular} \\ \hline 
    $\mu > 0$ & $R \to L$ & No bright soliton solutions
               &  \begin{tabular}[c]{@{}l@{}}Tunnelling if\\
                    $a_+ > a_{cr} = 2(u_-  -u_+)$\end{tabular} \\ \hline
    $\mu < 0$ &$L \to R$ & \begin{tabular}[c]{@{}l@{}}Tunnelling
                             if \\  $a_- > a_{cr} = 2(u_+ -
                             u_-)$\end{tabular}
               & \begin{tabular}[c]{@{}l@{}}Tunnelling if \\
                   $a_- > a_{cr} =  -2(u_+ + u_-)$\end{tabular}
  \end{tabular}%
  \caption{Results for the bright soliton tunnelling problem through
    RWs. $R\to L$ means that $x_0 = x_+$ and the soliton propagates
    from right ($R$) to left ($L$), otherwise $x_0 = x_-$
    ($L \to R$).}
  \label{table:summaryRW}
\end{table} 
      
\subsection{Soliton tunnelling through DSWs: Regions I and V}

For $\mu > 0$, the DSW leading and trailing edge velocities for both
regions I and V of Fig. \ref{fig:mKdVRiemannsolns} are given by
$s_+ = 6u_-^2 - 3u_+^2$ and $s_- = u_-^2 + 2u_+^2$
\cite{el_dispersive_2017}. Comparing the soliton's initial velocity to
the edge velocities, we see that there can be interaction in both
directions in region V ($u_-<u_+<0$).  In region I ($0 < u_+ < u_-$),
bright soliton solutions do not exist.
	
First, we consider the backward interaction in region V in which
$x_0 = x_+$. For the interaction to occur, the soliton speed
$c(a_+,u_+)$ must be smaller than $s_+$, which gives the condition
\begin{equation}
  \label{eq_cond_hyp}
  2(u_- - u_+) < a_+ < -2u_+.
\end{equation}
This condition is satisfied for any admissible, initial bright soliton
$a_+$. It and the transmission condition
\eqref{eqn:transmissionphasecond} imply that $0 < a_- < -4u_-$ so,
invoking monotonicity of the DSW mean flow, we can see that strict
hyperbolicity and amplitude positivity is maintained along the soliton
trajectory.  The soliton will always tunnel.
		
Next, we consider forward interaction where $x_0 = x_-$. In order for
the soliton to overtake the DSW, we require $c(a_-,u_-) > s_-$, which
implies that
\begin{equation}
  q^2 > u_+^2 . \label{eqn:muposregionV}
\end{equation}
Then one of
\begin{equation}
  \label{eq:18}
  a_- > -2(u_+ +u_-), \quad \text{ or } \quad a_- < 2(u_+ - u_-)
\end{equation}
holds.  The first inequality in \eqref{eq:18} cannot be satisfied for
an admissible, initial bright soliton constrained by
\eqref{eqn:mupos_solitonexistence}. The second inequality in
\eqref{eq:18} can be satisfied by an initial bright soliton, but
\eqref{eqn:transmissionphasecond} implies that $a_+ < 0$ so the
soliton is trapped. We can also see this by comparing the
characteristic velocities $c(a_-,u_-)$ and $3u_-^2$ where valid
initial soliton amplitudes $a_-$ satisfying
\eqref{eqn:mupos_solitonexistence} result in $q^2 < u_-^2$. Since
$q^2 > u_+^2$ is necessary for the interaction to occur (see
\eqref{eqn:muposregionV}), $u_+^2 > u_-^2$, and $\bar{u}$ is
continuous, the velocities must intersect and therefore the soliton is
trapped by the DSW.

For $\mu < 0$, the DSW leading and trailing edge speeds are given by
$s_+ = u_+^2 + 2u_-^2$ and $s_- = 6u_+^2 - 3u_-^2$
\cite{el_dispersive_2017}. Initial solitons with either $x_0 = x_+$ or
$x_0 = x_-$ can exist in both Regions I and V and the tunnelling
criterion can again be determined by comparing velocities and then
looking at the admissibility criterion for tunnelling.  Table
\ref{table:summaryDSW} summarises the results for bright soliton-DSW
interaction.
	
\begin{table}
  \centering
  \begin{tabular}{l|l|l|l}
    Dispersion & Direction & \begin{tabular}[c]{@{}l@{}}Region I -
                               DSW\\ ($u_- > u_+ > 0$)\end{tabular}
               & \begin{tabular}[c]{@{}l@{}}Region V - DSW\\ ($u_- <
                   u_+ < 0$)\end{tabular} \\ \hline
    \multirow{2}{*}{$\mu > 0$} & $R \to L$ & No soliton solutions
               &  \begin{tabular}[c]{@{}l@{}}Tunnelling always
                    occurs\\\end{tabular} \\  \cline{2-4} 
               & $L \to R$ & No soliton solutions
               & \begin{tabular}[c]{@{}l@{}}Interaction and trapping
                   if \\ $a_- < 2(u_+ - u_-)$  \end{tabular}\\ \hline
			\multirow{2}{*}{$\mu < 0$} & $R \to L$ & No interaction
               &  \begin{tabular}[c]{@{}l@{}}Interaction and trapping
                    if \\ $a_+ < -2(u_+ + u_-)$\end{tabular} \\ \cline{2-4}
			& $L \to R$ & \begin{tabular}[c]{@{}l@{}}Tunnelling always
                            occurs\end{tabular}
               & \begin{tabular}[c]{@{}l@{}}Tunnelling if \\ $a_- >
                   a_{cr} = 2(u_+ - u_-)$\end{tabular}
  \end{tabular}%
  \caption{Results for the bright soliton-DSW interaction problem.}
  \label{table:summaryDSW}
\end{table}

\begin{figure}
  \centering
  \begin{subfigure}{0.27\textwidth}
    \centering
    \includegraphics[width=\textwidth]{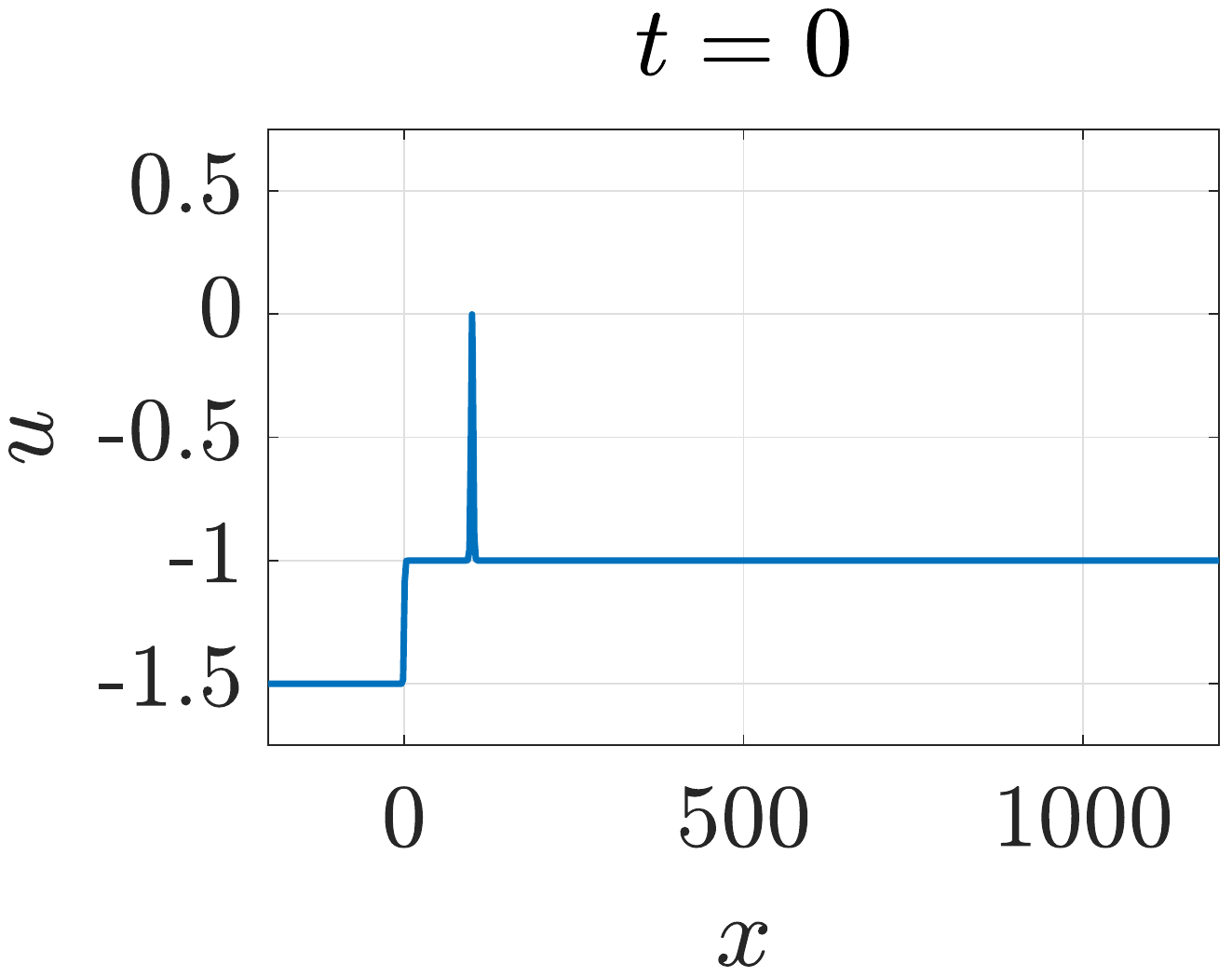}
  \end{subfigure}
  \begin{subfigure}{0.4\textwidth}
    \centering
    \includegraphics[width=\textwidth]{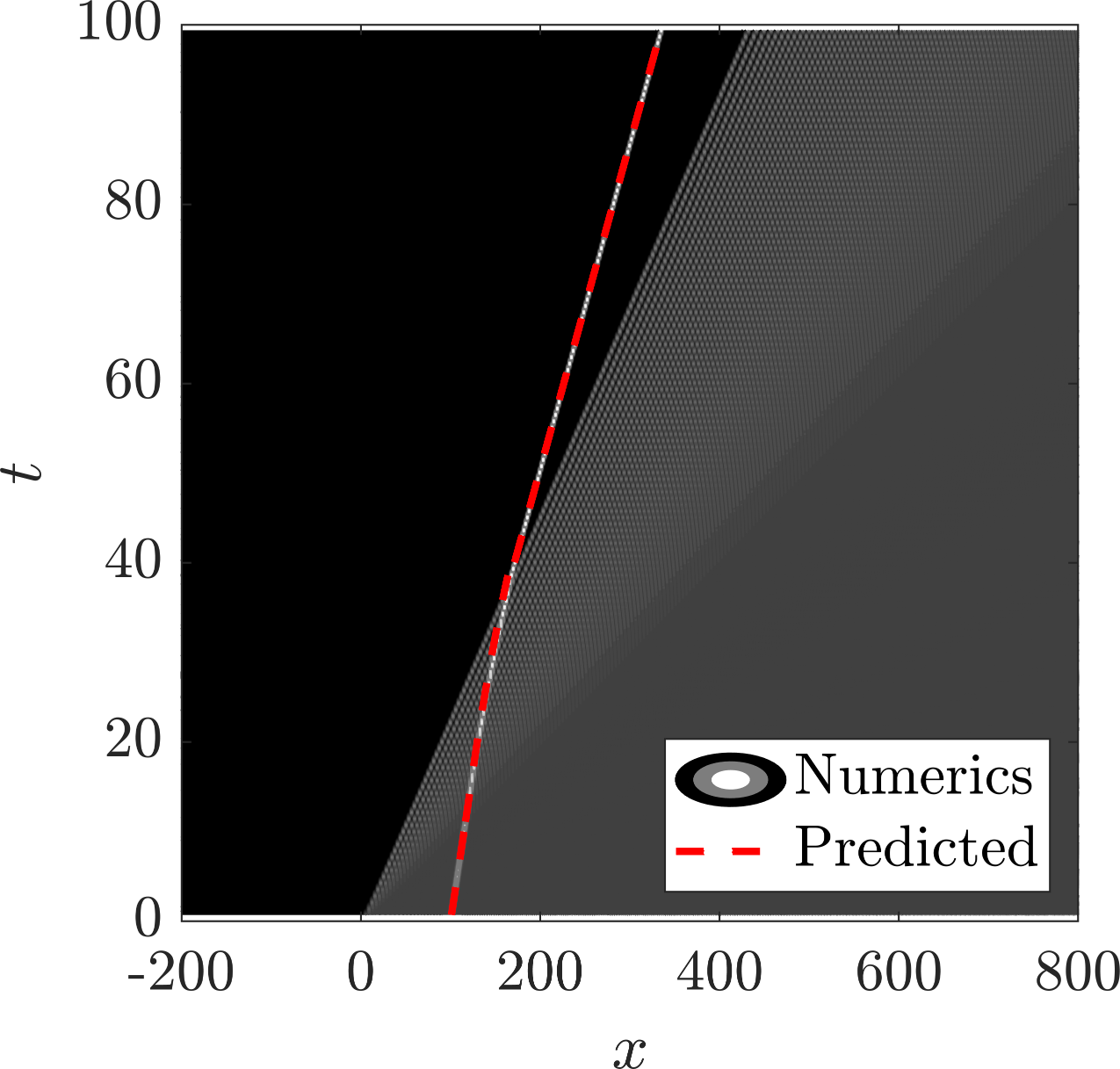}
  \end{subfigure}		
  \begin{subfigure}{0.27\textwidth}
    \centering
    \includegraphics[width=\textwidth]{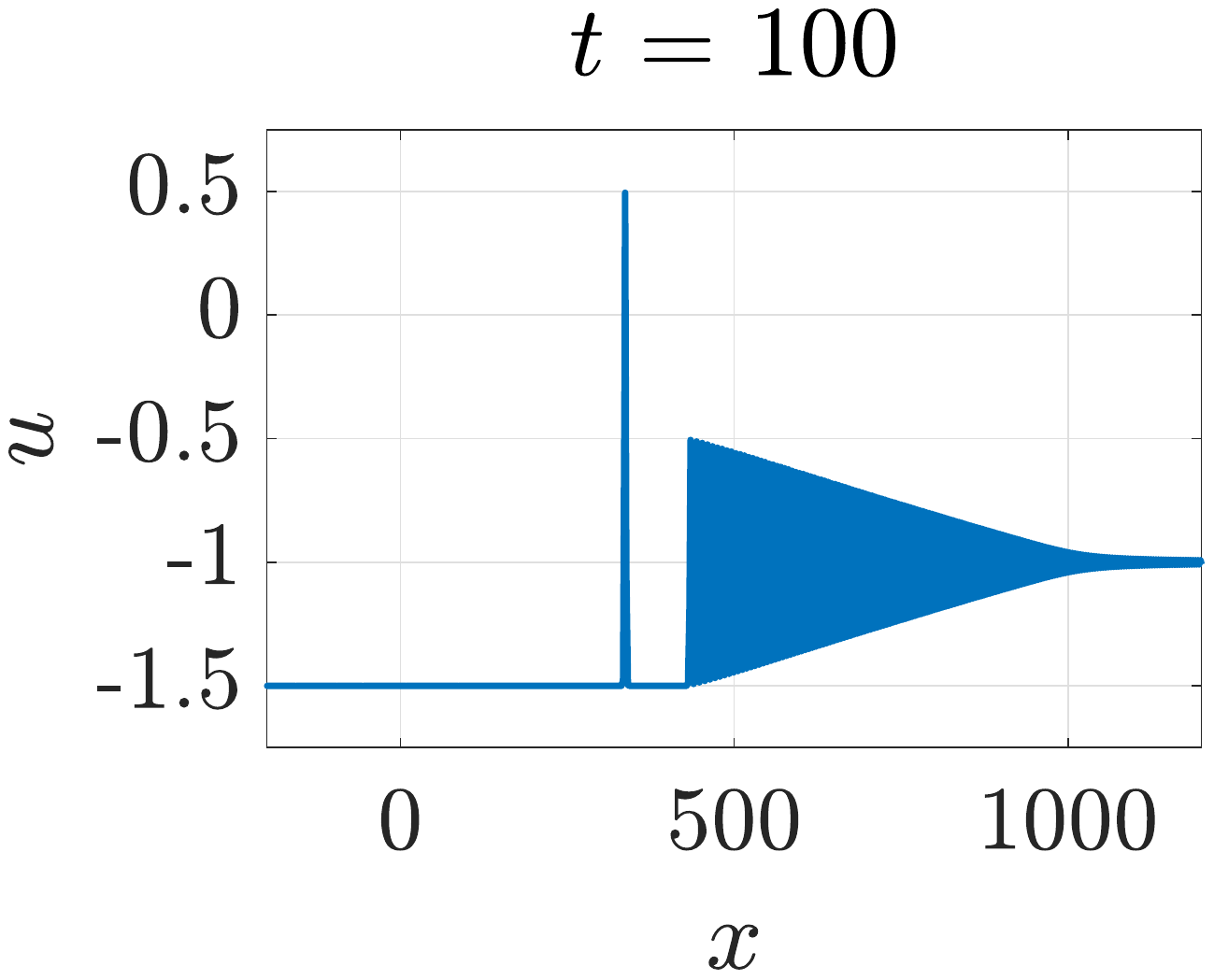}
  \end{subfigure}
  \caption{Soliton-DSW interaction with $\mu = 1$, $a_+ = 1$,
    $x_+ = 100$, $u_- = -1.5$, and $u_+ = -1$. The predicted $a_-$ is
    2 and the predicted $\Delta$ is $-38.76$. At $t=100$, the
    numerical solution gives $a_- = 1.9960$ and $\Delta = -38.33$.}
  \label{fig:case1_region5}
\end{figure}

Numerical experiments on soliton tunnelling through DSWs were
conducted for both signs of $\mu$ with results given in Table
\ref{table:DSWnumericsCase1} and Table
\ref{table:DSWnumericsCase3}. Good agreement was found between the
predicted amplitude and phase shift (equations
\eqref{eqn:transmissionphasecond}, \eqref{mkdv_shift}) and the
numerical results, confirming that hydrodynamic reciprocity is
maintained. Figure \ref{fig:case1_region5} shows one sample numerical
solution.
	
\begin{table}
  \centering
  \begin{tabular}{cccccccc}
    $u_+$ & $u_-$ & $a_+$ & $T_{final}$
    & $a_-$ (pred) & $a_-$ (num) & $\Delta  x/x_-$ (pred)
    & $\Delta x/x_-$  (num)\\
    \hline
    -1 & -1.5 & 0.1 & 300 & 1.1 & 1.0968 & -0.7310 & -0.7226 \\ 
    -1 & -1.5 & 0.5 & 200 & 1.5 & 1.4914 & -0.4908 & -0.4880 \\
    -1 & -1.5 & 1 & 100 & 2 & 1.9960 & -0.3876 & -0.3833 
  \end{tabular}
  \caption{Numerical tests of backward ($R \to L$) bright
    soliton-DSW interaction for $\mu = 1$ and $x_+ = 100$ in Region V.}
  \label{table:DSWnumericsCase1}
\end{table}
	
\begin{table}
  \centering
  \begin{tabular}{cccccccc}
    $u_-$ & $u_+$ & $a_-$ & $T_{final}$
    & $a_+$ (pred) & $a_+$ (num) & $\Delta  x/x_-$ (pred)
    & $\Delta x/x_-$ (num)\\
    \hline
    1.5 & 1 & 0.1 & 170 & 1.1 & 1.0989 & -0.6703 & -0.6577 \\
    1.5 & 1 & 0.5 & 170 & 1.5 & 1.4740 & -0.3724 & -0.3590 \\
    1.5 & 1 & 1 & 70 & 2 & 1.9818 & -0.2362 & -0.2320 \\
    -0.6 & -0.1 & 2.6 & 100 & 1.6 & 1.5966 & -0.4796 & -0.4539 \\		 
  \end{tabular}
  \caption{Numerical tests of forward ($L \to R$) bright soliton-DSW
    interaction, $\mu = -1$ and $x_- = 100$ in Region I and V.}
  \label{table:DSWnumericsCase3}
\end{table}
	
For a RW, the soliton trajectory $x(t)$ and amplitude $a(\bar{u})$
throughout the soliton-RW interaction is known. In contrast, the mean
flow $\ub(x,t)$ for a DSW is given by the modulation of a periodic
travelling wave and is more complicated. Although the soliton
amplitude and phase shift on either side of the DSW can be predicted
without knowing the space-time variation of the mean flow in the DSW's
interior by invoking hydrodynamic reciprocity (see Section
\ref{sec:hydro_recipr}), the Riemann invariants $q$ and $kp$ of the
augmented solitonic system \eqref{eqn:ubarWhitham},
\eqref{eqn:consofwaves1} are not held constant throughout the DSW.
What is desired is some way to estimate the wave curve $a(\ub)$ within
the DSW.  Since $\ub$ is known and now described by equation
\eqref{ubarPDE} rather than $\ub_t + f'(\ub)\ub_x=0$, we require an
alternative approach to approximating the wave curve $a(\ub)$.

To obtain $a(\bar{u})$ along the soliton trajectory
$\mathrm{d}x/\mathrm{d}t = c(a, \ub)$, we make an assumption that
soliton-DSW interaction can be approximated by the interaction of a
soliton with the DSW mean flow, and take advantage of the
characteristic ODE \eqref{char_ODE} in which we replace the
characteristic velocity $f'(\ub)$ in the RW solution with the
characteristic velocity $W_2(\ub)$ of the simple wave equation
\eqref{ubarPDE} for the DSW modulation solution.  The velocity
$W_2(\ub)$ is specified parametrically by \eqref{eqn:mupos_W2},
\eqref{eqn:mupos_dswmeanflow} (or equivalently, \eqref{eqn:muneg_W2},
\eqref{mean2}). Thus, we obtain the ODE
\begin{equation}
  \frac{\mathrm{d}\tilde{k}}{\mathrm{d}\bar{u}} = \frac{\partial
    _{\ub}\tilde{\omega}_0}{W_2(\bar{u})- \partial_{\tilde{k}}
    \tilde{\omega}_0}, \quad \tilde{k}(u_-) = \tilde{k}_-  \label{eqn:dkdu},
\end{equation}
where, as earlier, $\tilde{\omega}_0 = -i\omega_0(i\tilde{k},\bar{u})$
and $\omega_0(k,\bar{u})$ is the mKdV linear dispersion relation
\eqref{eqn:mKdVdispersionrelation}.  The relation between the
conjugate wavenumber $\tilde k$ and the soliton amplitude $a$ is given
by equation \eqref{conjug}, which is
\begin{equation}
  \tilde{k}^2 = -\frac{1}{\mu}\left(\frac{1}{2}a^2+2a\bar{u}\right)  \label{eqn:mKdVconjugatewavenumber},
\end{equation}
so that $\tilde{k}_-$ is \eqref{eqn:mKdVconjugatewavenumber} evaluated
at $a = a_-$, $\ub = u_-$.  Substituting the expression for
$W_2(\bar{u})$ given by \eqref{eqn:mupos_W2} or \eqref{eqn:muneg_W2}
and the dispersion relation \eqref{eqn:mKdVdispersionrelation} into
\eqref{eqn:dkdu}, we can numerically integrate for
$\tilde{k}(\bar{u})$ and invert \eqref{eqn:mKdVconjugatewavenumber} to
solve for the approximate wave curve $a(\bar{u})$. Since
\eqref{eqn:mKdVconjugatewavenumber} is double valued, we use the
existence conditions \eqref{eqn:mupos_solitonexistence} for $\mu > 0$
and \eqref{eqn:muneg_solitonexistence} for $\mu < 0$ in order to
determine the correct value for $a$.

We now consider the implications of this mean flow approach toward
understanding soliton-DSW interaction for each sign of $\mu$
separately.

\begin{figure}
  \centering
  \begin{subfigure}{0.48\textwidth}
  	\centering
    \includegraphics[width=\textwidth]{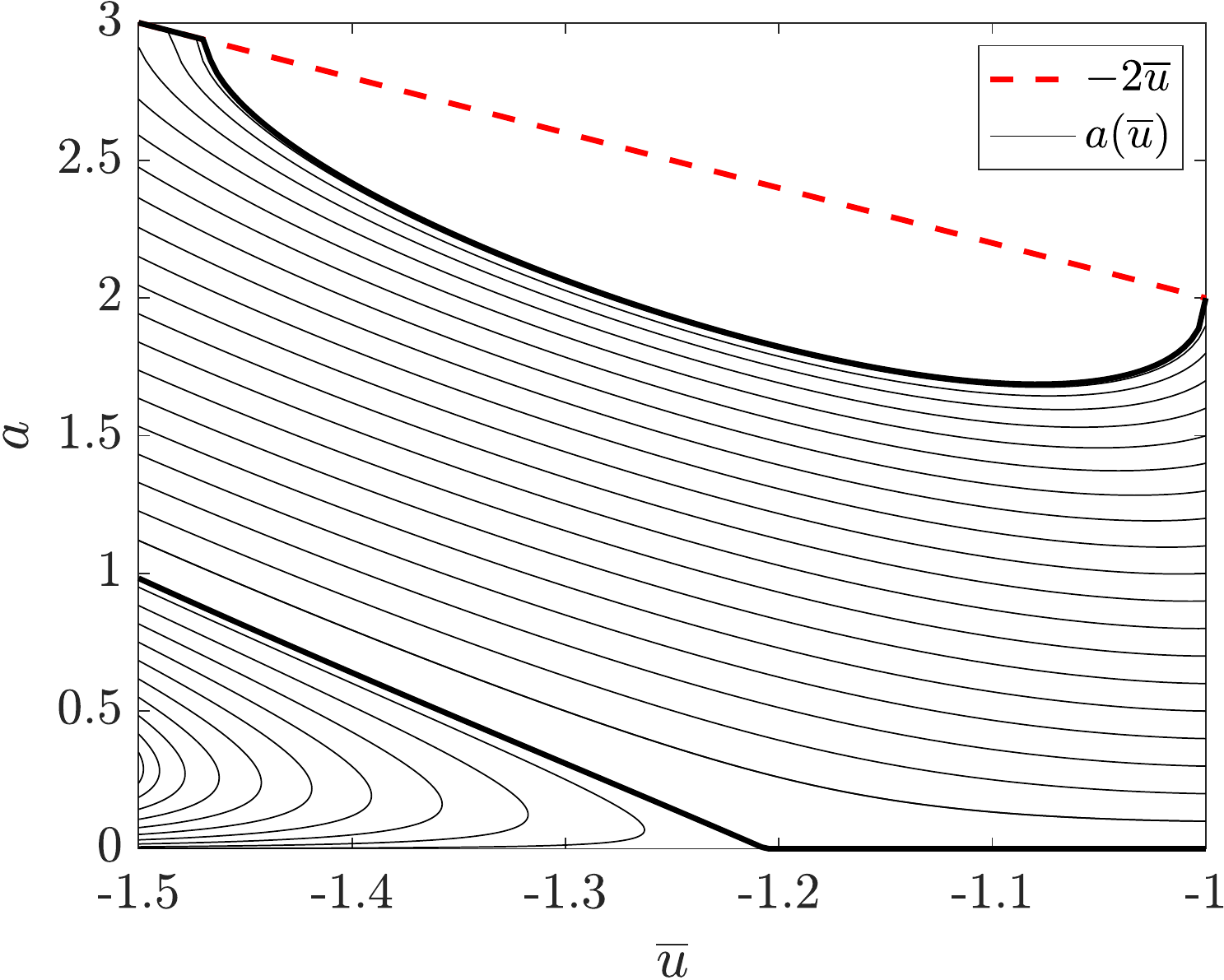}
    \caption{Soliton amplitudes.}
    \label{fig:case1region5_avsubar}
  \end{subfigure}
  \begin{subfigure}{0.48\textwidth}
    \centering
    \includegraphics[width=\textwidth]{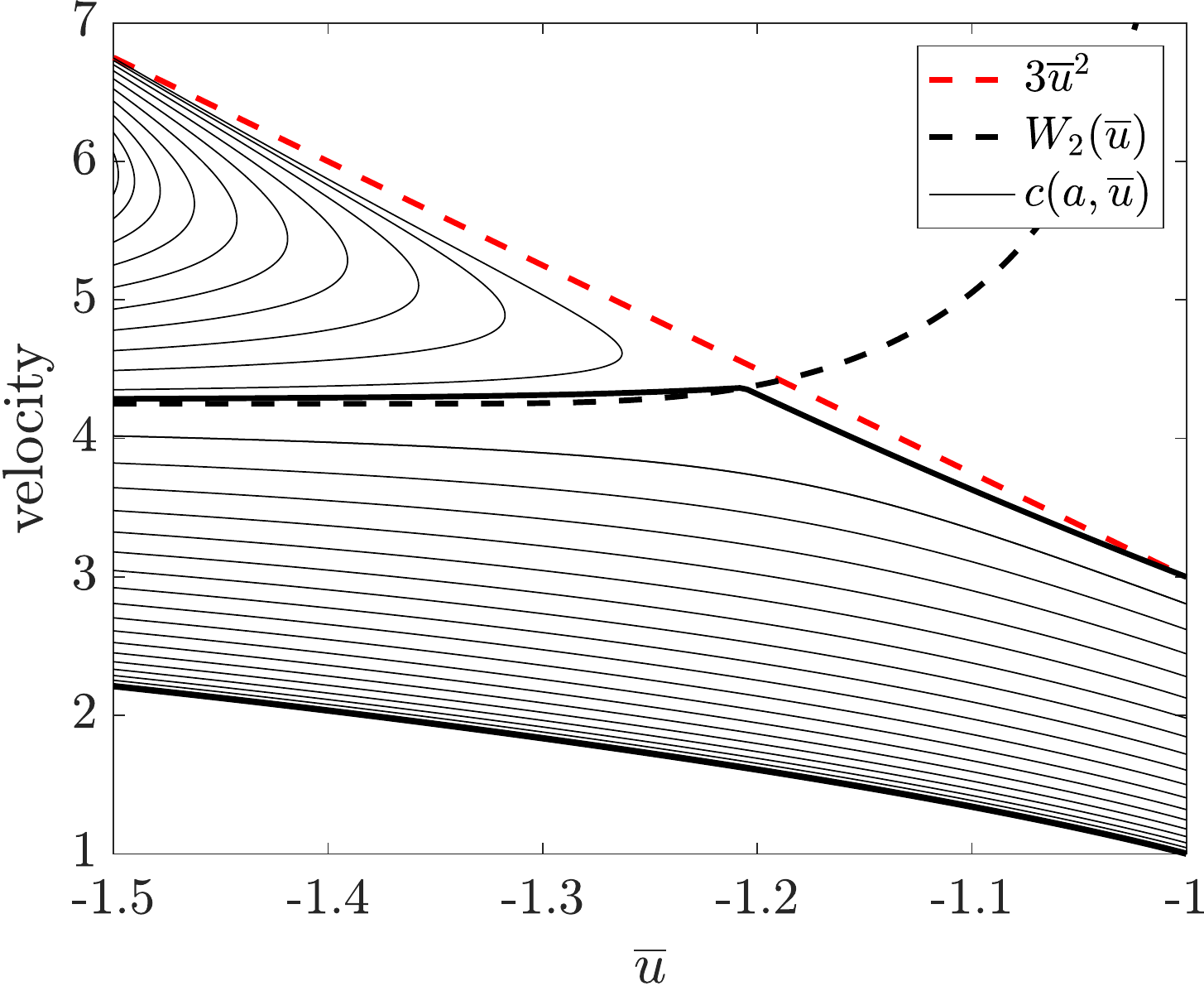}
    \caption{Soliton velocities.}
    \label{fig:case1region5_w2cs}
  \end{subfigure}
  \caption{Wave curves (a) and corresponding local velocities (b) of a
    soliton interacting with a DSW$^+$ for $u_- = -1.5$, $u_+ = -1$
    and $\mu = 1$. When traversing curves from $u_+$ to $u_-$ with
    $a_+ < a_{cr} = $, the curves intersect the characteristics
    corresponding to loss of strict hyperbolicity. In this case, these
    amplitudes decay to zero.}
  \label{fig:case1region5}
\end{figure}

First, for $\mu > 0$ and a backward soliton ($x_0 = x_+ > 0$)
tunnelling through a DSW$^+$ (region V),
Fig.~\ref{fig:case1region5_avsubar} shows the wave curves $a(\bar{u})$
representing the soliton amplitude as it passes through the DSW mean
flow with corresponding local trajectory velocity in
Fig.~\ref{fig:case1region5_w2cs}. As expected from Table
\ref{table:summaryDSW}, the soliton always tunnels through the DSW
from right to left ($x_0 = x_+ >0$). For any valid initial positive
amplitude satisfying $0 < a_+ < -2 u_+$
\eqref{eqn:mupos_solitonexistence}, the soliton's amplitude neither
crosses the critical line $-2\bar{u}$ nor decays to zero during
propagation. Correspondingly, the velocities of solitons starting at
$u_+ = -1$ and moving to $u_- = -1.5$ mostly remain below the DSW
velocity $W_2(\bar{u})$, save for very small $a_+$. However, examining
solitons travelling from left to right ($x_0 = x_- <0$), if the
initial amplitude is below the critical value
$a_- < a_{cr} = 2(u_+-u_-)$, the soliton is trapped and the amplitude
decays to zero. The initial amplitudes below $a_{cr}$ in
Fig.~\ref{fig:case1region5_avsubar} correspond to the velocities in
Fig.~\ref{fig:case1region5_w2cs} that lie between $3\bar{u}^2$ and the
characteristic $W_2(\bar{u})$, indicating that the soliton is
trapped. Initial amplitudes satisfying $a_- > a_{cr}$ correspond to
soliton velocities that never catch up to the DSW,
$c(a_-,u_-) < W_2(u_-)$, so soliton-DSW interaction doesn't occur.
	
\begin{figure}
  \centering
  \includegraphics[scale=0.25]{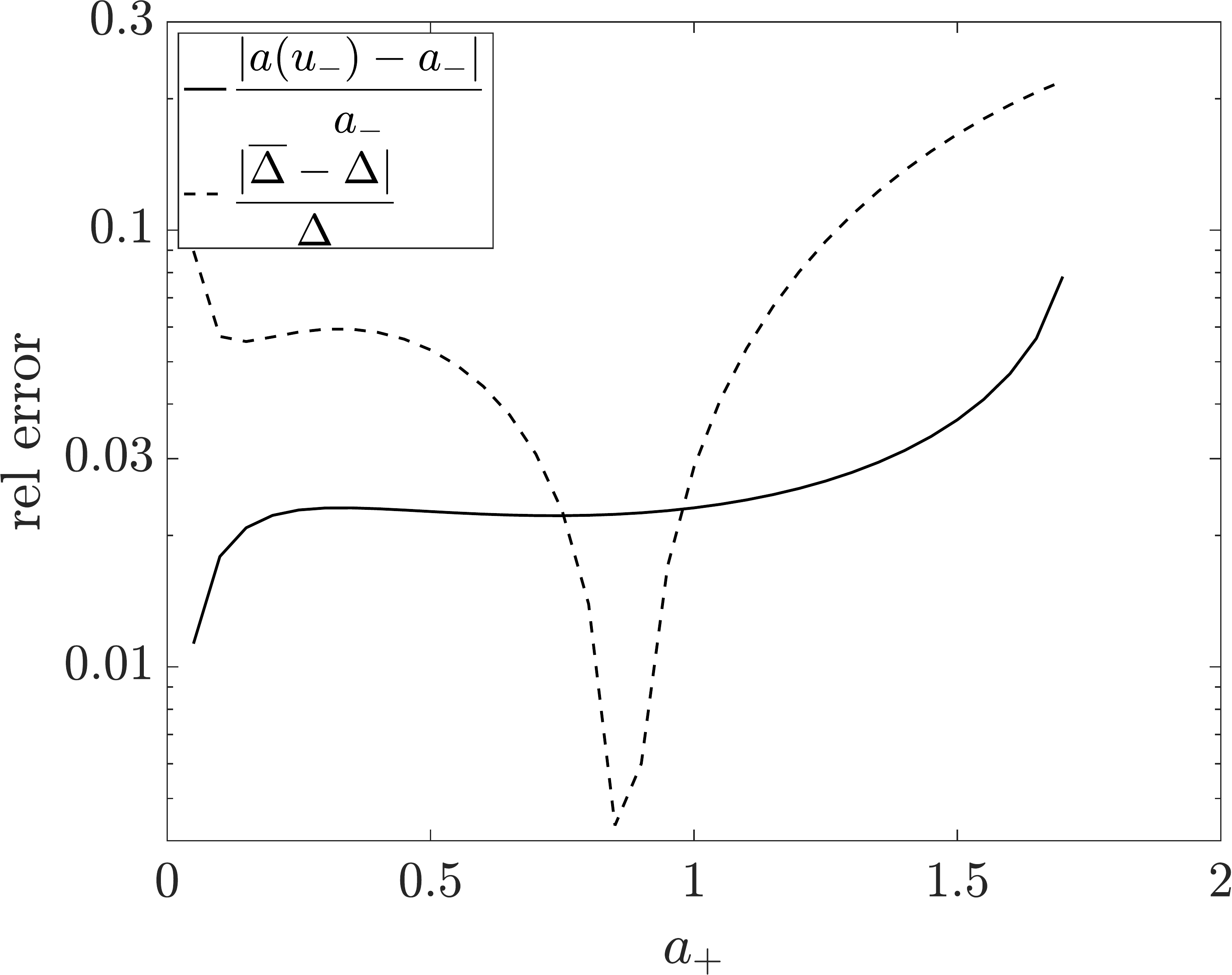}
  \caption{Relative error in the predicted transmitted soliton
    amplitude $a(u_-)$ and soliton phase shift $\overline{\Delta}$ for
    soliton-DSW$^+$ interaction using the mean flow description from
    right to left.  Parameter values are $\mu = 1$, $x_0 = x_+ = 100$,
    $u_- = -1.5$, and $u_+ = -1$.}
  \label{fig:case1_region5_relerror}
\end{figure}

We make two consistency checks in
Fig.~\ref{fig:case1_region5_relerror}.  The soliton amplitude $a(u_-)$
computed from the wave curve $a(\bar{u})$ that includes the point
$a(u_+) = a_+$ remains within 8\% of $a_- = a_+ + 2(u_+-u_-)$
predicted by the transmission condition
\eqref{eqn:transmissionphasecond} when $u_- = -1.5$, $u_+ = -1$.  This
result holds for all initial amplitudes satisfying
$0 < a_+ \lessapprox 1.7134$.  The upper bound on admissible initial
amplitudes is below the critical value $a_{cr} = 2$.  When
$1.7134 < a_+ < 2$, the wave curve terminates at $a(\ub) = -2\ub$ for
$\ub > u_- = -1.5$ as shown in
Fig.~\ref{fig:case1region5_avsubar}. The error is shown in
Fig.~\ref{fig:case1_region5_relerror} as a function of initial soliton
amplitude.

Another consistency check is the predicted phase shift
\begin{equation}
  \label{eq:19}
  \overline{\Delta} \equiv x_+ - \Big ( x(t_-) - c\big (a(u_-),u_-
  \big) t_-  \Big ) , \quad \frac{\mathrm{d}x}{\mathrm{d}t} = 
  \frac{\tilde{\omega}(\tilde{k}(x,t),\ub(x,t))}{\tilde{k}(x,t)},
  \quad x(t_+) = s_+ t_+ ,
\end{equation}
where $t_\pm$ are the times that the soliton's trajectory crosses the
DSW's leading ($s_+t$) and trailing ($s_- t$) edges.  This is compared
with the phase shift $\Delta$ determined by the transmission condition
\eqref{mkdv_shift} in Fig.~\ref{fig:case1_region5_relerror}.

\begin{figure}
  \centering
  \begin{subfigure}{0.45\textwidth}
    \includegraphics[width=\textwidth]{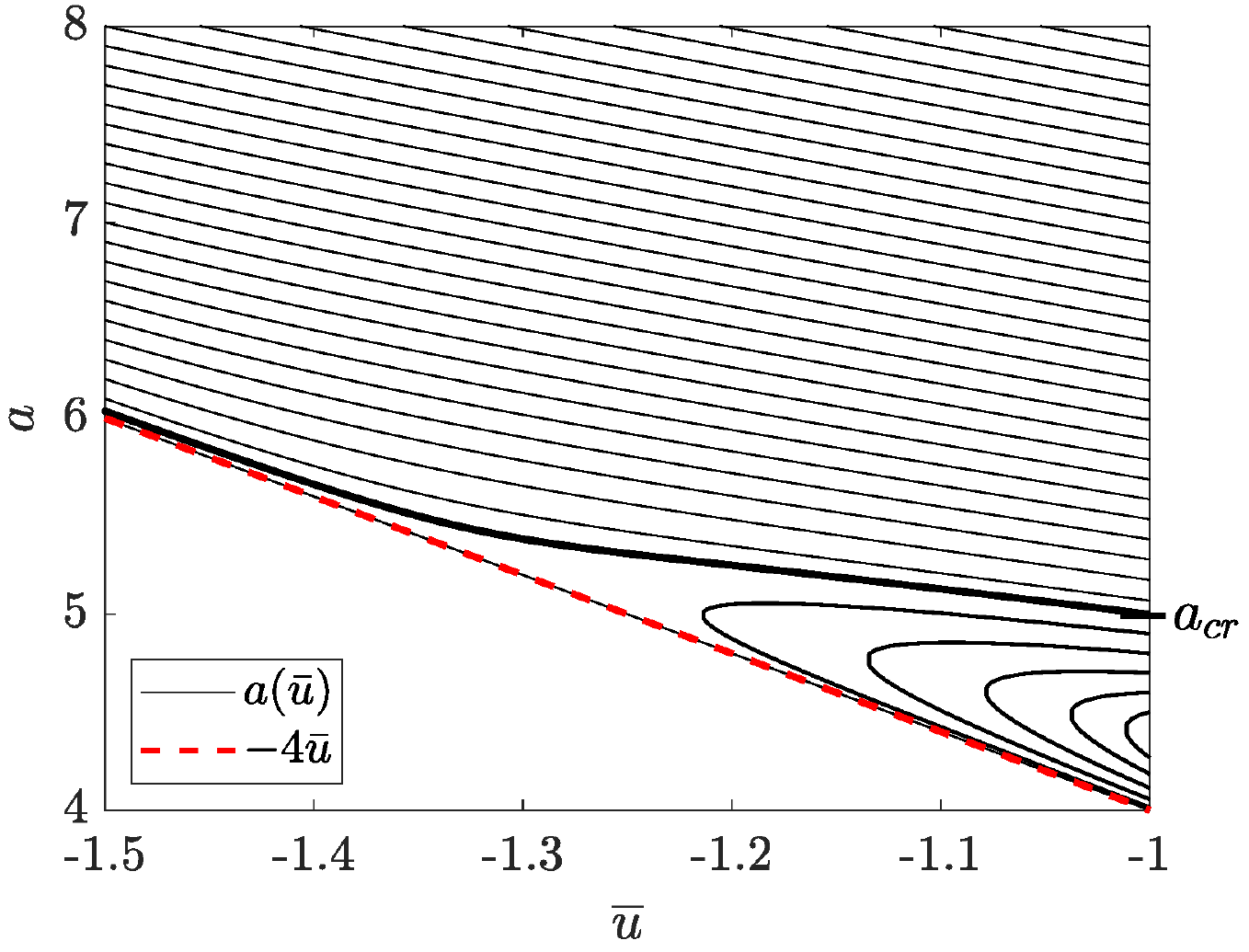}
    \caption{Soliton amplitudes.}
    \label{fig:case3region5_avsubar}
  \end{subfigure}
  \begin{subfigure}{0.45\textwidth}
    \centering
    \includegraphics[width=\textwidth]{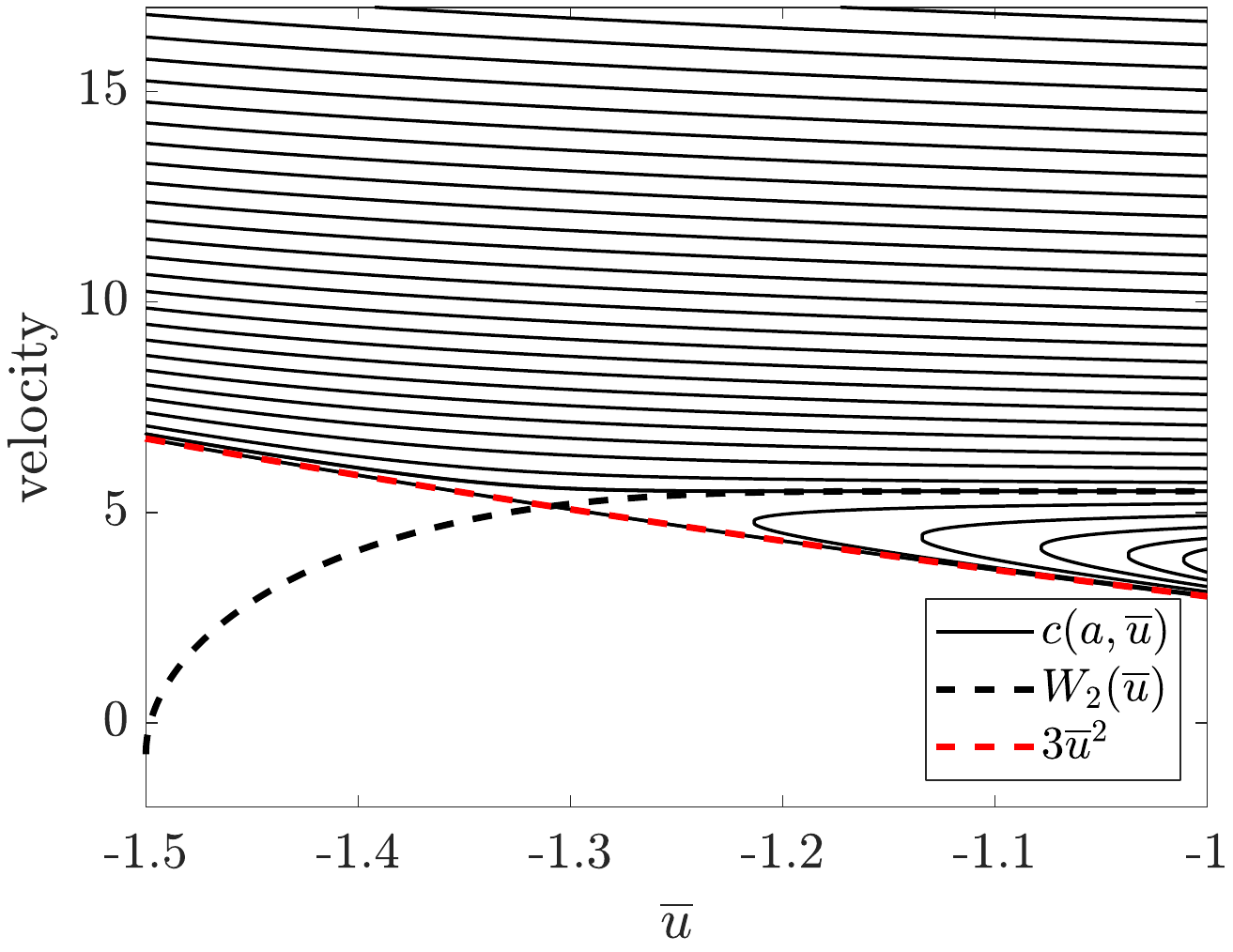}
    \caption{Soliton velocities.}
    \label{fig:case3region5_w2cs}
  \end{subfigure}
  \caption{Soliton interaction with a DSW$^-$ with $u_- = -1.5$,
    $u_+ = -1$ and $\mu = -1$ for various amplitudes. When initialised
    on the right at $u_+$ with $a_+<a_{cr}$, the wave curves do not
    reach $u_-$ so the soliton is trapped.}
  \label{fig:case3region5}
\end{figure}

An interesting case where convexity is lost, but the trapped soliton
amplitude does not decay to zero is for $\mu < 0$ and a soliton
interacting with a DSW$^-$ (Region V). Figure
\ref{fig:case3region5_avsubar} shows the soliton amplitude as it
passes through the DSW and Fig.~\ref{fig:case3region5_w2cs} shows the
corresponding soliton velocity. When travelling from left to right
($x_0 = x_- < 0$) with $u_- = -1.5$, admissible solitons satisfying
\eqref{eqn:muneg_solitonexistence} always have velocities faster than
$W_2(\bar{u})$, so interaction occurs but the velocities never cross
$W_2(\ub)$ so strict hyperbolicity is maintained. This can also be
seen in the smooth amplitude curves from $u_-$ to $u_+$, which never
intersect $-4\bar{u}$. However, when the soliton is slow enough to
interact from right to left through the DSW from $u_+ = -1$ to
$u_- = -1.5$, then the soliton becomes trapped and the velocities lie
between $3\bar{u}^2$ and $W_2(\bar{u})$. This corresponds to
amplitudes below the critical amplitude $a_{cr} = 2(u_+-u_-)$. For a
soliton with amplitudes below this critical amplitude, the velocity
crosses $W_2$ and limits to $3\bar{u}^2$, while the amplitude limits
to $-4\bar{u}$, indicating that it behaves like an algebraic soliton.
	
\begin{figure}
  \centering
  \includegraphics[width=0.55\textwidth]{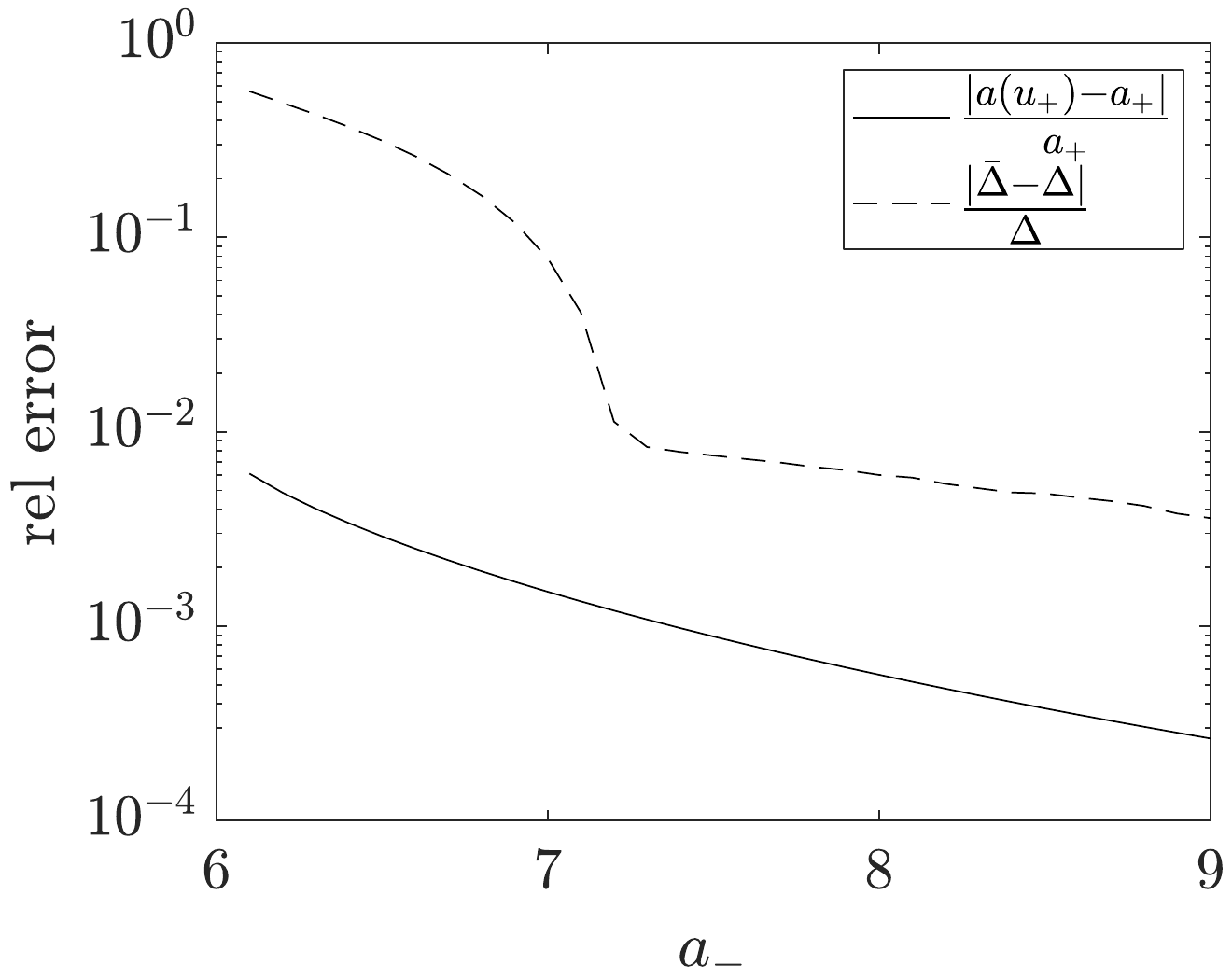}
  \caption{Relative error in the predicted transmitted soliton
    amplitude $a(u_+)$ and soliton phase shift $\overline{\Delta}$ for
    soliton-DSW$^+$ interaction using the mean flow description from
    left to right.  Parameter values are $\mu = -1$,
    $x_0 = x_- = -100$, $u_- = -1.5$, and $u_+ = -1$.}
  \label{fig:case3region5_relerror}
\end{figure}

When tunnelling occurs, the transmitted amplitude from numerical
integration of the ODE \eqref{eqn:dkdu} is compared to the amplitude
predicted by transmission conditions and the error is shown in
Fig.~\ref{fig:case3region5_relerror}. For initial amplitudes that satisfy $a_- > 6.1$ with $a_{cr} = 6$ for $u_- = -1.5$ and $u_+ = -1$, the computed soliton amplitude $a(u_+)$ is within 1\% of the predicted amplitude. The predicted phase shift based on \eqref{eq:19} with subscripts $-$ replaced with $+$ is also compared to the phase shift as determined from the transmission condition \eqref{mkdv_shift} in Fig~\ref{fig:case3region5_relerror}. 

\section{Soliton-nonconvex mean flow interaction}
\label{sec:soliton-nonconvex}

In this section, we study interactions of solitons with nonconvex mean
flows arising from the mKdV GP problem with $u_- u_+ <0$. We consider
interactions with ``pure'' nonconvex mean flows generated for the
symmetric conditions $u_-=-u_+$. These are kinks ($\mu>0$) and CDSWs
($\mu<0$); see Section~\ref{sec:noncon_mean_flow}.
	
\subsection{Soliton-kink interaction}

A kink solution to the GP problem when $\mu > 0$ is realised when
$u_+ = -u_-$. To be definite, we assume that $u_-<0$. The kink
velocity $u_-^2=u_+^2$ is slower than the soliton velocity for any
amplitude so interaction happens from left to right with
$x_0 = x_- < 0$. By the soliton existence conditions
\eqref{eqn:mupos_solitonexistence}, when $u_- < 0$, we must initialise
with a bright soliton $(a_- > 0)$ on the left side. Since bright
solitons cannot exist on the right side of the kink where $u_+ > 0$,
we expect that the soliton polarity undergoes a switch as a result of
kink interaction in order for the soliton to be a valid solution. To
determine the transmitted soliton amplitude, we observe that, under
the quadratic transformation \eqref{eq:rtolambda}, the mKdV
soliton-kink interaction problem in the limit $\mu \to 0$ is mapped
onto the problem of KdV soliton train propagation on a background
$-3\ub^2$ described by the modulation system (cf.~\eqref{eq:la13})
\begin{equation}\label{eq:la13_kdv}
  \begin{aligned}
    &\frac{\partial r_1}{\partial t} - r_1 \frac{\partial
      r_1}{\partial x} = 0,   \\ 
    &\frac{\partial r_3}{\partial t} - \frac13(r_1+2r_3)
    \frac{r_3}{\partial x} = 0,
  \end{aligned} 
\end{equation}	
where $r_1=-3\lambda_1^2$ and $r_3=-3\lambda_3^2$. Since
$\lambda_1=\ub$, the above quadratic transformation maps the
discontinuous solution \eqref{eq:kink_weak} for $\ub$ to the
\emph{constant solution} $r_1 = -3 \ub^2= -3 u_-^2$ of the first
equation in \eqref{eq:la13_kdv}. The second equation also admits the
constant solution $r_3=const$, which is mapped to $\lambda_3 = const$,
and therefore $|a|=2(\lambda_3-\lambda_1)= const$ for the mKdV
equation.

The transformation of the soliton amplitude in soliton-kink tunnelling
can be obtained from the transmission condition
\eqref{eqn:transmissionphasecond} by assuming that the normalisation
\eqref{norm_q} of the Riemann invariant $q=Q(a, \ub)$ changes to
$Q(0, \ub)=-\ub$ when crossing the zero convexity point $\ub=0$ at
$x=0$, yielding the transmission condition
$ u_++\frac12 a_+ = -(u_-+\frac12 a_-)$. Since $u_-=-u_+$, we obtain
$a_+ = - a_-$. Inserting this result into the soliton phase shift
formula \eqref{mkdv_shift}, we obtain $\Delta=0$. To be clear, the
predicted zero phase shift is an approximate result within the context
of modulation theory in the limit $\mu \to 0$. For nonzero but small
$\mu$, a small phase shift due to soliton-kink interaction is
expected. Such an interaction for the mKdV equation can be
investigated using the inverse scattering transform (IST) with
non-zero boundary conditions developed for both signs of $\mu$ in
\cite{zhang_focusing_2020}. Within the IST formalism, the conservation
of the absolute value of the soliton amplitude pre and post
interaction is a consequence of the discrete spectrum's conservation.

\begin{figure}
  \centering
  \begin{subfigure}{0.27\textwidth}
    \centering
    \includegraphics[width=\textwidth]{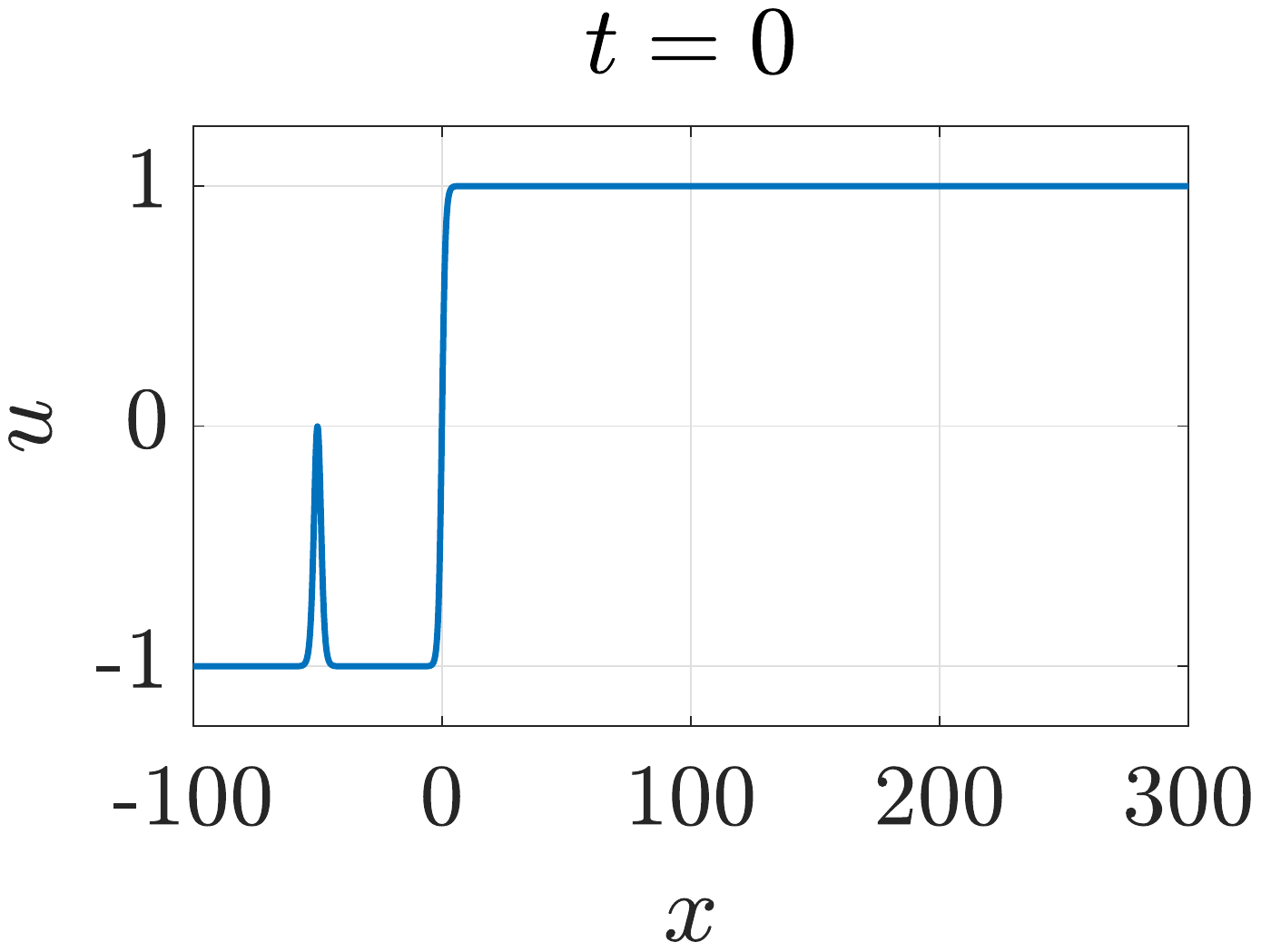}
  \end{subfigure}
  \begin{subfigure}{0.4\textwidth}
    \centering
	\includegraphics[width=\textwidth]{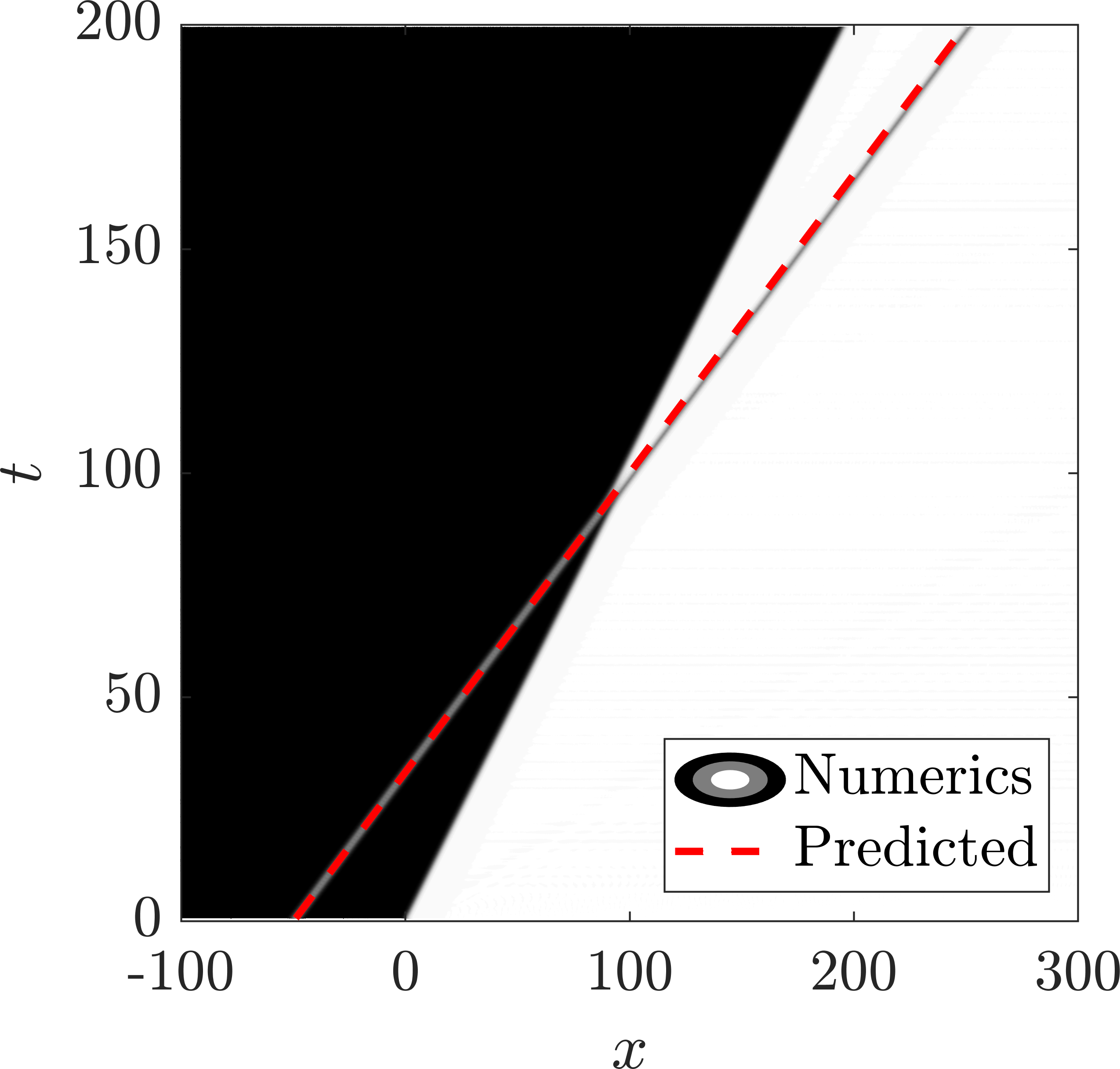}
  \end{subfigure}		
  \begin{subfigure}{0.26\textwidth}
    \centering
    \includegraphics[width=\textwidth]{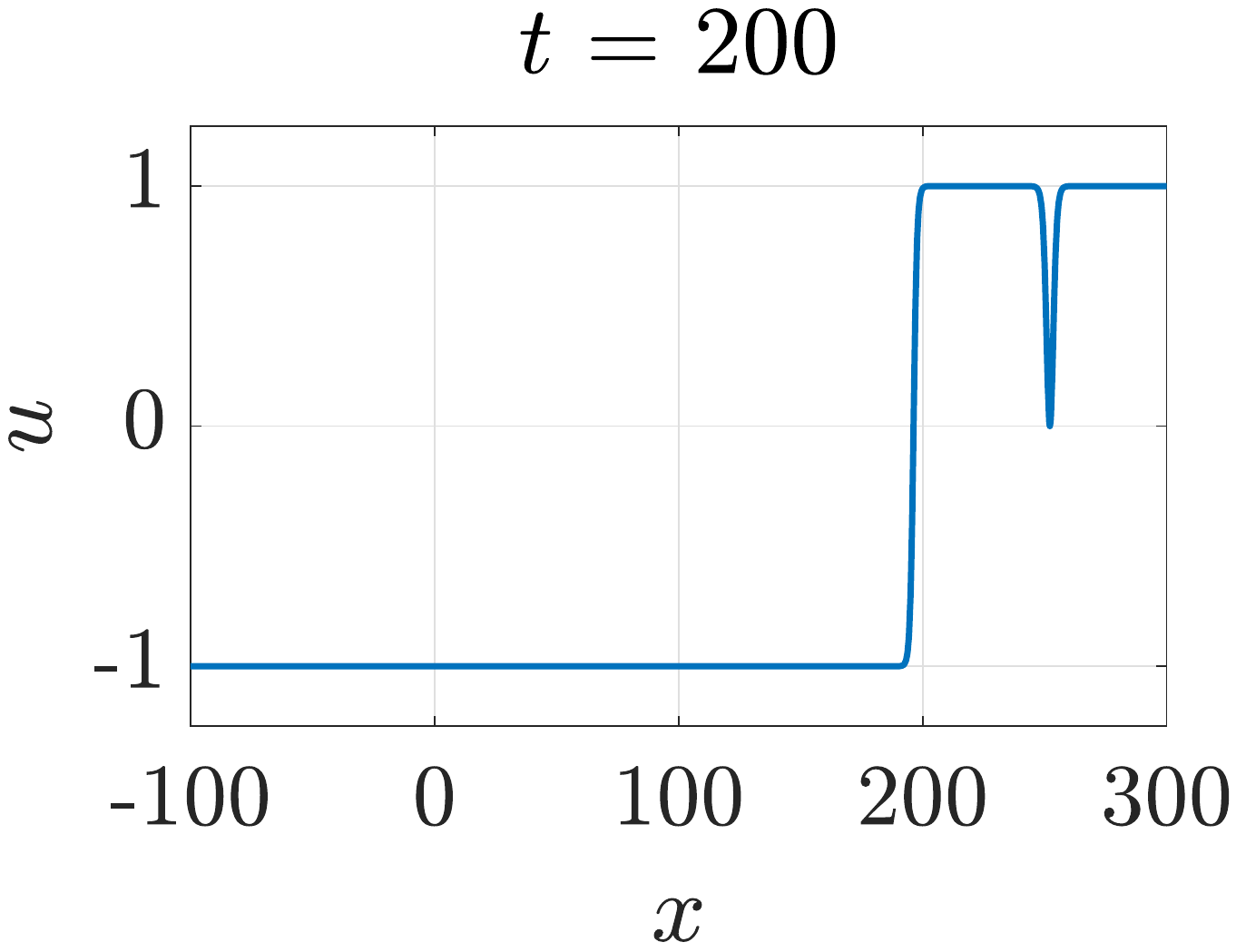}
  \end{subfigure}
  \caption{Soliton-kink interaction for $\mu = 1$, $a_- = 1$,
    $x_- = -50$, $u_- = -1$, and $u_+ = 1$. The predicted $a_+$ is -1,
    after flipping polarity in the interaction, and the predicted phase shift $\Delta$ is 0. The numerical
    solution gives $a_+ = -1.0000$ and $\Delta = 2.1484$ at $t = 200$.}
  \label{fig:soliton-kink}
\end{figure}

Numerical experiments with results in Table
\ref{table:solitonkinknumerics} confirm that, as the soliton
propagates through the kink, it switches polarity while preserving the
absolute value of the amplitude and that the phase shift is very
small, as seen in Fig. \ref{fig:soliton-kink}. The observed small
phase shift from numerical experiments is due to the nonzero value of
$\mu$ used in numerical simulations.  Note that the kink itself
undergoes a phase shift in the direction opposite to the soliton.
\begin{table}
  \centering
  \begin{tabular}{cccccccccc}
    $u_-$ & $u_+$ & $a_-$ & $t_{final}$ & $a_+$ & $a_+$ (num) &
    $\Delta x$ & $\Delta x$ (num) & $\Delta x/|x_-|$ (num)
    & $\Delta x_{kink}$ \\ 
    \hline
    -1 & 1 & 0.5 & 200 & -0.5 & -0.4991 & 0 & 1.6530 & 0.0331 & -2.3438\\
    -1 & 1 & 1.0 & 200 & -1.0 & -1.0000 & 0 &  2.1484 & 0.0430 & -3.8086\\
    -1 & 1 & 1.5 & 500 & -1.5 & -1.4999 & 0 & 3.0273 & 0.0605 & -5.8594 \\
    -2 & 2 & 1.5 & 50 & -1.5 & -1.4988 & 0 & 0.9766 & 0.0195 & -1.6113 \\		 
  \end{tabular}
  \caption{Numerical tests of bright soliton-kink interaction from
    left to right for $\mu = 1$ with $x_- = 50$. Values for $a_+$ and
    $\Delta x$ according to the transmission conditions are compared
    with numerical (num) results. }
  \label{table:solitonkinknumerics}
\end{table}
	
\subsection{Soliton-CDSW interaction}
	
When $\mu < 0$ and $u_- = -u_+$, the resulting solution that emerges
from the GP problem is a CDSW. The leading and trailing edge travel at
characteristic velocity $s_+ = 3u_-^2$ and $s_- = -3u_-^2$, so that
solitons can only interact with the CDSW from left to right,
$x_0 = x_- < 0$.  Tunnelling always occurs in this interaction and the
transmitted amplitude is obtained from
\eqref{eqn:transmissionphasecond} as
\begin{equation}\label{eq:apm}
  a_+=a_-+4u_- .
\end{equation}
By the existence conditions \eqref{eqn:muneg_solitonexistence}, an
initial soliton satisfies $a_- > -4u_-$. To maintain strict
hyperbolicity, $a_+ > -4u_+$ as well. Note that the CDSW mean flow is
monotone along the characteristic $\mathrm{d}x/\mathrm{d}t=W_2(\ub)$;
see \eqref{cdsw_mod}. Using the transmission condition, we can see
that
\begin{align}
  a_+ &> -4u_+ \\
  \implies a_- &> 2u_+ - 2u_- = -4u_-,
\end{align} 
and this relation is always satisfied by $a_-$ so strict hyperbolicity
is maintained.  
	
\begin{figure}
  \centering
  \begin{subfigure}{0.27\textwidth}
    \centering
    \includegraphics[width=\textwidth]{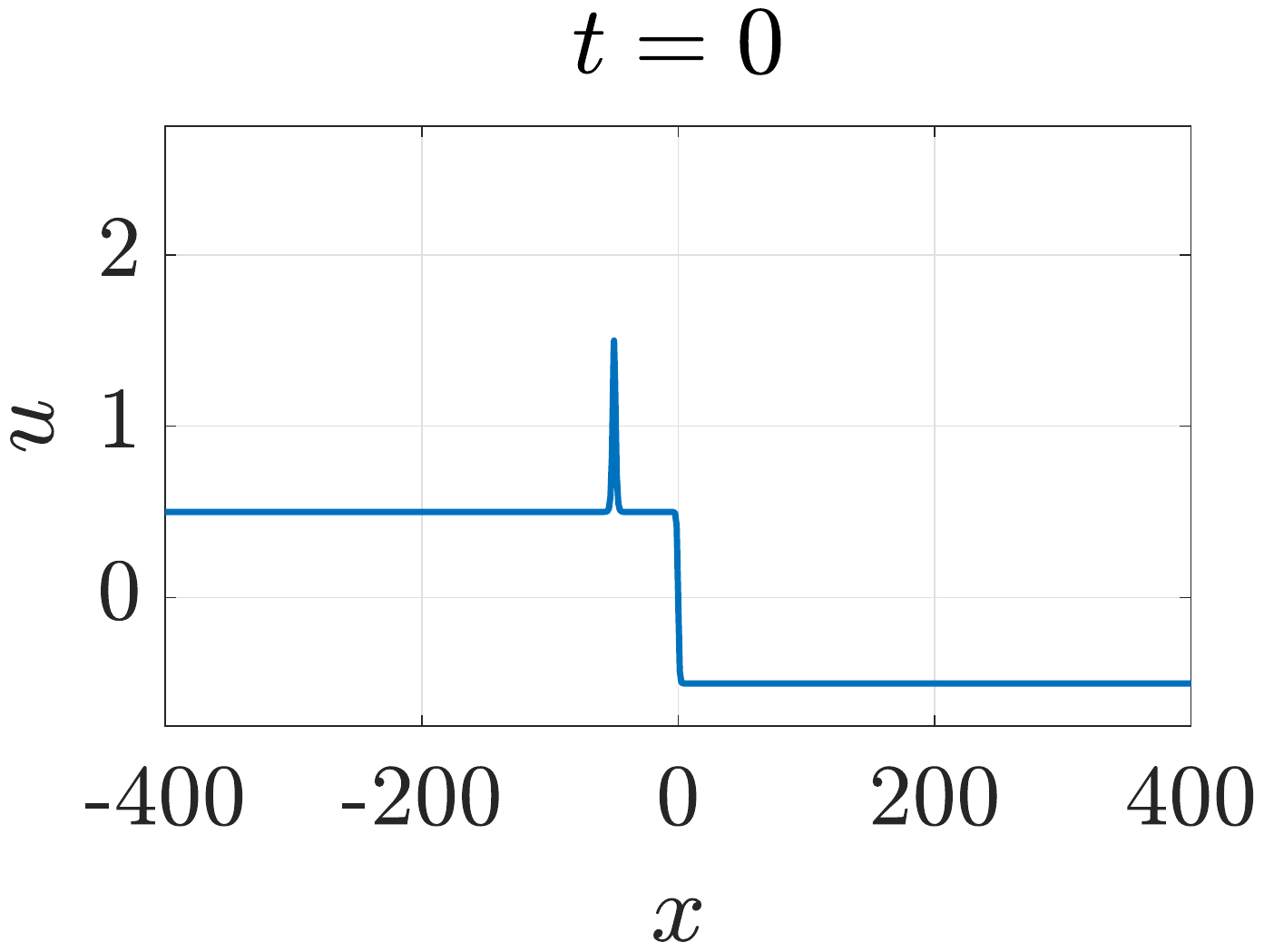}
  \end{subfigure}
  \begin{subfigure}{0.4\textwidth}
    \centering
    \includegraphics[width=\textwidth]{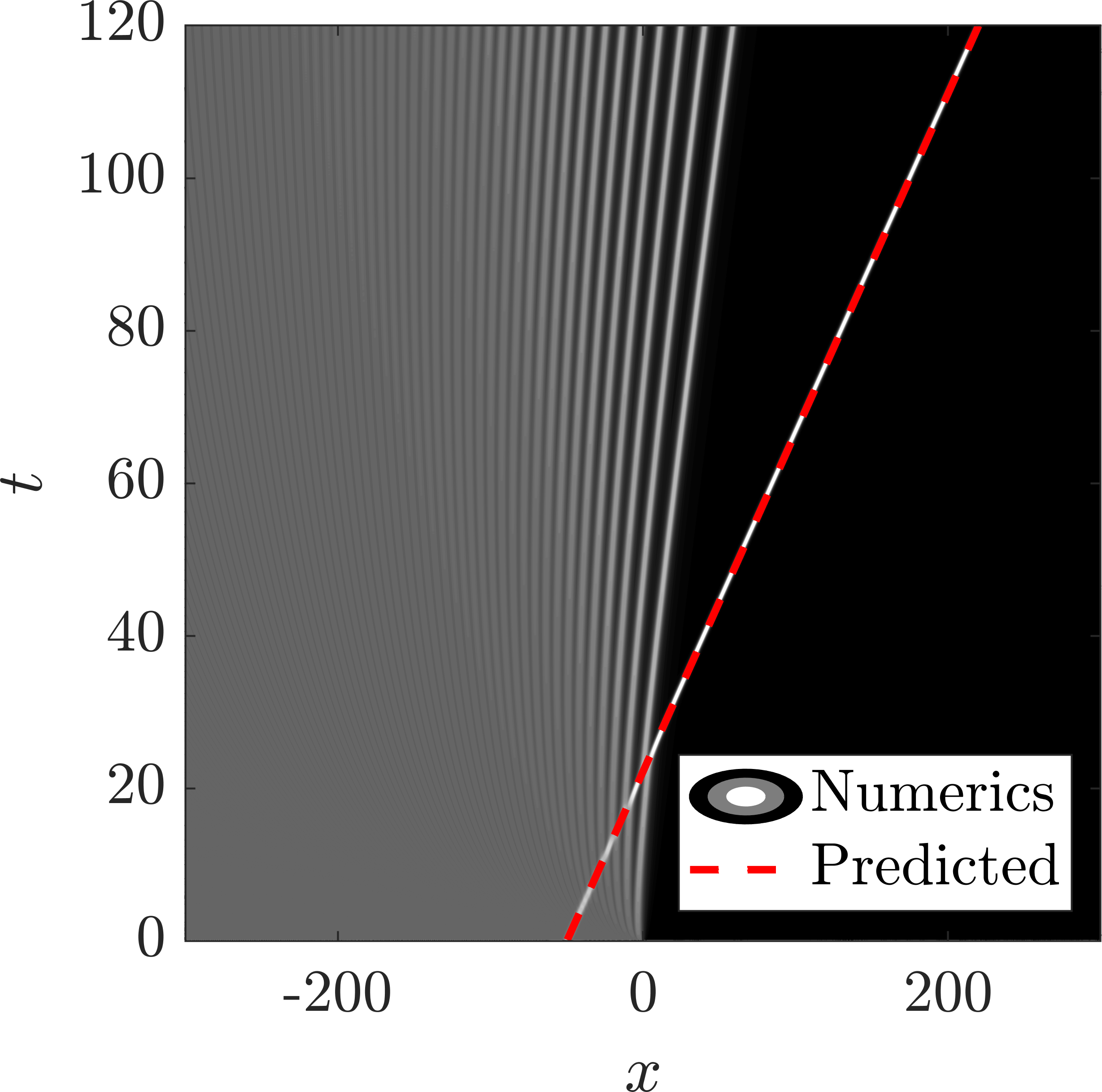}
  \end{subfigure}		
  \begin{subfigure}{0.26\textwidth}
    \centering
    \includegraphics[width=\textwidth]{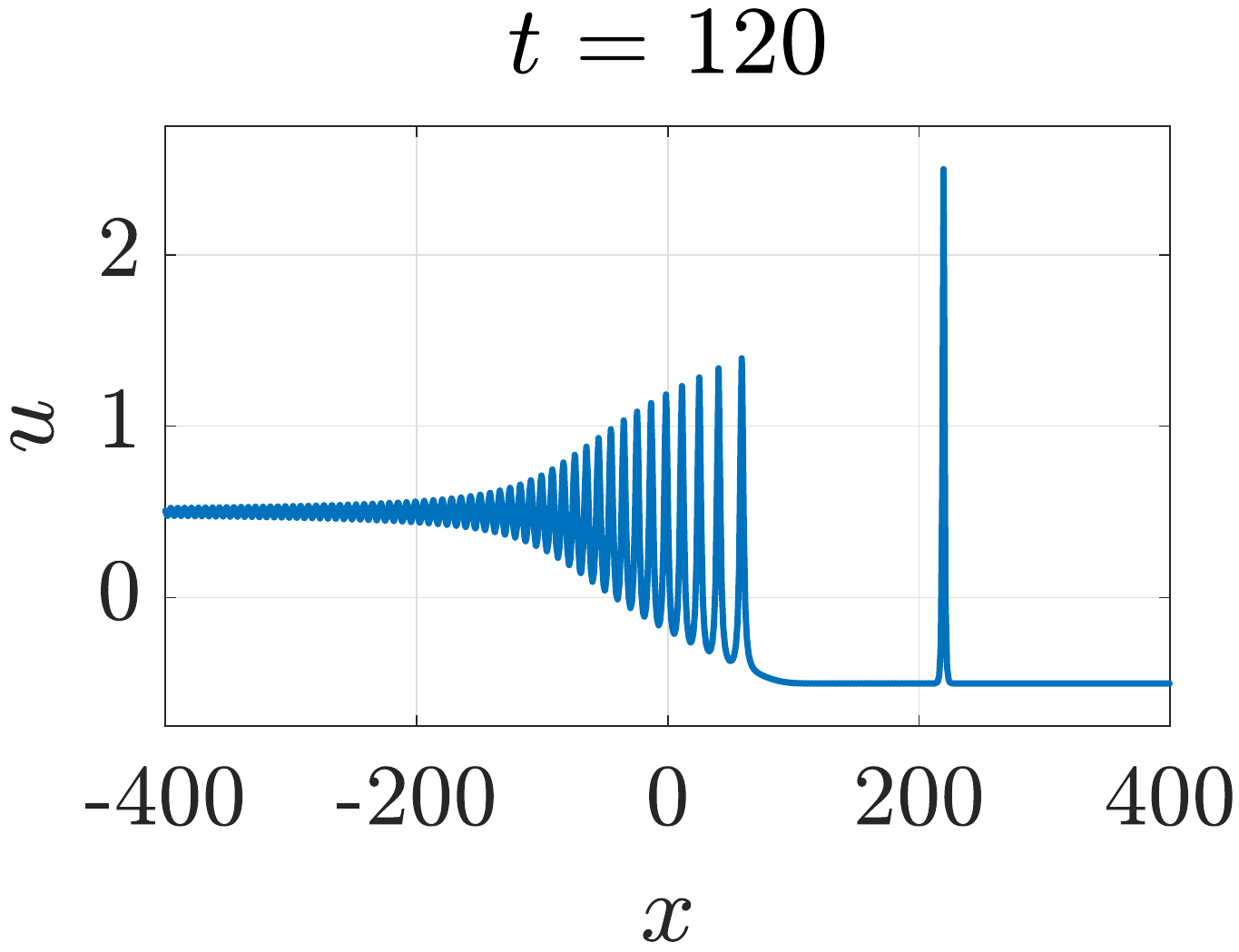}
  \end{subfigure}
  \begin{subfigure}{0.27\textwidth}
    \centering
    \includegraphics[width=\textwidth]{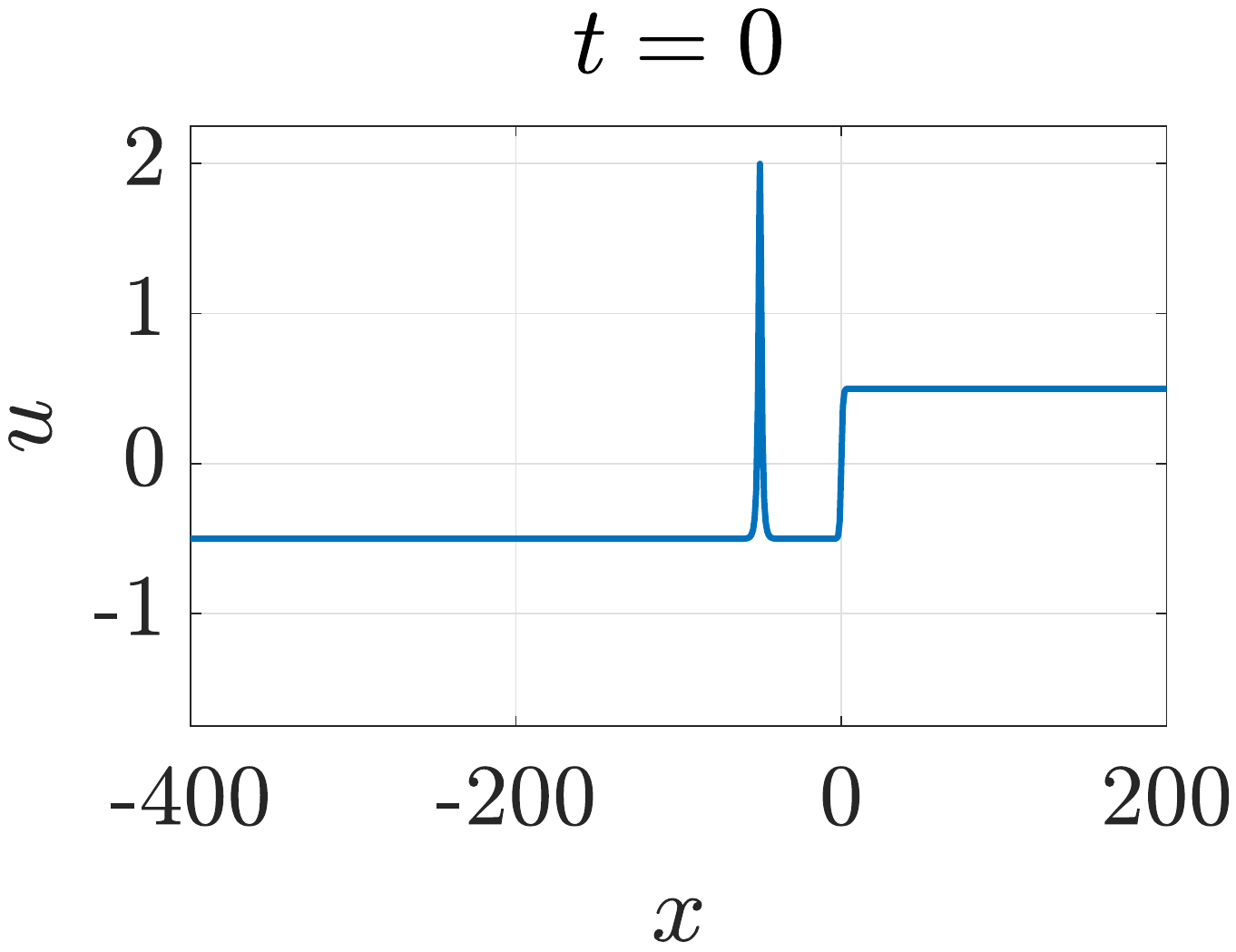}
  \end{subfigure}
  \begin{subfigure}{0.4\textwidth}
    \centering
    \includegraphics[width=\textwidth]{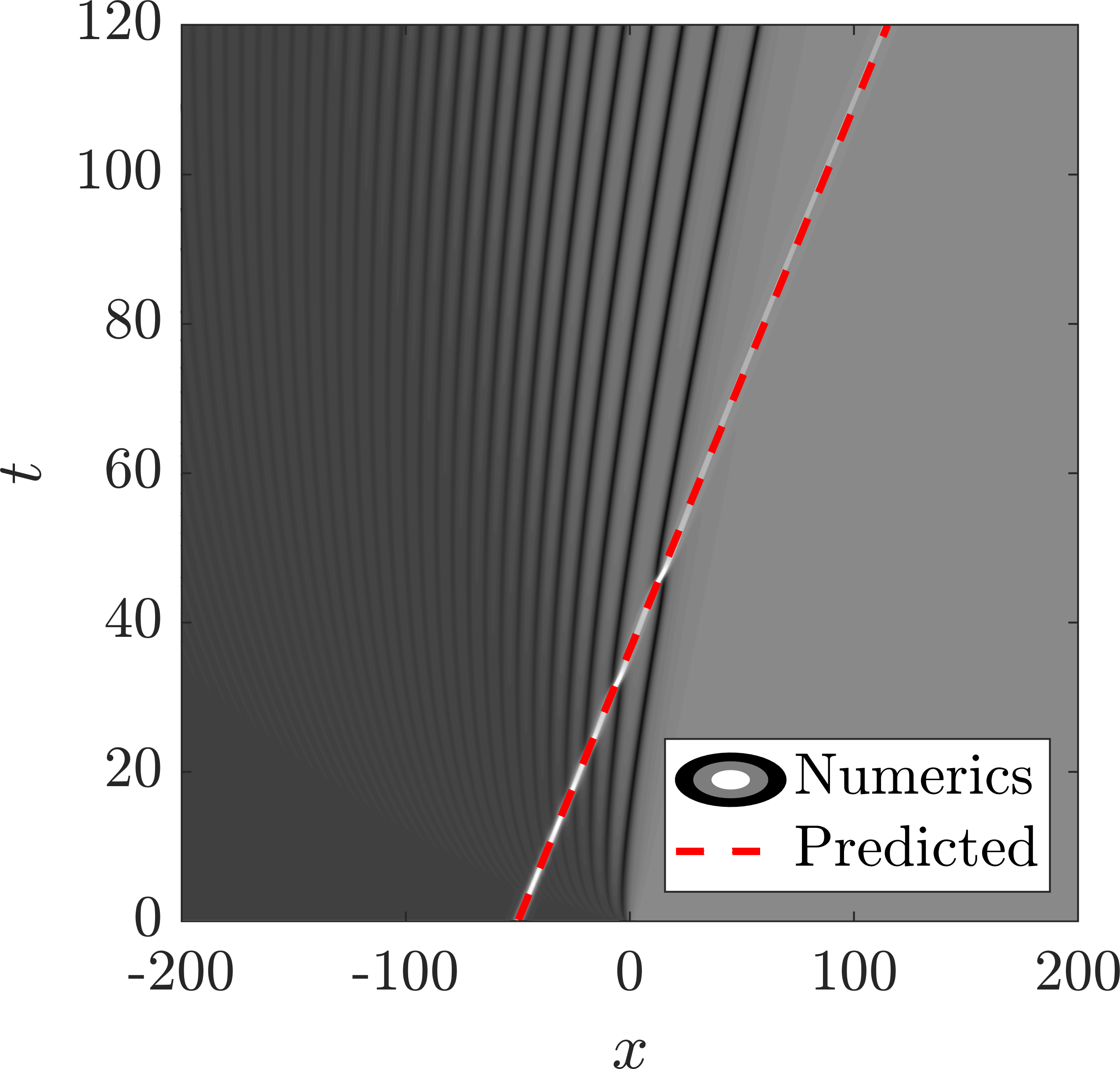}
  \end{subfigure}		
  \begin{subfigure}{0.26\textwidth}
    \centering
    \includegraphics[width=\textwidth]{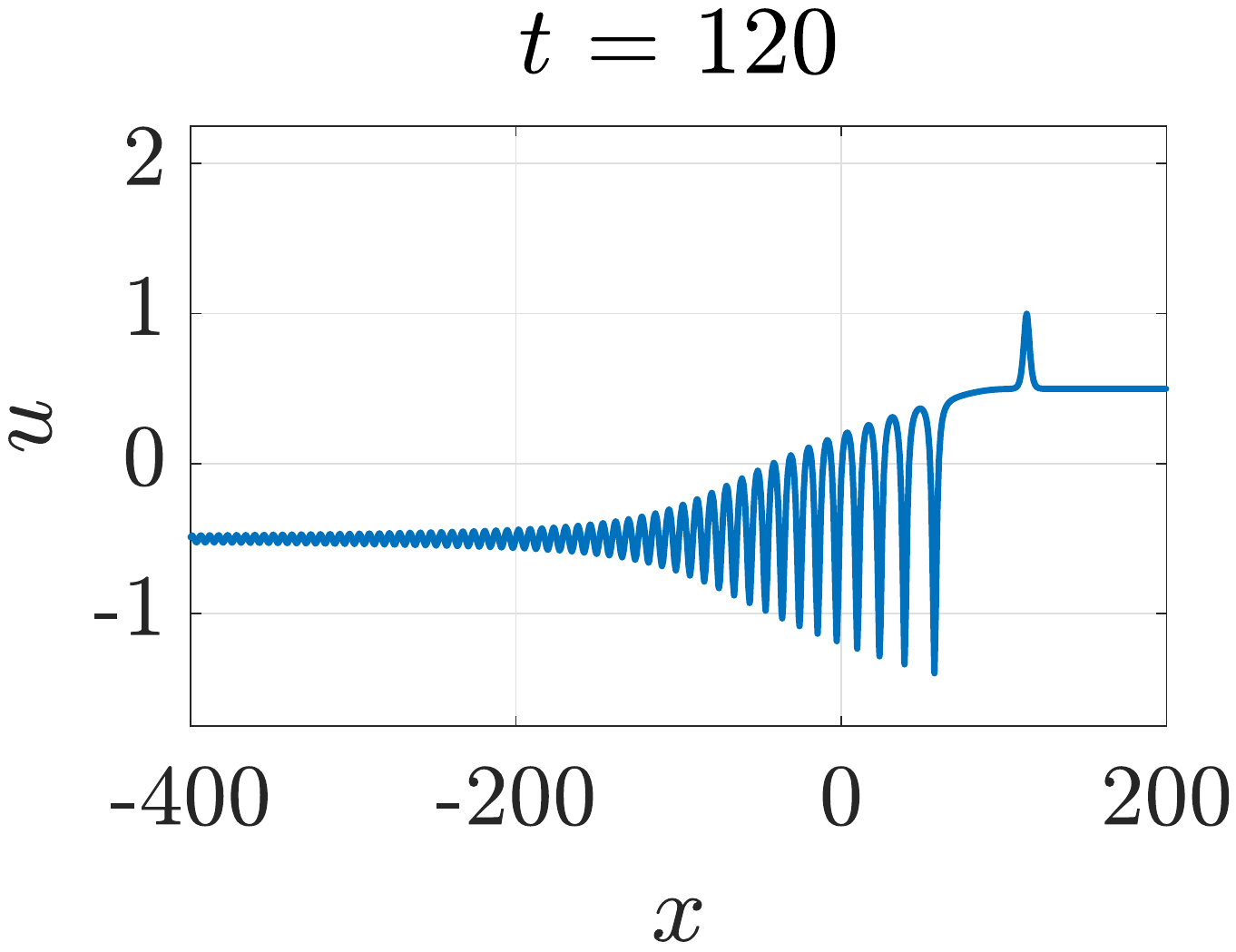}
  \end{subfigure}
  \caption{Top, a soliton and CDSW$^+$ interaction where $\mu = -1$,
    $a_- = 1$, $x_- = -50$, $u_- = 0.5$ and $u_+ = -0.5$. Bottom,
    soliton and CDSW$^-$ where $\mu = -1$, $a_- = 2.5$, $x_- = -50$,
    $u_- = -0.5$ and $u_+ = 0.5$.}
  \label{fig:solitonCDSW}		
\end{figure}

The predicted phase shift is found from equation \eqref{mkdv_shift}
with $u_+ = -u_-$ and $a_+$ given by \eqref{eq:apm}. As with the pure
kink interaction, $\Delta = 0$. Numerical experiments verify the
conservation of $q$ and $kp$, although we do see a small phase shift,
likely due to higher order effects. See Fig.~\ref{fig:solitonCDSW} for
depictions of soliton interaction with CDSWs of both polarities at the
boundaries of regions VII and VIII (positive polarity) and III and IV
(negative polarity). The predicted soliton trajectory (dashed) in
Fig.~\ref{fig:solitonCDSW} was generated by assuming that it was
unchanged by the mean flow, rather than considering the CDSW mean flow
given by \eqref{eq:mean_CDSW} and the ODE \eqref{eqn:dkdu}.  This
approximation is justified by the predicted zero phase shift and
unchanged soliton velocity post CDSW interaction.
	
In contrast to the soliton-kink interaction, the soliton-CDSW
interaction does not result in the soliton's polarity change. This is
because strict hyperbolicity is maintained throughout and the
existence condition \eqref{eqn:muneg_solitonexistence} for solitons
with $\mu < 0$ allow for bright solitons on either side of the mean
flow.

\begin{table}
  \centering
  \begin{tabular}{ccccccccc}
    $u_-$ & $u_+$ & $a_-$ & $T_{final}$ & $a_+$ (pred) & $a_+$ (num) &      $\Delta x$ (pred) & $\Delta x$ (num) & $\Delta x/|x_-|$ (num) \\
    \hline
    -0.5 & 0.5 & 2.5 & 120 & 0.5 & 0.4993 & 0 & -1.0352 & 0.0207 \\
    0.5 & -0.5 & 1.0 & 120 & 3.0 & 2.9960 & 0 &  -0.4688 & 0.0094 \\
  \end{tabular}
  \caption{Numerical tests of bright soliton-CDSW interaction from
    left to right for $\mu = -1$ and $x_- = -50$.}
  \label{table:solitonCDSWnumerics}
\end{table}

\subsection{Hybrid mean-flows}

Regions III, IV, VII and VIII for the Riemann problem result in hybrid
mean-flow dynamics involving a CDSW or kink coexisting with a RW or
DSW. The analysis for RWs, DSWs and pure kinks and CDSWs serve as the
building blocks for determining the soliton tunnelling criterion
through these combination flows. We summarise these results for
$\mu > 0$ in Table \ref{table:summaryhybridmupositive} and for
$\mu < 0$ in Table \ref{table:summaryhybridmunegative}.

\begin{table}
  \resizebox{\textwidth}{!}{%
    \begin{tabular}{l|l|l|l|l}
      Direction & \begin{tabular}[c]{@{}l@{}}Region III - kink$|$RW \\ ($u_+ > -u_- > 0$)\end{tabular} & \begin{tabular}[c]{@{}l@{}}Region IV - kink$|$DSW\\ ($u_- < -u_+ < 0$)\end{tabular} & \begin{tabular}[c]{@{}l@{}}Region VII - kink$|$RW\\ ($u_+ < -u_- < 0$)\end{tabular} & \begin{tabular}[c]{@{}l@{}}Region VIII - kink$|$DSW\\ ($-u_- < u_+ < 0$)\end{tabular} \\ \hline
      $R \to L$ & No soliton solutions & No soliton solutions & \begin{tabular}[c]{@{}l@{}} Tunnelling through RW if \\ $a_+ > -2u_- - 2u_+$, \\ trapped to the right of the kink\end{tabular} & \begin{tabular}[c]{@{}l@{}}Tunnelling through DSW always, \\ trapped to the right of the kink \end{tabular} \\ \hline 
      $L \to R$ & \begin{tabular}[c]{@{}l@{}}Tunnelling through kink, \\ polarity flips, \\ trapped to the left of the RW \end{tabular} & \begin{tabular}[c]{@{}l@{}}Tunnelling through kink, \\ polarity flips, \\ trapping in DSW if \\ $a_- < 2u_+ - 2u_-$ \end{tabular} & No soliton solutions & No soliton solutions 
    \end{tabular}%
  }
  \caption{Results for $\mu > 0$ with bright solitons interacting with
    hybrid mean flows.}
  \label{table:summaryhybridmupositive}
\end{table}	
	
\begin{table}
  \resizebox{\textwidth}{!}{%
    \begin{tabular}{l|l|l|l|l}
      Direction & \begin{tabular}[c]{@{}l@{}}Region III - CDSW$|$RW \\ ($u_+ > -u_- > 0$)\end{tabular} & \begin{tabular}[c]{@{}l@{}}Region IV - CDSW$|$DSW\\ ($u_- < -u_+ < 0$)\end{tabular} & \begin{tabular}[c]{@{}l@{}}Region VII - CDSW$|$RW\\ ($u_+ < -u_- < 0$)\end{tabular} & \begin{tabular}[c]{@{}l@{}}Region VIII - CDSW$|$DSW\\ ($-u_- < u_+ < 0$)\end{tabular} \\ \hline
      $R \to L$ & No interaction & \begin{tabular}[c]{@{}l@{}} Interaction and trapping if \\ $a_+ < -2u_+ - 2u_-$ \end{tabular} & No interaction & No interaction \\ \hline 
      $L \to R$ & \begin{tabular}[c]{@{}l@{}}Tunnelling if \\ $a_- > 2u_+ - 2u_-$ \end{tabular} & \begin{tabular}[c]{@{}l@{}} Tunnelling if \\ $a_- > 2u_+ - 2u_-$ \end{tabular} & \begin{tabular}[c]{@{}l@{}}Tunnelling if \\ $a_- > -2u_+ - 2u_-$ \end{tabular} & Tunnelling for any amplitude
    \end{tabular}%
  }
  \caption{Results for $\mu < 0$ with bright solitons interacting with
    hybrid mean flows.}
  \label{table:summaryhybridmunegative}
\end{table}

\section{Kink-mean flow interaction}
\label{sec:kink-mean}
	
So far, we have considered the case of solitons interacting with mean
flows. Kinks are another localised wave structure that arise as
solutions in nonconvex systems, so it is natural to consider their
interaction with mean flows. Unlike in soliton-mean flow tunnelling
where the mean flow is essentially unchanged by the interaction, in
this case, both the kink and the mean flow are significantly altered
by the interaction. In addition, we find that the admissibility
condition for tunnelling is always satisfied by the kink so there is
no trapping.

Kinks occur when $u_3 \to u_2 = u_1$, causing all three modulation
equations in the system \eqref{eqn:mKdVWhitham} to collapse into the
dispersionless mean flow equation $\ub_t + 3 \ub^2 \ub_x=0$. The
amplitude of the kink is $a = -2\bar{u}$, resulting in $q = 0$. Since
the kink velocity is slower than the RW or DSW speed, it can only
interact from right to left. The kink trajectory is given by
\begin{equation}
  \frac{\mathrm{d}x}{\mathrm{d}t} = \bar{u}^2, \qquad x(0) = x_+, \label{eqn:kinktraject}
\end{equation}	
which satisfies the Rankine-Hugoniot condition.  The kink propagates
like a shock.
	
\bigskip
{\it (i) Kink-RW interaction }
	
\begin{figure}
  \centering
  \begin{subfigure}{0.27\textwidth}
    \centering
    \includegraphics[width=\textwidth]{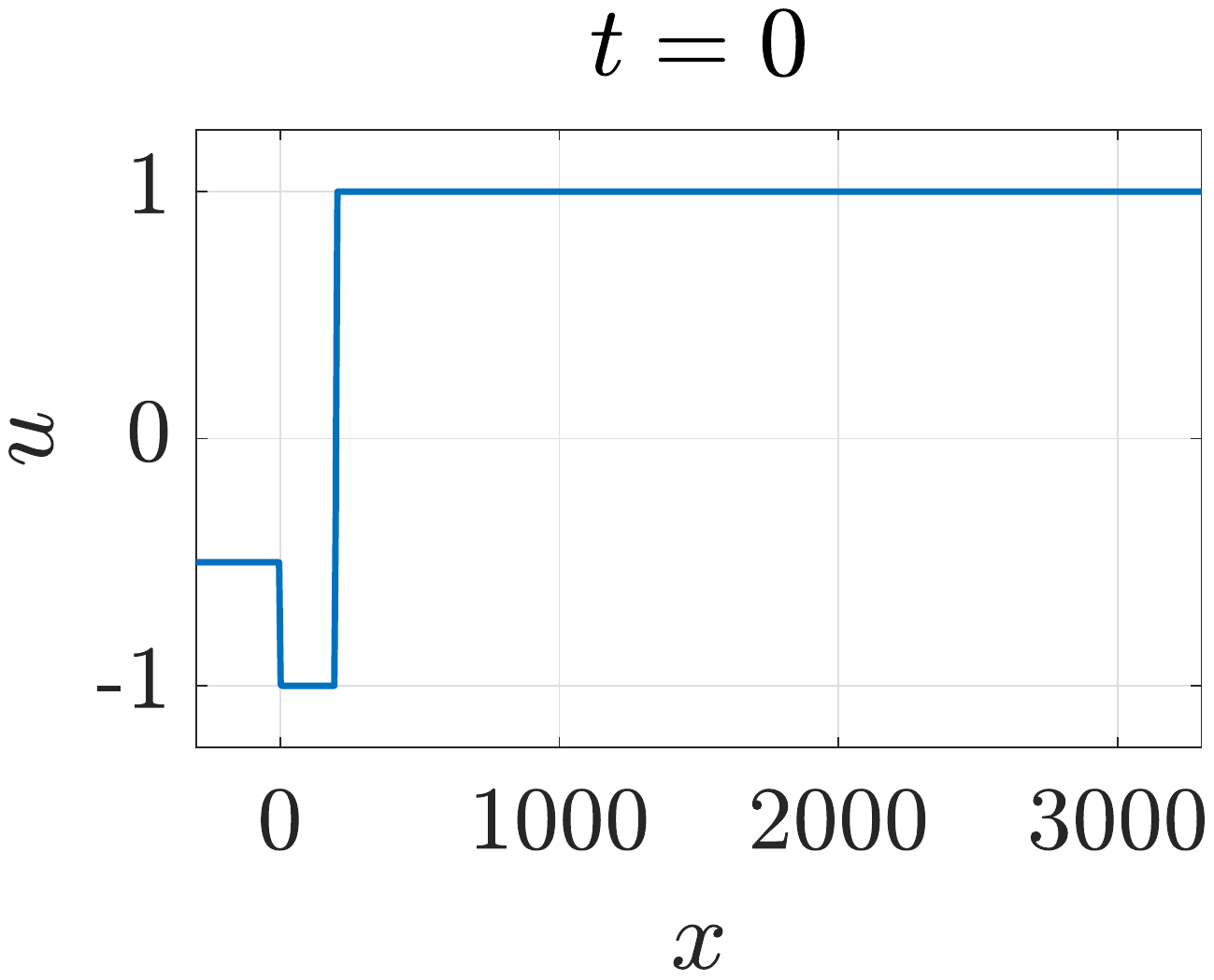}
  \end{subfigure}
  \begin{subfigure}{0.4\textwidth}
    \centering
    \includegraphics[width=\textwidth]{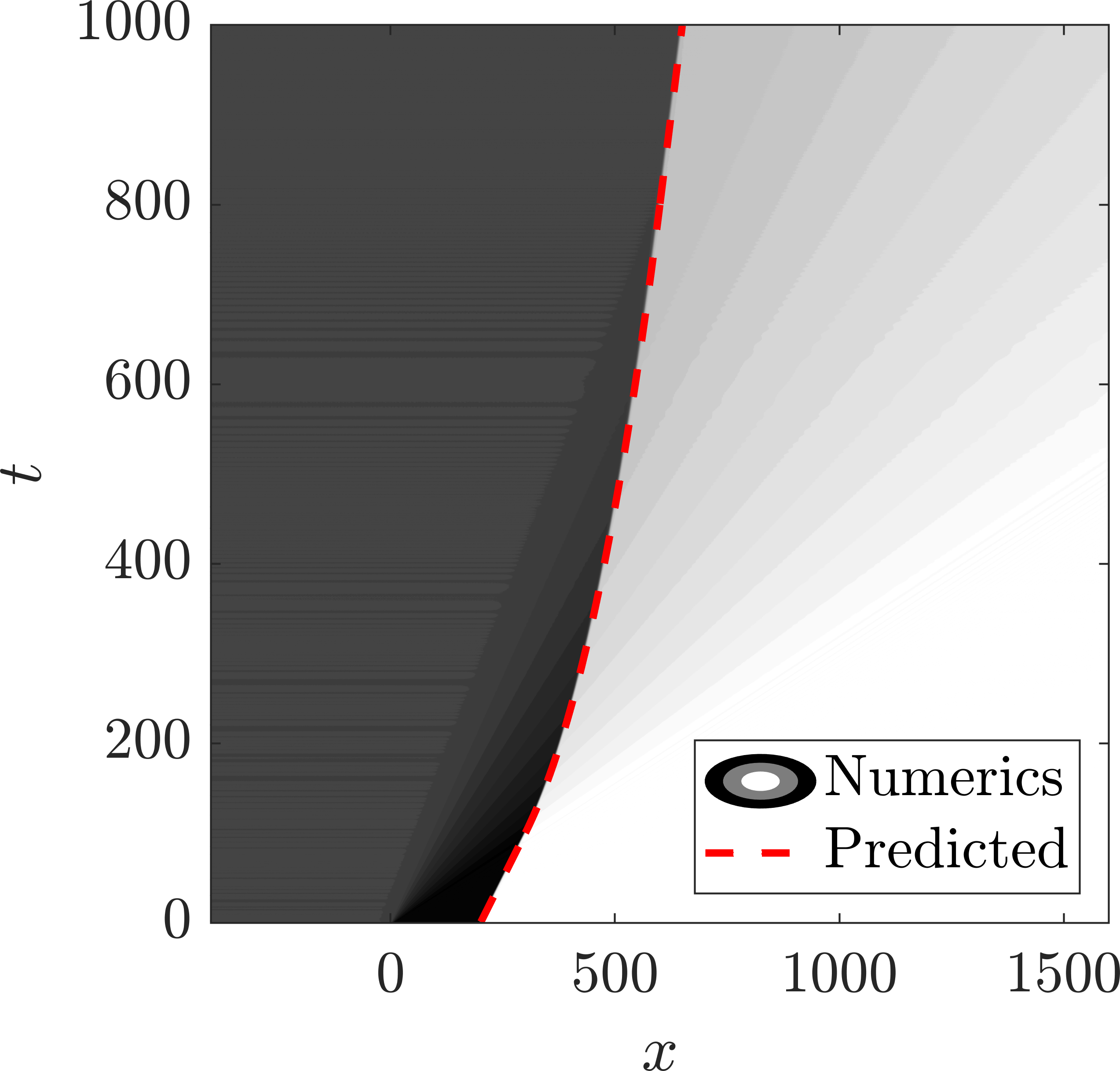}
  \end{subfigure}		
  \begin{subfigure}{0.27\textwidth}
    \centering
    \includegraphics[width=\textwidth]{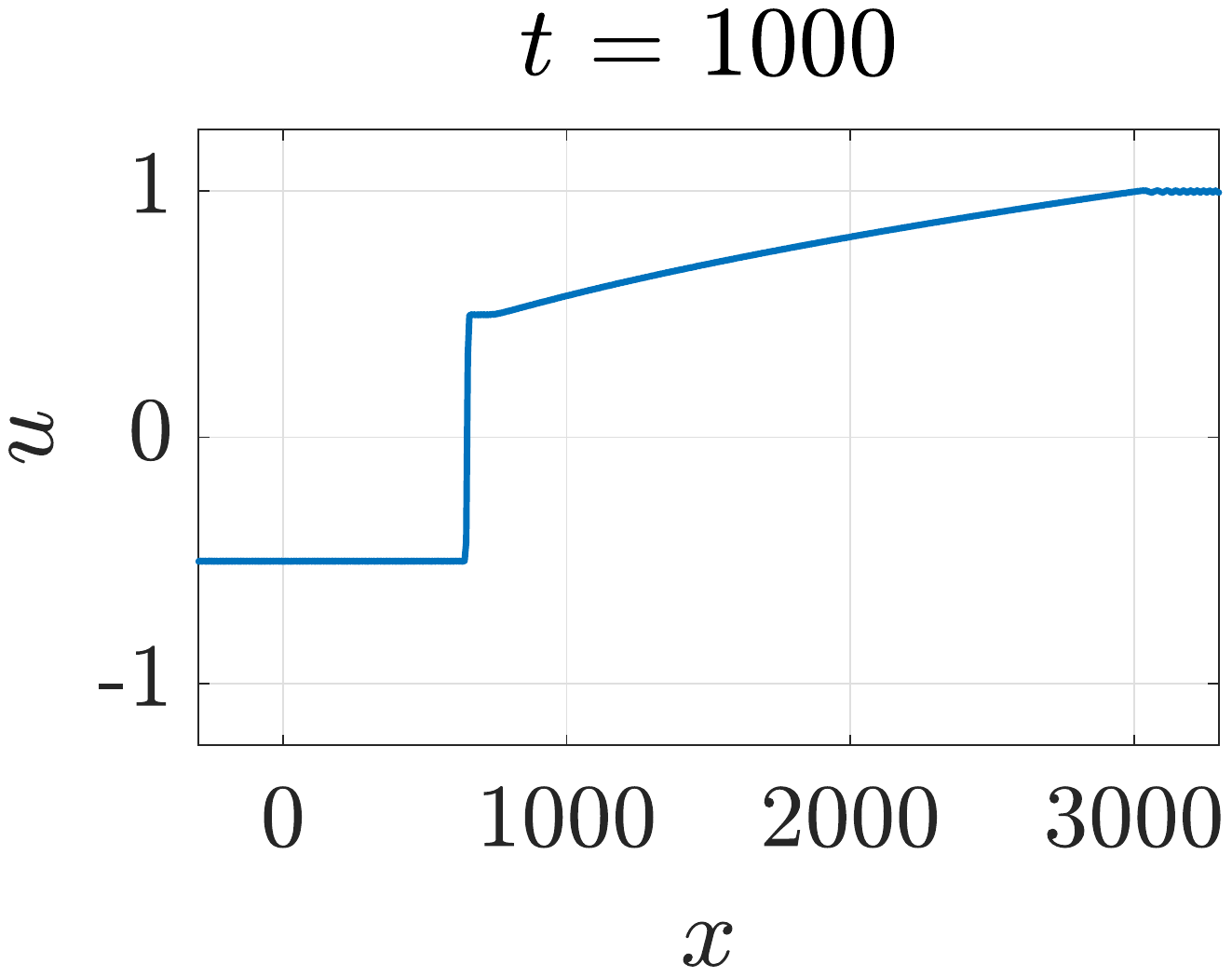}
  \end{subfigure}
  \caption{Backward kink-RW interaction with $\mu = 1$, $a_+ = 2$,
    $x_+ = 200$, $u_- = -0.5$ and $u_+ = -1$. The predicted $a_-$ is
    1 and the predicted $\Delta$ is 200. The numerical solution gives $a_- = 0.9984$ and $\Delta = 199.02$ at $t = 1000$. }
  \label{fig:kinkRW}
\end{figure}

\medskip The kink trajectory during interaction with a RW is the shock
trajectory and can be solved for explicitly using \eqref{eqn:RWsoln}
and \eqref{eqn:kinktraject}, giving
\begin{equation}
  \frac{\mathrm{d}x}{\mathrm{d}t} = \frac{x}{3t}, \qquad 3u_-^2 t < x
  < 3u_+^2 t, 
\end{equation}
with the initial position given by $x_0 = x_+ > 0$.  Numerical
experiments verify that this is true, as seen in
Fig.~\ref{fig:kinkRW}. As the kink travels through, the RW switches
polarity.
	
\begin{figure}
  \centering
  \begin{subfigure}{0.27\textwidth}
    \centering
    \includegraphics[width=\textwidth]{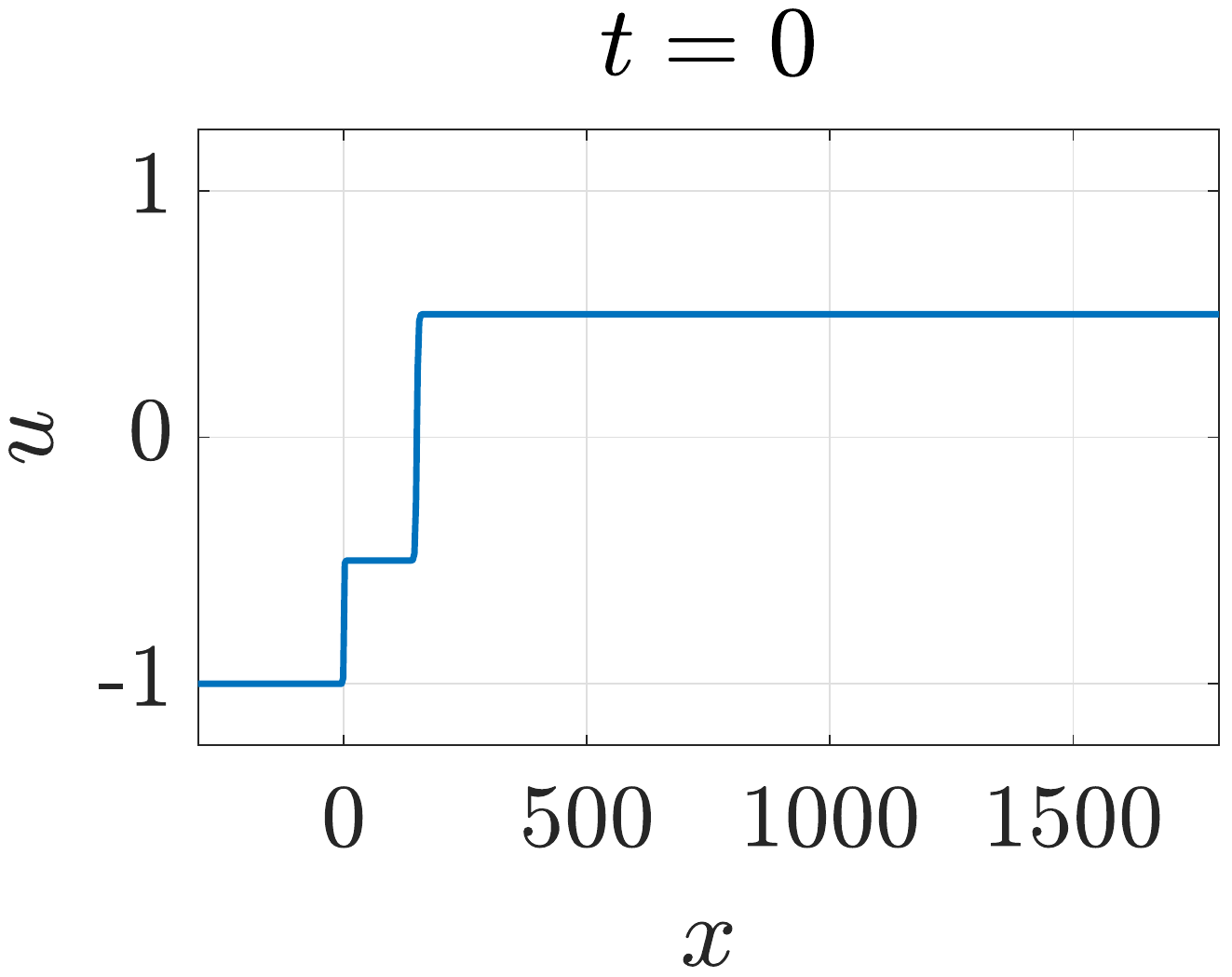}
  \end{subfigure}
  \begin{subfigure}{0.4\textwidth}
    \centering
  \includegraphics[width=\textwidth]{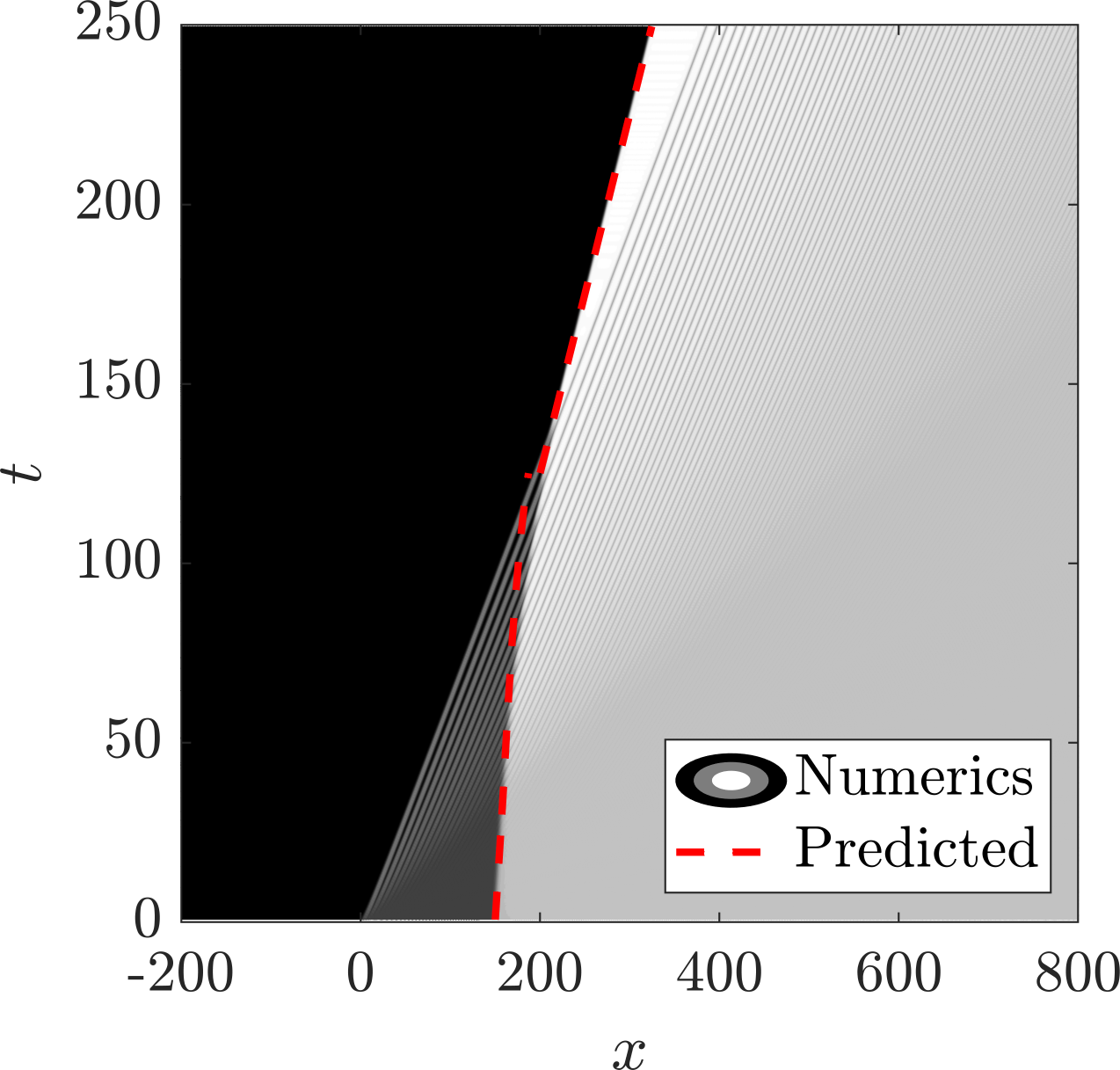}
\end{subfigure}		
\begin{subfigure}{0.27\textwidth}
  \centering
  \includegraphics[width=\textwidth]{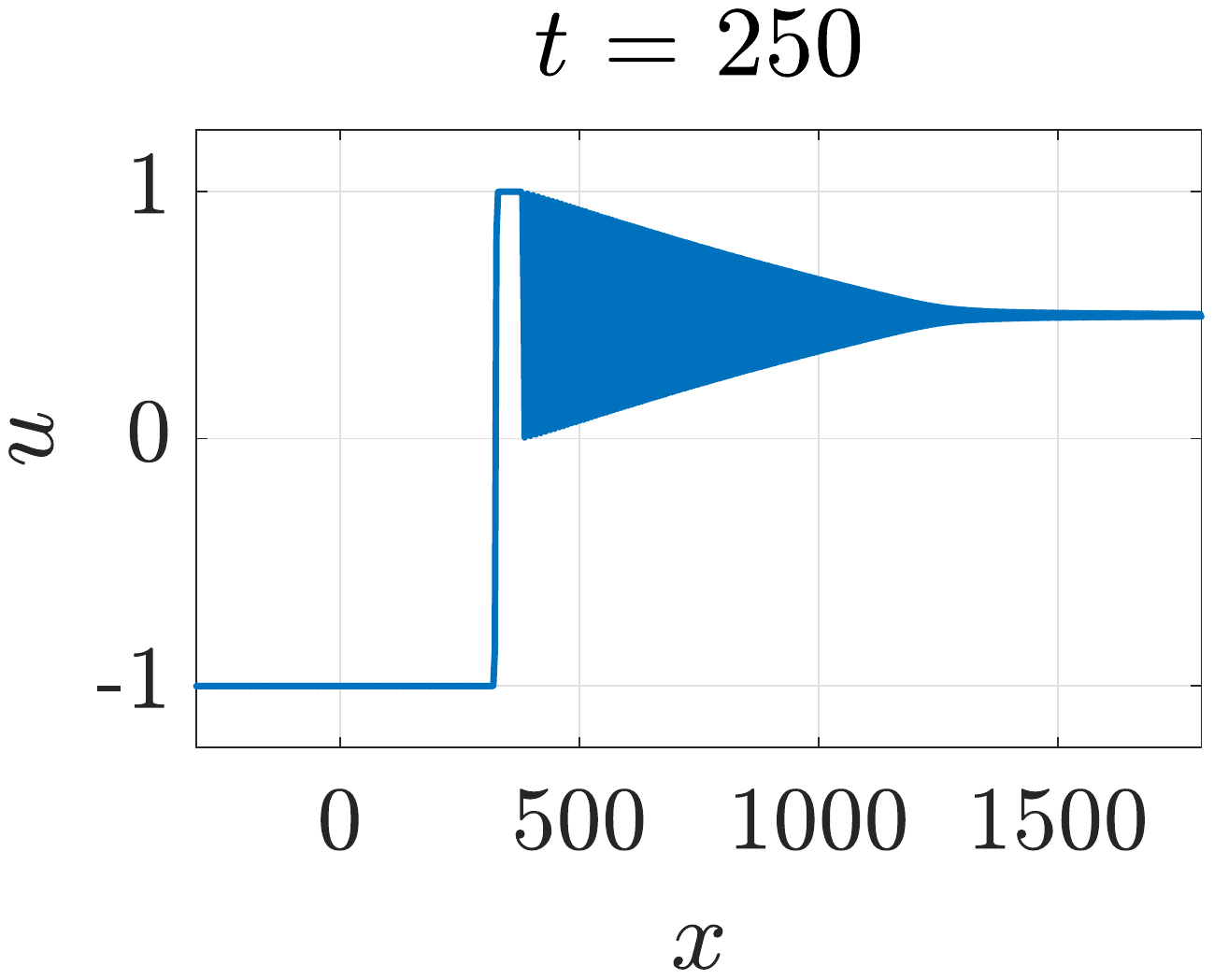}
\end{subfigure}
\caption{Kink-DSW interaction with $\mu = 1$, $a_+ = 1$. $x_+ = 150$,
  $u_- = -1$ and $u_+ = -0.5$. The predicted $a_-$ is 2 and the predicted $\Delta$ is -75. The
  numerical solution gives $a_- = 2.0022$ and $\Delta = -74.56$ at $t = 250$. }
\label{fig:kinkDSW}
\end{figure}

Again for DSWs, interaction with the kink causes the DSW to switch
polarity as seen in the numerical experiment of
Fig.~\ref{fig:kinkDSW}. These polarity switches are only possible due
to nonconvexity. The kink-DSW trajectory is given by
eq.~\eqref{eqn:kinktraject}, where the DSW mean flow
$\bar{u} = \bar{u}(x/t)$ is determined by
\eqref{eqn:mupos_dswmeanflow}, \eqref{GP1} and \eqref{eqn:mupos_W2}.

Note that kink-kink interaction is not possible as multiple kinks will
co-propagate.
		
\section{Generalisation to arbitrary soliton-convex mean flows}
\label{sec:arbitrarymeanflow}
	
We have described soliton tunnelling interactions specifically with
mean flows that emerge from a Riemann step-type initial
condition. However, the tunnelling problem can be generalised to
determine the phase shift and amplitude of a soliton that tunnels
through an arbitrary mean hydrodynamic flow. If tunnelling occurs,
only the far field mean flow conditions $u_-$ and $u_+$ are needed to
predict the transmitted soliton amplitude. The phase shift can be
calculated by approximating the initial mean flow $\bar{u}(x,0)$ with a
series of step functions and taking a limit that results in the
Riemann integral
\begin{equation}
  \Delta = \left\{
    \begin{array}{ll}
      \displaystyle \int_{x_-}^{x_0}
      \left(\sqrt{\frac{\bar{u}(x,0)^2 - q^2}{u_-^2 -q^2}}
      -1 \right)dx, &  \text{ assuming } R \to L \\
      \displaystyle \int_{x_0}^{x_+} \left(
      \sqrt{\frac{\bar{u}(x,0)^2 - q^2}{u_+^2 -q^2}} -1
      \right)dx, & \text{  assuming } L \to R 
    \end{array}
  \right.
\end{equation}
where $x_0$ is the initial soliton position and $x_-$ or $x_+$ is a
point where the mean flow has equilibrated to the far field constant.
	
This phase shift calculation holds only when the system remains
strictly hyperbolic, otherwise we have $\bar{u} \to q$ as the soliton
travels through the mean flow and trapping will occur. However, this
can also be predicted from the far field mean flow conditions, as we
have done throughout this work.
	
\begin{figure}
  \centering
  \begin{subfigure}{0.27\textwidth}
    \centering
    \includegraphics[width=\textwidth]{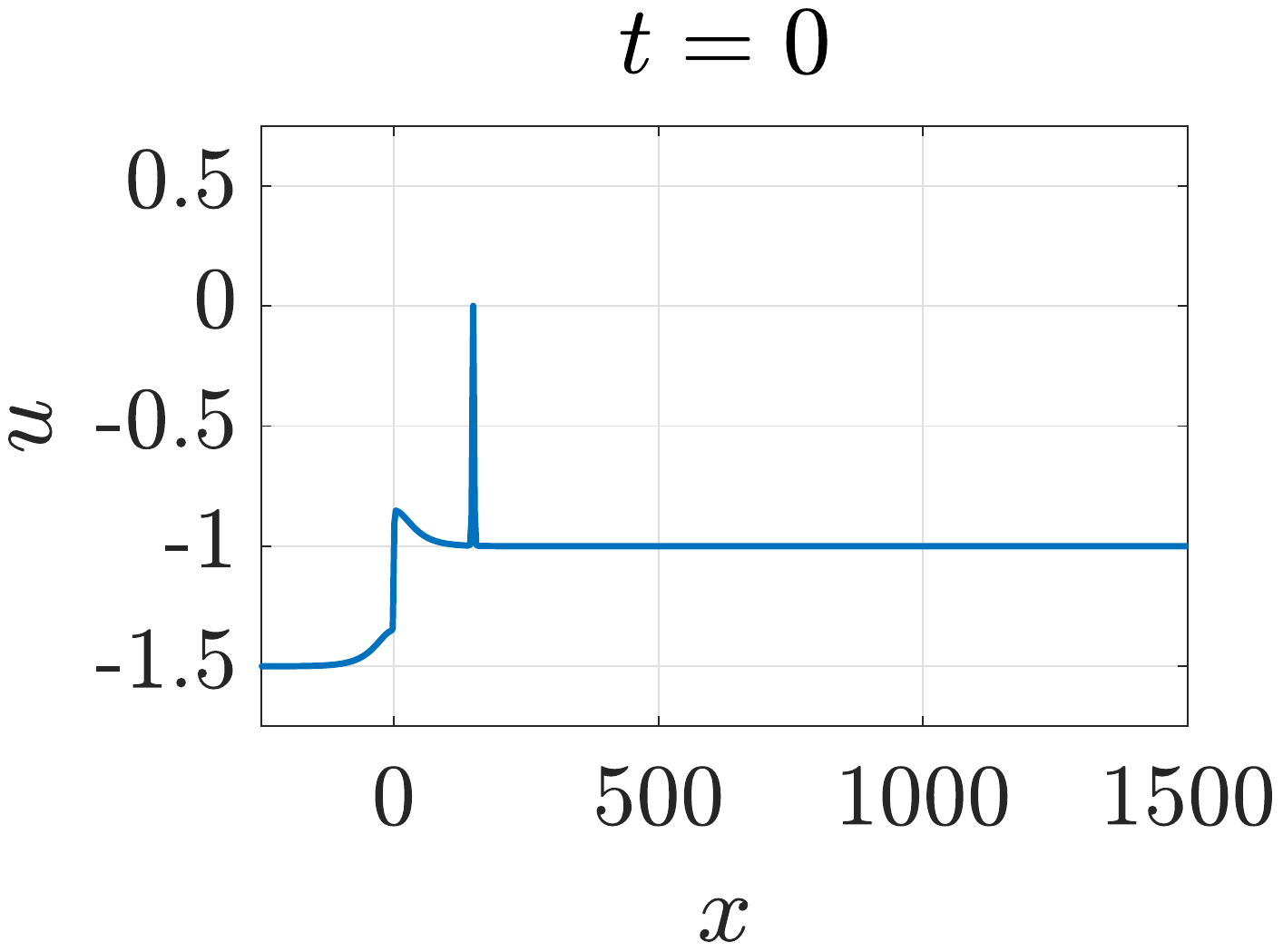}
  \end{subfigure}
  \begin{subfigure}{0.4\textwidth}
    \centering
    \includegraphics[width=\textwidth]{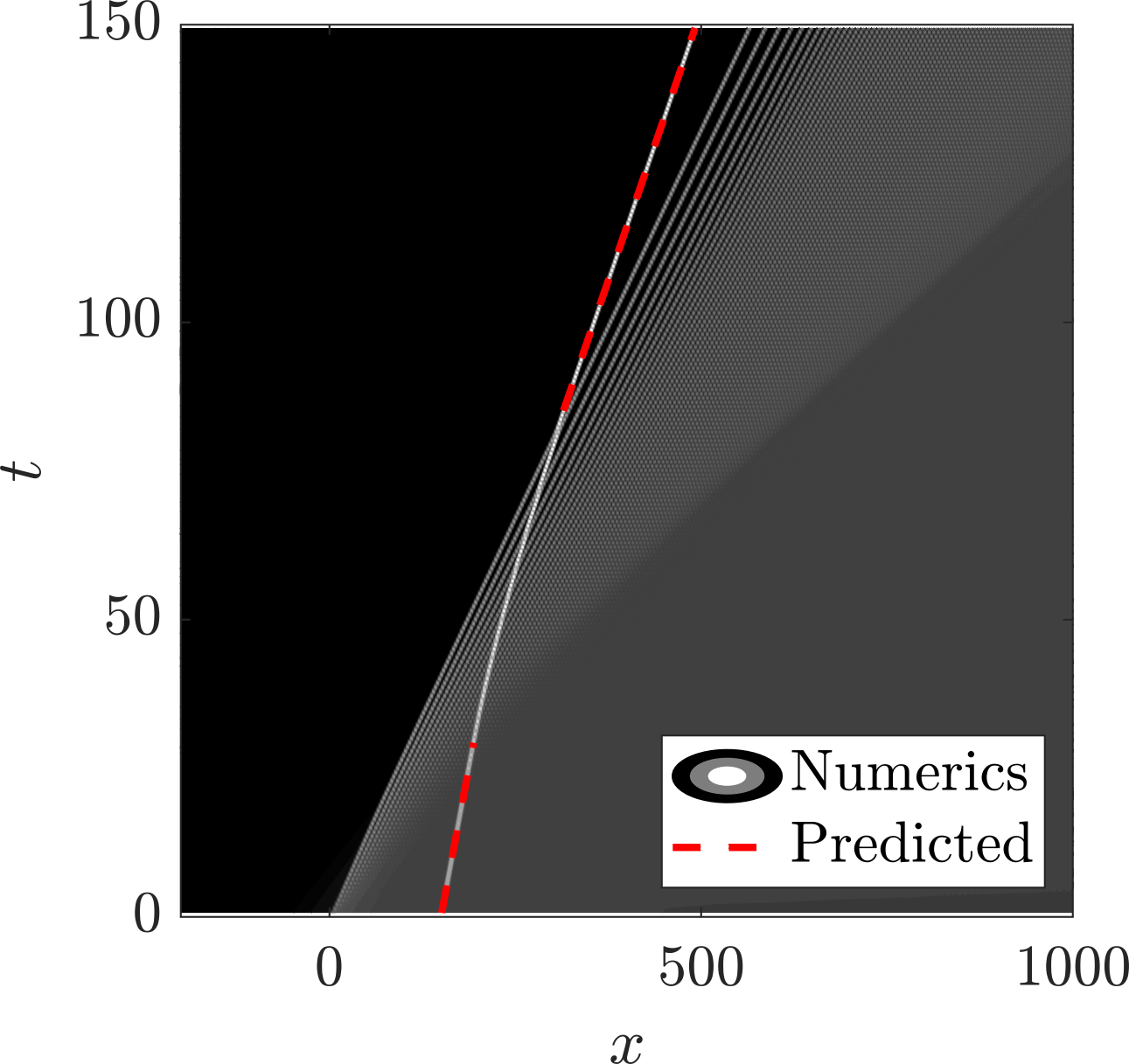}
  \end{subfigure}		
  \begin{subfigure}{0.27\textwidth}
    \centering
    \includegraphics[width=\textwidth]{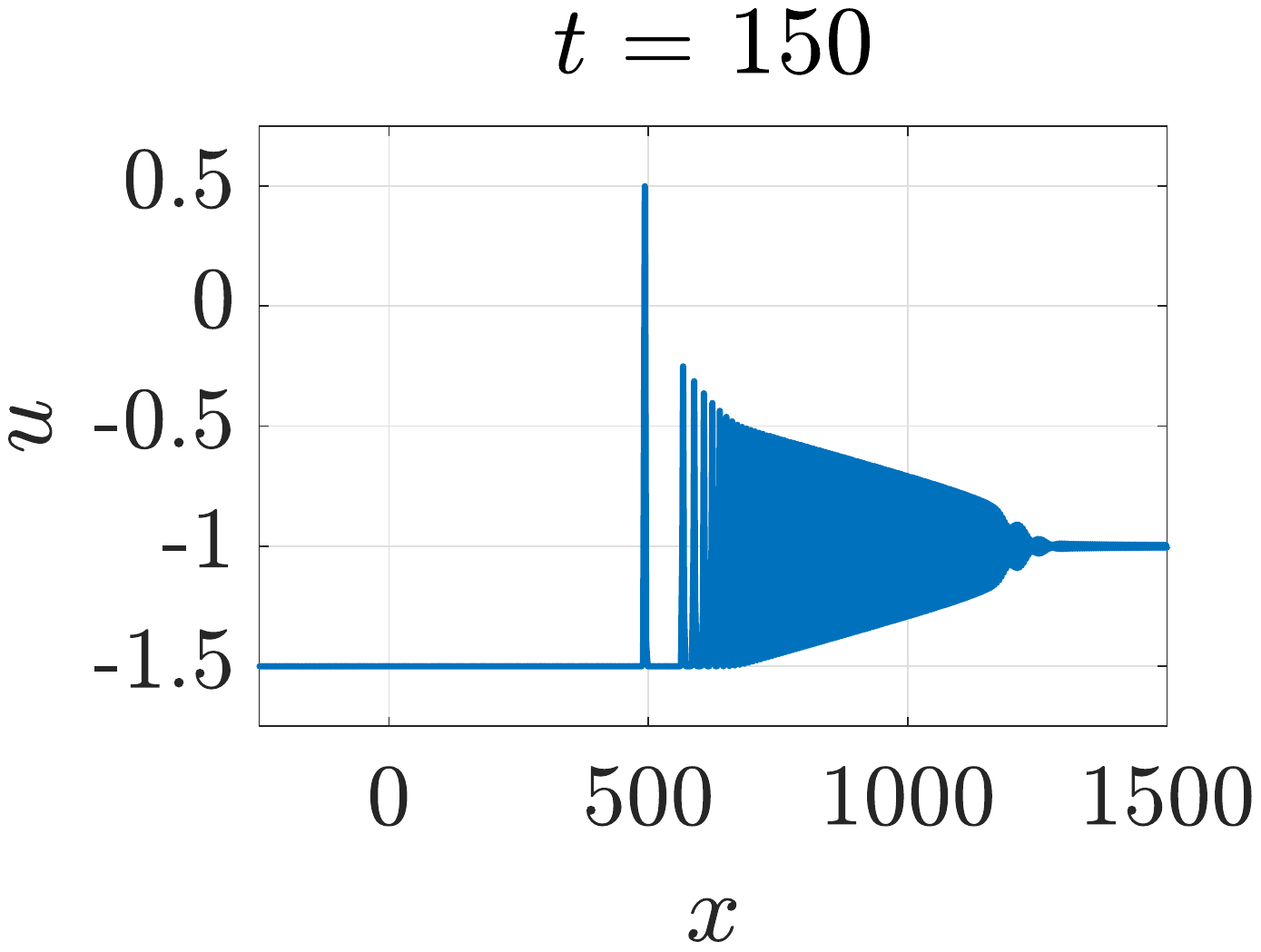}
  \end{subfigure}
  \caption{Soliton interaction with an arbitrary mean flow with
    $\mu = -1$, $x_+ = 150$, $a_+ = 1$, $u_- = -1.5$, and $u_+ =
    -1$. The predicted amplitude after tunnelling is 2 and the predicted phase shift is -69.28. The
    numerical result is $a_- = 2.028$ and $\Delta = -69.24$ at $t = 150$.}
  \label{fig:generalmeanflow}
\end{figure}

Figure \ref{fig:generalmeanflow} depicts a numerical example of a
soliton tunnelling through a mean flow that is a combination of a
Gaussian and a Riemann step. The predicted trajectory pre and post
mean flow interaction shows good agreement with the numerics, as does
the transmitted amplitude.
	
\section{Conclusions and Outlook}
\label{sec:conclusion} 	

In this work, we have investigated the impact of hydrodynamic flux
nonconvexity on the relatively new type of soliton-mean flow
interaction in which both the soliton and the slowly varying mean flow
are governed by the same nonlinear dispersive PDE.  The role of mean
flows in such solitonic dispersive hydrodynamics is played by either
rarefaction waves (RWs) or dispersive shock waves (DSWs). In the
latter case, the slowly varying mean flow occurs as a result of
averaging over fast, locally periodic nonlinear oscillations of the
DSW wave field. This kind of wave-mean flow interaction has been
studied previously in the framework of dispersive hydrodynamics
described by equations of the Kortweg-de Vries (KdV) type, whose
hydrodynamic flux satisfies the convexity property $f''(u) \ne 0$
\cite{maiden_solitonic_2018, sprenger_hydrodynamic_2018}.

As a basic model acutely capturing the effects of nonconvex flux on
solitonic dispersive hydrodynamics, we used the modified KdV (mKdV)
equation, a canonical equation describing, in particular, internal
waves in stratified fluids.  We considered both focusing and
defocusing variants of the mKdV equation, which exhibit rather
differing properties in the nonconvex propagation regime. Along with
KdV type solitary waves, RWs and DSWs, the mKdV equation supports
nonclassical wave structures including kinks, algebraic solitons and
contact DSWs \cite{el_dispersive_2017}. Due to the variety of solitary
waves and mean flow configurations, the mKdV equation exhibits a much
richer classification of soliton-mean flow interactions than the KdV
equation, and more, generally, dispersive hydrodynamics with convex
flux.

Multiscale, soliton-mean flow interactions are studied using Whitham
modulation theory. An accurate and effective description of
soliton-mean flow interactions is achieved by the soliton reduction of
the Whitham modulation system, which we formulate in a general form
for scalar dispersive hydrodynamics with nonconvex flux.  Expanding on
the results of \cite{maiden_solitonic_2018}, we show that the
solitonic modulation system with nonconvex flux admits two invariants
that yield the amplitude and phase conditions relating the soliton
parameters pre and post interaction with the expanding mean flow
between distinct boundary states at $\pm \infty$.

One of the main results of our study is the general admissibility
criterion for a soliton to tunnel through a mean flow, which is
formulated in terms of maintenance of strict hyperbolicity (distinct
characteristic speeds) of the solitonic modulation system
\eqref{eqn:2by2modulationsystem} through the interaction. Conversely,
trapping occurs when the characteristics of the solitonic system
coalesce.

The general construction of the solitonic modulation system for
dispersive hydrodynamics with nonconvex flux is realised for the mKdV
equation for both dispersion signs. The nonconvexity of the
hydrodynamic flux in the mKdV equation allows for both polarities of
solitons and DSWs to exist, no matter the sign of dispersion.  This is
demonstrated to lead to new, interesting interaction behaviour, which
includes soliton polarity reversal as a result of tunnelling through a
kink mean flow. On the other hand, kinks are heteroclinic solitary
waves that tunnel through DSWs and RW mean flows accompanied by the
mean flow's polarity change.

An important application of this analysis is to the nonlinear dynamics
of internal ocean waves, where solitons of both polarities and mean
flows such as RWs and DSWs (undular bores) are often observed
\cite{apel_oceanic_2002}.  Along with the mKdV equation, internal
waves can be modelled by the Gardner equation that combines KdV and
mKdV hydrodynamic fluxes \cite{grimshaw_internal_2002}. The
generalisation of our results to the Gardner equation is
straightforward.  We stress that our approach does not make use of the
integrability of the mKdV equation, so can be applied to
nonintegrable, nonconvex dispersive hydrodynamics. In particular, a
new, intringuing nonconvex scalar model has recently been derived for
long-wave potential-vorticity dynamics of coastal fronts in
\cite{jamshidi_long-wave_2020}. The study of soliton-mean flow
interaction for this model and its fluid dynamic implications would be
an interesting and relevant application of the theory presented here.

Our theory can also be extended to bidirectional wave systems
describing internal gravity waves in fluids with the
Miyata-Choi-Camassa equations \cite{choi_fully_1999} being an obvious
candidate model.  Other applications include polarisation waves in
two-component Bose-Einstein condensates \cite{congy_nonlinear_2016,
  ivanov_solution_2017}, nonlinear optics described by nonconvex
equations \cite{ivanov2020_chenleeliu, ivanov2017_riemann} and
collisionless plasma \cite{nakamura_experiments_1985,
  chanteur_formation_1987, ruderman_dynamics_2008}. Yet another
possible application is in the hydrodynamic interpretation of
far-from-equilibrium nonlinear magnetisation dynamics such as in
\cite{iacocca_breaking_2017}, where solitons and DSWs can also emerge.
	
Future directions include considering dispersion as an additional
source of nonconvexity. Such systems are abundant in geophysical
fluids, describing magma and glacier flows
\cite{scott_observations_1986, stubblefield_solitary_2020} or wave-ice
sheet interactions \cite{ilichev_soliton-like_2015}. Indeed, recent
works on DSWs in systems with nonconvex dispersion
\cite{lowman_dispersive_2013, el_radiating_2016, sprenger_shock_2017,
  hoefer_modulation_2019, baqer_modulation_2020,
  congy_dispersive_2020} reveal a plethora of unusual behaviours that
can lead to new, interesting soliton-mean flow interactions.

\section*{Acknowledgements}

The work of GAE was partially supported by the EPSRC grant
EP/R00515X/2. The work of MAH and KVDS was partially supported by NSF
grant DMS-13012308.



\bibliographystyle{plain}

\end{document}